\documentclass[11pt]{article}
\pdfoutput = 1
\usepackage[utf8]{inputenc}
\usepackage{amsmath,amsfonts,amssymb}
\usepackage{amsthm}
\usepackage{bbold}
\usepackage{latexsym}
\usepackage[noadjust]{cite}
\usepackage{graphicx}
\usepackage{xcolor}
\usepackage{dirtytalk}
\usepackage{mathtools}
\usepackage[margin=1in]{geometry}
\usepackage{thm-restate}
\usepackage{enumitem}
\usepackage[linesnumbered,ruled]{algorithm2e}
\usepackage[colorlinks=true, allcolors=blue]{hyperref}
\usepackage[nameinlink,capitalise]{cleveref}
\hypersetup{
    citecolor={violet}
}

\newtheorem{theorem}{Theorem}[section]
\newtheorem{definition}{Definition}[section]
\newtheorem{corollary}[theorem]{Corollary}
\newtheorem{lemma}[theorem]{Lemma}
\newtheorem{claim}[theorem]{Claim}
\newtheorem{remark}[theorem]{Remark}

\theoremstyle{definition}

\newtheorem{fact}[theorem]{Fact}

\newcommand{\F}{\mathbb{F}}

\newcommand{\E}{\mathbb{E}}
\newcommand{\R}{\mathbb{R}}
\newcommand{\Z}{\mathbb{Z}}
\newcommand{\calD}{\mathcal{D}}
\newcommand{\calC}{\mathcal{C}}
\newcommand{\calc}{\calC}

\newcommand{\calL}{\mathcal{L}}
\newcommand{\Q}{\mathbb{Q}}
\newcommand{\dual}{\text{dual}}
\newcommand{\rank}{\text{rank}}
\newcommand{\supp}{\text{supp}}

\newcommand{\zo}{\{0, 1\}}

\newcommand{\Density}{\mathrm{Density}}
\newcommand{\Contract}{\mathrm{Contract}}
\newcommand{\ContractAbelian}{\mathrm{ContractAbelian}}

\newcommand{\eps}{\epsilon}

\newcommand{\ra}{\rightarrow}

\newcommand{\wt}{\mathrm{wt}}

\newcommand{\cut}{\text{cut}}

\newcommand{\zivny}{\v Zivn\'y}

\newcommand{\Supp}{\mathrm{Supp}}

\newcommand{\polylog}{\mathrm{polylog}}
\newcommand{\Deven}{\calD_{\text{even}}}
\newcommand{\Dodd}{\calD_{\text{odd}}}
\newcommand{\teveni}{n_{\text{even}, i}}

\newcommand{\CSP}{\textrm{CSP}}
\newcommand{\AND}{\textrm{AND}}

\newif\ifdraft
\newif\ifanon

\drafttrue
\anonfalse

\title{Efficient Algorithms and New Characterizations for CSP Sparsification}

\ifanon
\else
\author{Sanjeev Khanna\thanks{School of Engineering and Applied Sciences, University of Pennsylvania, Philadelphia, PA. Email: {\tt sanjeev@cis.upenn.edu}. Supported in part by NSF awards CCF-1934876 and CCF-2008305.} \and Aaron (Louie) Putterman\thanks{School of Engineering and Applied Sciences, Harvard University, Cambridge, Massachusetts, USA. Supported in part by the Simons Investigator Awards of Madhu Sudan and Salil Vadhan, NSF Award CCF 2152413 and a Hudson River Trading PhD Research Scholarship. Email: \texttt{aputterman@g.harvard.edu}.} \and Madhu Sudan\thanks{School of Engineering and Applied Sciences, Harvard University, Cambridge, Massachusetts, USA. Supported in part by a Simons Investigator Award and NSF Award CCF 2152413. Email: \texttt{madhu@cs.harvard.edu}.}}
\date{November 5, 2024}
\fi

\begin{document}

\pagenumbering{gobble}
\thispagestyle{empty}

\maketitle

\vspace*{-.3in}

\begin{abstract}
CSP sparsification, introduced by Kogan and Krauthgamer (ITCS 2015), considers the following question: how much can an instance of a constraint satisfaction problem be sparsified (by retaining a reweighted subset of the constraints) while still roughly capturing the weight of constraints satisfied by {\em every} assignment. CSP sparsification captures as a special case several well-studied problems including graph cut-sparsification, hypergraph cut-sparsification, hypergraph XOR-sparsification, and corresponds to a general class of hypergraph sparsification problems where an arbitrary $0/1$-valued {\em splitting function} is used to define the notion of cutting a hyperedge (see, for instance, Veldt-Benson-Kleinberg SIAM Review 2022).
The main question here is to understand, for a given constraint predicate $P:\Sigma^r \to \{0,1\}$ (where variables are assigned values in $\Sigma$), the smallest constant $c$ such that $\widetilde{O}(n^c)$ sized sparsifiers exist for every instance of a constraint satisfaction problem over $P$. 
A recent work of Khanna, Putterman and Sudan (SODA 2024) [KPS24] showed {\em existence} of near-linear size sparsifiers for new classes of CSPs. 
In this work 
(1) we significantly extend the class of CSPs for which nearly linear-size sparsifications can be shown to exist while also extending the scope to settings with non-linear-sized sparsifications; 
(2) we give a polynomial-time algorithm to extract such sparsifications for all the problems we study including the first efficient sparsification algorithms for the problems studied in [KPS24]. 
 
Our results captured in item (1) lead to two new classifications: First we get a complete classification of all symmetric Boolean predicates $P$ (i.e., on the Boolean domain $\Sigma = \{0,1\}$) that allow nearly-linear-size sparsifications. This classification reveals an inherent, and previously unsuspected, number-theoretic phenomenon that determines near-linear size sparsifiability. Second, we also completely classify the set of Boolean predicates $P$ that allow non-trivial ($o(n^r)$-size) sparsifications, thus answering an open question from the work of Kogan and Krauthgamer. 

The constructive aspect of our result is an arguably unexpected strengthening of [KPS24]. Their work roughly seemed to suggest that sparsifications can be found by solving problems related to finding the minimum distance of linear codes. These problems remain unsolved to this date and our work finds a different path to achieve poly-time sparsification, resolving an open problem from their work. As a consquence we also get the first efficient algorithms to spectrally sparsify Cayley graphs over $\F_2^n$ in time polynomial in the number of generators. Our techniques build on [KPS24] which proves the existence of nearly-linear size sparsifiers for CSPs where the unsatisfying assignments of the underlying predicate $P$ are given by a linear equation over a finite field. Our main contributions are to extend this framework to \emph{higher-degree equations} over \emph{general Abelian groups} (both elements are crucial for our classification results) as well as designing polynomial-time sparsification algorithms for all problems in our framework.
\end{abstract}

\newpage
{\small \tableofcontents}
\newpage

\pagenumbering{arabic}

\section{Introduction}

In this work we study the problem of ``CSP sparsification'' and give the first efficient algorithms and characterizations for sparsifying many classes of CSPs. Our contributions yield immediate, novel results in efficient Cayley graph sparsification \cite{KPS24}, efficient hedge-graph sparsification \cite{GKP17} and sparsifying hypergraphs with cardinality-based splitting functions \cite{VBK22}. We start by introducing CSP sparsification more precisely before moving on to the motivation, our results, and the applications of these results.

\subsection{CSP sparsification}

CSP sparsification was introduced by Kogan and Krauthgamer~\cite{KK15} as a broad extension of the notion of cut sparsification in graphs. A cut sparsifier of a graph is a weighted subgraph of a given graph on the same set of vertices that roughly preserves the size of every cut. The seminal works of Karger~\cite{Kar93} and Bencz\'ur and Karger~\cite{BK96} showed that every undirected graph admits a cut sparsifier of size nearly linear in the number of vertices, thus potentially compressing graph representations by a nearly linear factor while preserving some significant information. CSP sparsification aims to extend this study to broader classes of structures than just graphs, and aims to preserve a broader class of queries than cuts. (We remark here that this is just one of several directions of generalizations. Other notable directions such as spectral sparsification (\cite{BSS09, ST11, Lee23, JLS22, JLLS23}) and general submodular hypergraph sparsification (\cite{KK23, JLLS23}) are not covered in this work.)

In this work, a CSP problem, $\CSP(P)$,  is specified by a predicate $P:\Sigma^r \to \{0,1\}$. An instance $\Phi$ of $\CSP(P)$ on $n$ variables is given by $m$ weighted constraints $(w_1,C_1),\ldots,(w_m,C_m)$ where each constraint $C_j$ applies the constraint $P$ to a specified sequence $(i_1(j),\ldots,i_r(j)) \in [n]^r$ 
of $r$ out of the $n$ variables. An assignment to the variables is given by $a \in \Sigma^n$ and $a$ satisfies constraint $C_j$ if $P(a_{i_1(j)},\ldots,a_{i_r(j)}) = 1$. In CSP sparsification, an instance $\Phi$ is viewed as the specification of a structure that we wish to compress with the goal that the compressed representation approximately preserves the total weight of satisfied constraints for {\em every assignment} $a \in \Sigma^n$. 
A {\em sparsification} of $\Phi$ is obtained by retaining a subset of the constraints, possibly assigning them new weights, and the size of the sparsification is the number of constraints retained.
\subsection{Motivation}

Beyond its immediate interest as a natural combinatorial question, CSP sparsification is motivated by the need to simplify large complex networks represented by hypergraphs.
While graphs model pairwise interactions, hypergraphs are needed to model interactions between larger subsets of entities. The work of Veldt, Benson and Kleinberg~\cite{VBK22} describes a broad collection of settings where the central interactions require this greater flexibility to model.
Furthermore, while graphs admit only one interesting notion of cutting an edge,
 with hyperedges one can imagine more complex notions. Typically, this is referred to as \emph{generalized hypergraph sparsification} (see for instance, \cite{VBK22}), where each hyperedge $e \subseteq V$ is equipped with a splitting function $g_e: 2^e \rightarrow \R^+$. Given a cut $S \subseteq V$, the contribution of a hyperedge $e$ to the cut is exactly $g_e(S \cap e)$, and the total cut size is $\sum_{e \in E} g_e(S \cap e)$. For instance, by setting the splitting function to be $0$ on inputs $e$ and $\emptyset$, and $1$ otherwise, this models the standard notion of cuts in hypergraphs.
By allowing more general notions of hyperedge cuts, say, that is, $g_e$ is an arbitrary function of $S \cap e$, this model captures a variety of applications, ranging from scientific computing on sparse matrices \cite{BDKS16}, to clustering and machine learning \cite{YNYNLT19, ZHS06}, to modelling transistors and other circuitry elements in circuit design \cite{AK95, DLaw73}, and even to human behavior and biological interactions (see \cite{VBK22}). In each of these applications, the ability to sparsify the hypergraph while still preserving cut-sizes is a key building block, as these dramatically decrease the memory footprint (and complexity) of the networks being optimized. 

Within this general framework, boolean CSP sparsification captures the important and expressive case where the splitting functions are $\zo$-valued, (i.e., each $g_e: 2^e \rightarrow \{0,1\}$). This richness of expressibility in CSPs leads to the question (raised in \cite{KK15}) of for which CSPs (i.e., for which predicates $P$) is $\CSP(P)$ sparsifiable to nearly linear size, or more generally, what is the best (possibly non-integral) exponent $\ell$ such that $\CSP(P)$ is sparsifiable to instances of size roughly $n^\ell$ on instances with $n$ variables.\footnote{In the literature on CSP decision or maximization problems, a number of further subtleties arise. Ideally one would like sparsifications for {\em families} or predicates as opposed to a single predicate. There is a difference between whether constraints must be application of the predicate to distinct variables or we allow repetition of variables in a constraint. In the Boolean setting there is a difference between predicates being applied to variables versus literals. None of these issues is relevant in the case of CSP {\em sparsification} since the task is rich enough to allow easy reductions among all these problems.}  In Kogan and Krauthgamer's work \cite{KK15}, they showed that for any $r$-CNF constraint, CSPs admit sparsifiers of size $\widetilde{O}(rn / \eps^2)$. Filtser and Krauthgamer~\cite{FK17} later gave a complete characterization of Boolean CSPs on two variables ($|\Sigma| = r = 2$) establishing a dichotomy result that shows that each predicate $P$ either allows for near-linear size sparsification or requires quadratic size sparsifiers (which is trivially achievable since that is the number of distinct constraints). Their result was extended to binary ($r$=2) CSPs over all finite alphabets $\Sigma$ by Butti and \zivny~\cite{BZ20}, thus giving the first classification for an infinite subclass of CSPs. A recent work of \ifanon Khanna, Putterman and Sudan~\cite{KPS24} \else the authors~\cite{KPS24} \fi extended the classification of \cite{KK15} to the case of ternary Boolean CSPs ($r=3$, $\Sigma = \{0,1\}$) but only identified the predicates that allow nearly linear sparsification while showing the rest required at least quadratic size. Note that the trivial sparsification in this case has size $O(n^3)$. The central notion developed in their work is sparsifiers of linear systems of equations over finite fields (aka ``code sparsifiers''). Their results are non-constructive in that they only show existence of nearly-linear size sparsifiers, and did not give a polynomial time algorithm to find them. In fact, their work highlighted some natural obstacles to achieving efficient sparsification.

In this work we extend the work of \cite{KPS24} in two directions: (1) We give a polynomial time algorithm for constructing sparsifiers of linear systems of equations, thereby making the results of \cite{KPS24} constructive. (2) We extend the (constructive) sparsification from linear equations over finite fields to linear and higher degree equations over abelian groups. These generalizations allow us to get new (efficiently constructive) dichotomies for sparsifiability of some (infinite) subclasses of CSPs.

\subsection{Our Results}

We start by describing our new results, and defer the new technical ingredients to the following section. 

\paragraph{Efficient Code Sparsification}

Our main technique extends a technique called ``code sparsification'' introduced in \cite{KPS24}. The primary structure of interest in code sparsification is a linear code $C \subset \F_q^m$ over some finite field $\F_q$. A \emph{sparsifier} for $C$ is a restriction $C|_S$ (also called ``puncturing'') of the code $C$ to a subset $S \subseteq [m]$ of the coordinates along with weights on the coordinates so that the weighted Hamming weight of each codeword in $C|_S$ is approximately the same as the weight of the corresponding codeword in $C$. \cite{KPS24} show that for every code of dimension $n$ there \emph{exists} a sparsifier of size nearly-linear in $n$. Code sparsification turns out to be a powerful tool for CSP sparsification and indeed is used in \cite{KPS24} to get nearly linear-size sparsifiers for many classes of CSPs. However, the result of \cite{KPS24} is only {\em existential} and indeed seems to require solving NP-hard problems to get algorithmic code sparsification. We remedy this problem by giving an \emph{efficient} algorithmic implementation of code-sparsification:

\begin{theorem}
    For any code $C \subset \F_q^m$ of dimension $n$ and a parameter $\eps \in (0,1)$, there is a polynomial time (in $n,m, \log(q), \eps^{-1}$) randomized algorithm for computing (with high probability) a $(1 \pm \eps)$ code-sparsifier $C|_S$ of $C$, with $|S| = \widetilde{O}(n / \eps^2)$.
\end{theorem}

This settles a key open question of \cite{KPS24} regarding tractability of computing code-sparsifiers and leads to efficient algorithms for the sparsifiers constructed in their work. For instance it leads to the first efficient algorithm to sparsify a Cayley graph over $\F_2^n$ into another Cayley graph with roughly the same spectrum, in the sense of \cite{ST11}, where efficiency is with respect to the number of generators (and not the size of the graph)! 

\begin{corollary}
    For any Cayley graph $G$ over $\F_2^n$ with associated generating set $Q \subseteq \F_2^n$ with $|Q| = m$ and a parameter $\eps \in (0,1)$ of our choosing, there is a polynomial time (in $n,m, \eps^{-1}$) randomized algorithm for computing a $(1 \pm \eps)$ \emph{Cayley-graph spectral-sparsifier} $\tilde{G} = \mathrm{Cay}(\F_2^n, \hat{Q})$, where $|\hat{Q}| = \widetilde{O}(n / \eps^2)$ is a re-weighted subset of the generators.
\end{corollary}

We elaborate on this result in \cref{sec:cayleyGraph}, and also show a novel generalization to Cayley graphs over $\Z_q^n$ in \cref{sec:cayleyGraphGeneral}.

Another corollary of this result is an efficient algorithm for ``hedge-graph sparsification'', a topic of extensive study in the literature~\cite{GKP17, JLMPS23, FGKLS24}, where lack of sub-modularity had thus far been a barrier to efficient sparsfication. We resolve this in \cref{sec:hedgegraphs}.  

\paragraph{CSP Sparsification}
Next, we discuss our contributions to the CSP sparsification regime. We start by formalizing the notion of sparsifying CSPs. An instance of $\CSP(P)$ on $n$ variables is given by a collection of $m$ weighted constraints $(w_1,C_1), \dots (w_m,C_m)$ where $C_j$ is given by a sequence $(i_1(j),\ldots,i_r(j))\in [n]^r$. For $a \in \Sigma^n$, let $C_j(a) = P(a_{i_1(j),\ldots,i_r(j)})$ denote whether $a$ satisfies the $j$th constraint. We let $\Phi(a)$ denote the total weight of constraints satisfied by $a$, i.e., 
    \[
    \Phi(a) = \sum_{j = 1}^m w_j P(a_{i_1(j)},\ldots,a_{i_r(j)}).
    \]

\begin{definition}[CSP Sparsification]
    For $P:\Sigma^r \to \zo$, $\epsilon>0$ and $s \in \Z^+$ we say that an instance $\hat{\Phi}\in \CSP(P)$ is an $(\eps, s)$-\emph{sparsifier} for $\Phi \in \CSP(P)$ if there are at most $s$ constraints $\left \{ \hat{C}_i\right \}_{i \in [s]}$ in $\hat{\Phi}$, each of which satisfies $\hat{C}_i \in \{C_1, \dots C_m \}$ \emph{and} for every $a \in \Sigma^n$ we have $(1 - \eps)\Phi(a) \leq \hat{\Phi}(a) \leq (1 + \eps)\Phi(a)$. (Note that the constraints of $\hat{\Phi}$ may have very different weights than corresponding constraints in $\Phi$.)
    
    For $P:\Sigma^r \to \zo$, $\epsilon>0$ and $s: \Z^+ \to \Z^+$, we say that $\CSP(P)$ is {\em $(\epsilon,s(\cdot))$-sparsifiable} if for every $n \in \Z^+$, every instance $\Phi \in \CSP(P)$ with $n$ variables has an $(\epsilon,s(\eps, n))$-sparsifier. Furthermore we say that $\CSP(P)$ is {\em $(\epsilon,s(\cdot))$-efficiently sparsifiable} if there is a probabilistic polynomial time algorithm to compute such a sparsification.
\end{definition}

Essentially, the requirement of a sparsifier is that the constraints in the sparsifier $\hat{\Phi}$ are a (suitably re-weighted) subset of the original constraints in $\Phi$ such that the value attained on each assignment is approximately preserved. The only freedom we get is in assigning new weights to these constraints.
For brevity, we will say that $\CSP(P)$ is sparsifiable to {\em near-linear size} if $\CSP(P)$ is $(\eps, s(\eps, n))$-sparsifiable for some function $s(\eps, n) = \widetilde{O}_{\eps}(n)$.

Our first theorem completely characterizes which symmetric Boolean predicates admit nearly linear-size sparsification. Recall that a predicate $P:\Sigma^r\to\{0,1\}$ is Boolean if $\Sigma = \{0,1\}$. We say $P$ is {\em symmetric} if $P(a_1,\ldots,a_r) = P(a_{\pi(1)},\ldots,a_{\pi(r)})$ for every $a_1,\ldots,a_r \in \Sigma$ and every permutation $\pi:[r]\to[r]$. Hence a Boolean predicate $P$ is symmetric if and only if there exists $P_0:\{0,\ldots,r\}\to \{0,1\}$ such that $P(a_1,\ldots,a_r) = P_0(a_1 + \cdots + a_r)$ for every $a_1,\ldots,a_r \in \{0,1\}$. Define a symmetric Boolean predicate $P$ to be {\em periodic} if the zeroes of $P_0$ form an arithmetic progression. 
That is, $P$ given by $P(a_1,\ldots,a_r) = P_0(a_1 + \cdots + a_r)$ is periodic if and only if there exists $c,d$ such that $P_0(\alpha) = 0$ for some $\alpha \in \{0,\ldots,r\}$ if and only if $\alpha \in \{c+i\cdot d \mid i \in \Z\}$, where the addition and multiplication in $c+i\cdot d$ are over the integers. We say $P$ is \emph{aperiodic} otherwise.

\begin{restatable}{theorem}{introsym}
\label{thm:intro-sym}
Let $P:\{0,1\}^r \to \{0,1\}$ be a symmetric predicate. Then if $P$ is periodic, $\CSP(P)$ is $(\epsilon, \widetilde{O}(n / \eps^2))$-efficiently sparsifiable for every $\epsilon \in (0,1)$.
On the other hand, if $P$ is not periodic then for every $0 < \epsilon < 1$, $\CSP(P)$ is not $(\epsilon, o(n^2))$-sparsifiable. 
\end{restatable}

\begin{remark}
Note that symmetric CSPs exactly capture the $\zo$-valued case of so-called ``cardinality-based splitting functions'' where the splitting function $g_e(S \cap e)$ depends only on the carindality of $|S \cap e|$. Such splitting functions appear broadly in the clustering literature \cite{LM17, LM18, VBK20, VBK21,LVSLG21, ZLS22} and in summarization of complex data sets \cite{GK10, LB11, TIWB14}. This result leads to an exact (and efficient) characterization of when such splitting functions allow for sparsifiers of near-linear size.
\end{remark}

As noted earlier, this generalizes all known results on nearly-linear CSP sparsification while adding efficiency to previous results. 
Note that the class of symmetric predicates is quite general, capturing cut functions in graphs and hypergraphs, as well as parity functions.
In particular, this theorem provides a concise reason explaining why hypergraphs, graphs, and parity functions all admit linear-size sparsifiers, whereas a priori, it may have seemed quite arbitrary for these natural choices to all yield linear-size sparsifiers.
Prior to this work there was no reason to believe that the sparsifiability of symmetric predicates would be inherently tied to the periodicity of the pattern of unsatisfying assignments. Our results thus reveal this phenomenon for the first time and establish a formal connection between periodicity and sparsifiability.

Another interesting consequence of this theorem is that it implies that there are \say{discrete jumps} in the sparsifiability of predicates. That is to say, either symmetric predicates are sparsifiable to near-linear size, or one can do no better than quadratic size; there is no intermediate regime with for instance sparsifiers of size $O(n^{1.5})$. This has been observed in the field of sparsification in many regimes, for instance near-linear size sparsifiers exist for graphs \cite{BK96}, undirected hypergraphs \cite{CKN20}, and codes \cite{KPS24}, while requiring quadratic size sparsifiers for directed graphs. No problems were known for which intermediate complexity helped. Our results formally validate this phenomenon for the class of symmetric CSPs. 

Our next theorem classifies Boolean predicates that have a non-trivial sparsification. Note that every instance of $\CSP(P)$ on $n$ variables for $P:\{0,1\}^r \to \{0,1\}$ has $O(n^r)$ distinct constraints. Thus trivially $\CSP(P)$ is $(0,O(n^r))$-sparsifiable. We define a sparsification to be {\em non-trivial} if it achieves $s(n) = o(n^{r})$.
Our next theorem shows that $\CSP(P)$ is non-trivially sparsifiable if and only if $P$ does not have only one satisfying assignment.

\begin{restatable}{theorem}{intronontrivial}
\label{thm:intro-non-trivial}
    Let $P: \zo^r \ra \zo$. If $|P^{-1}(1)| \neq 1$ then $\CSP(P)$ is $(\epsilon,\widetilde{O}_r(n^{r-1}/\epsilon^2))$ efficiently-sparsifiable for every $\epsilon > 0$. Otherwise, for every $0 < \epsilon < 1$, $\CSP(P)$ is not $(\epsilon, o(n^r))$ sparsifiable. 
\end{restatable}

This settles an open question posed by \cite{KK15} who asked for such a classification. Surprisingly, this theorem may be viewed as a \say{mostly positive} statement by showing that almost every predicate, specifically every one with more than one satisfying assignment, is non-trivially sparsifiable. We also stress that in our result the existence of sparsifiability is accompanied by an efficient (randomized) algorithm to construct such sparsifiers. 

Finally we extend the characterization of nearly linear-size sparsifiable ternary Boolean CSPs from \cite{KPS24} to get a complete classification (in particular, separating CSPs that admit a near-quadratic sparsifier from those that require cubic size). To describe the theorem we need the notion of the projection of a predicate. Given predicates
$P:\{0,1\}^r \to \zo$ and $Q: \zo^c \to \zo$ we say that $P$ has a \emph{projection} to $Q$ if there exists a function $\rho:\{X_1,\ldots,X_r\} \to \{0,1\} \cup \{Y_1,\neg{Y_1},\ldots, Y_c, \neg{Y_c}\}$ such that $Q(Y_1,\ldots,Y_c) = P(\rho(X_1),\ldots,\rho(X_r))$. Let $\AND_c(Y_1,\ldots,Y_c) = Y_1 \wedge \cdots \wedge Y_c$. 

\begin{restatable}{theorem}{aryclassificationintro}
\label{thm:3-ary-classufication-intro}
    For a predicate $P: \zo^3 \ra \zo$ let $c$ be the largest integer such that $P$ has a projection to $\AND_c$. Then, $\CSP(P)$ is $(\epsilon,\widetilde{\Theta}(n^c))$ efficiently-sparsifiable for every $\epsilon \in (0,1)$, and moreover, for every $\epsilon \in (0,1)$, $\CSP(P)$ is not $(\epsilon,o(n^c))$ sparsifiable.
\end{restatable}

As mentioned before, this theorem continues adding evidence to the conjecture that sparsifier sizes always come in integral gaps. In particular, it shows that for predicates of arity $3$, optimal sparsifier sizes are either $\widetilde{\Theta}(n), \widetilde{\Theta}(n^2)$ or $\Theta(n^3)$.

\subsection{Technical Theorems}

As mentioned before, our main technique extends a technique called {\em code sparsification} introduced in \cite{KPS24}. We present two \say{orthogonal} improvements to this technique of code sparsification. The first is to make code sparsifiers \emph{algorithmic}, and the second is to \emph{extend} these notions \emph{beyond codes}. Getting an algorithm version of the existence result of \cite{KPS24} is non-trivial --- their work roughly decomposed the coordinates of a linear codes into ``dense'' and ``sparse'' coordinates, where the former support a subcode of relatively  high dimension (relative to their size). Applying this decomposition recursively until all coordinates are regular within their partition and then sampling an appropriate number of coordinates in each partition yields their sparsification. The partition of coordinates into ``dense'' and ``sparse'' ones is unfortunately not easy to compute (at least to our knowledge). For instance, the support of a minimum weight codeword can be dense in some codes and finding such codewords, or even approximating the size of their support is NP-hard in the worst case, and means there is no clear way to implement the sparsifiers of \cite{KPS24} in less than exponential time. One of the main contributions of this work is to find an alternate path to this decomposition. We elaborate more on this in \cref{ssec:KPS-review}, but the approach is to efficiently find a small superset of the dense coordinates of a code. We build on insights from Bencz\'ur and Karger~\cite{BK96} to develop this alternate path which is not as obvious in the coding setting as in the graph-theoretic one. Fortunately, not only does this approach extend to codes, but also to all the extensions of codes that we need to sparsify in this paper. In the rest of this section we focus only on the existential aspects of the results in the dicussion, while the theorems assert the algorithmic parts.

Turning to the second direction of improvements in this paper, as noted earlier, code sparsifiers already help with CSP sparsifications, but for our purpose of classifying symmetric Boolean CSPs they only go part of the distance. Specifically, towards our classification, it is relatively straightforward to show that aperiodic symmetric predicates $P$ do not allow for nearly linear sparsifiers. Applying the code sparsifiers of \cite{KPS24} we can show that periodic predicates $P$ with the {\em period being a prime} are sparsifiable to nearly linear size. However, this leaves a big gap where $P$ is periodic with a composite period -- we neither get near-linear sparsifiers in this case nor get to rule them out. To capture all periodic CSPs we extend the sparsifiers of \cite{KPS24} from codes over fields to roughly what might be called ``codes over groups''. We don't formalize this concept but instead describe the specific theorem we prove based on this extension. 

We say that a predicate $P:\zo^r \to \zo$ is an {\em affine predicate} over a group $(G,\odot)$ if there exist $a_1,\ldots,a_r,b \in G$ such that $P(x_1,\ldots,x_r) = 1$ if and only if 
$a_i x_1 \odot a_2 x_2 \odot \cdots \odot a_r x_r \neq b$. (Note that $a_i x_i = a_i$ if $x_i = 1$ and $0$ otherwise, where $0$ is the identity element of the group $G$.) We say that
$P$ is an {\em affine Abelian predicate} if there exists a finite Abelian group $A$ such that $P$ is an affine predicate over $A$. 

\begin{restatable}{theorem}{affineabelian}
\label{thm:affine-abelian}
    If $P:\zo^r \to \zo$ is an affine Abelian predicate over an Abelian group $A$, then $\CSP(P)$ is $(\epsilon, \widetilde{O}(n \cdot \min(r^4, \log^2(|A|)) / \eps^2)$-efficiently-sparsifiable for every $\epsilon > 0$.\footnote{We note that $A$ does not have to be a finite group in this theorem, though in all the applications to CSPs we only use finite Abelian groups.} 
\end{restatable}

In our language, the main technical result of \cite{KPS24} can be stated as being the special case corresponding to $A$ being $\Z_p$ the group of additions modulo a prime $p$. We note that the class of affine Abelian predicates is much richer and has nice closure properties. For instance if $P_1$ and $P_2$ are affine Abelian predicates then so is $P_1 \vee P_2$, thus establishing that the predicates covered by \cref{thm:affine-abelian} have a nice closure property. (Even an existential sparsification result would not follow from the \cite{KPS24} result if $P_1$ and $P_2$ were predicates over different prime groups $\Z_p$ and $\Z_q$, whereas with Abelian predicates we can now simply operate over $\Z_p \times \Z_q$.) In particular, \cref{thm:affine-abelian} immediately yields the characterization of symmetric Boolean CSPs that allow a near-linear sparsification (\cref{thm:intro-sym}), as well as their efficient construction. We also note here that \cref{thm:affine-abelian} does not follow immediately from the techniques of \cite{KPS24}. We elaborate more on this in \cref{sec:overview}, but briefly --- the entire analysis of \cite{KPS24} is coding-theoretic with notions like dimension and distance of codes playing a fundamental role, and linear algebra over $\F_q$ being the main engine. In our case we have to switch to a more lattice-theoretic approach and notions like dimension have to be replaced with more involved counting arguments, while some of the simpler linear algebra is replaced with gcd computations. We complement this with an efficient algorithm for computing a type of decomposition for the \say{code}, which yields the constructive aspect. While the final proof is not much more complicated each step requires extracting abstractions that were not obvious in their work.

We also show below that the requirement that the underlying group is Abelian can not be relaxed.

\begin{restatable}{theorem}{intrononabelian}
\label{lem:intro-non-abelian}
For every non-Abelian group $G$, there exists a predicate $P:\zo^4\to\zo$ and an $\epsilon_0 > 0$ such that $P$ is an affine predicate over $G$ and $\CSP(P)$ is not $(\epsilon_0,o(n^2))$-sparsifiable.
\end{restatable}

Turning to our second theorem (\cref{thm:intro-non-trivial}), observe that the main result of \cite{KPS24} and \cref{thm:affine-abelian} do not seem to yield non-trivial but superquadratic size sparsifiers. For many natural classes of problems, the best sparsifiers are superquadratic in size and so extending the techniques to address such questions is of importance (in particular to prove a statement like \cref{thm:intro-non-trivial}). To this end we give a different extension of \cref{thm:affine-abelian} to higher degree polynomials. 

For an Abelian group $A$ (possibly infinite) We say that $P:\zo^r \to \zo$ is a degree $\ell$ polynomial over $A$ if there exists a degree $\ell$ polynomial $Q  \in A[X_1,\ldots,X_r]$ such that for every $x \in \zo^r \subseteq \Z_q^r$ we have $P(x) = 1$ if and only if $Q(x)\ne 0$.\footnote{A polynomial in $A[X_1,\ldots,X_r]$ is just a formal sum of monomials in $X_1\ldots,X_r$ with coefficients from $A$. This polynomial naturally gives a function from $\Z^r \to G$ by evaluating the monomials over integers, and then adding/subtracting the appropriate number of copies of the coefficient.}

\begin{restatable}{theorem}{intropoly}
\label{thm:intro-poly}
    If $P:\zo^r \to \zo$ is a degree $\ell$ polynomial over an Abelian group $A$ then $\CSP(P)$ is
    $(\epsilon, \widetilde{O}(n^\ell \min(r^{4\ell}, \log^2(|A|)) / \eps^2)$-efficiently-sparsifiable for every $\epsilon > 0$. 
\end{restatable}

While the extension to higher degree polynomials is relatively straightforward, it vastly increases the expressive power of these algebraic CSPs. The positive result in \cref{thm:intro-non-trivial} turns out to be a natural extension obtained by proving that every predicate $P$ with more than one satisfying assignment can be written as the OR of polynomials of degree at most $r-1$ over $\F_2$. 

\subsection{Conclusions} 
CSP sparsification~\cite{KK15} is a powerful unifying abstraction that captures many central and often individually studied, sparsification questions. By studying them jointly we thus get a more global picture of the world of sparsifiability. Our work shows that it is possible to classify large subclasses completely and prove discrete jumps in the level of sparsifiability within these broad subclasses. The classifications also show an algorithmic jump, namely that when sparsifiers exist they can be found algorithmically while the non-existence results are information-theoretic. In the process, these sparsification results yield applications to sparsifying generalized hypergraphs (particularly, with cardinality-based splitting functions), and towards deriving efficient constructions of fundamental combinatorial objects such as Cayley-graph sparsifiers. 

Our work builds on the code sparsification technique of \cite{KPS24}, but extends it conceptually as well as algorithmically. 
In particular, we give a broad framework for upper-bounding the size of sparsifiers by interpreting predicates as polynomials over arbitrary groups. In fact, every known CSP sparsification result follows from our framework (specifically from \cref{thm:intro-poly}). Indeed already the code sparsification framework of \cite{KPS24} (which we generalize) captured the cut sparsifiers of \cite{BK96}, hypergraph cut sparsifiers, and $r$-CNF sparsifiers \cite{KK15, CKN20}, the sparsifiers for Boolean predicates of arity $2$ \cite{FK17} and even those over arbitrary alphabets \cite{BZ20}. (The final claim was not pointed out in \cite{KPS24} so we include a proof of the sparsification result in \cite{BZ20} using our framework in \cref{sec:GeneralAlphabet}.) Our framework further generalizes theirs and so captures all the above mentioned works and the code sparsifiers of \cite{KPS24} while giving new results for symmetric Boolean predicates, classifying general functions with non-trivial sparsifiability, as well as giving efficient algorithms for finding these sparsifiers.

For future directions we note that there is a noticeable gap between the upper bound and lower bound techniques, even for near-linear size sparsifiability of asymmetric predicates and for characterizing the size needs of symmetric CSPs, when the size lower bound is quadratic. The only known lower bounds on sparsifiability come from the notion of a projection to an AND predicate. A hope at this stage may be that every function that does not have a projection to $\AND_c$ can be expressed as a degree $c-1$ polynomial. Note that we proved this to be true when $c=2$ and $P$ is a symmetric predicate. Unfortunately this seems to fail beyond this setting. In \cref{sec:separatingAffineAND} we describe an asymmetric predicate that is not expressible as a degree $1$ polynomial but also does not have a projection to $\AND_2$ and in \cref{sec:symmetricSeparationPolyAND} we show that there is a symmetric predicate which does not have a projection to $\AND_3$ but also {\em seems} to not be expressible as a degree $2$ polynomial. Thus finding new lower bounds on sparsifiability (other than projection to $\AND_c$) or finding new reasons for sparsifiability (not captured by \cref{thm:intro-poly}) seem to be necessary for future progress. 

\paragraph{Addendum:} In a recent breakthrough subsequent to our work, Brakensiek and Guruswami~\cite{BrakensiekG} have given a complete characterization of the sparsifiability of every CSP (over all alphabets) in terms of the ``non-redundancy'' of the predicate, a quantity that has been the subject of some prior work in CSP dichotomy classifications. They do not give a general method to determine the non-redundancy of a predicate, or to analyze its growth as a function of $n$, the number of variables, however they do give bounds in specific instances. In particular their work shows the existence of predicates for which the best sparsifications are of size $n^\alpha$ for some $1.5 \leq \alpha \leq 1.6$ (and so not an integer). Their work is not algorithmic and sparsifications are only guaranteed to exist. 

\paragraph{Organization:}
In \cref{sec:overview}, we provide an in-depth overview of the techniques used to prove our results and show how these immediately imply \cref{thm:3-ary-classufication-intro}. \cref{sec:preliminaries} summarizes some useful definitions and previously known results that we utilize in our work. \cref{sec:decomposition} is dedicated to proving a decomposition theorem and counting bound for codewords of a specific weight, and we use this key result in \cref{sec:sparsificationAlgorithm} where we prove a version of \cref{thm:affine-abelian} only for groups of the form $\Z_q$ and a dependence on $\log^2(q)$. In \cref{sec:abelianConstraints}, we generalize this algorithm to all Abelian groups, proving \cref{thm:affine-abelian} in its entirety. In \cref{sec:impossibility}, we show a complement to this theorem, namely, there are affine predicates over a non-Abelian group that are \emph{not} sparsifiable which yields \cref{lem:intro-non-abelian}. In \cref{sec:symmetric}, we show how to extend our sparsification framework to Boolean symmetric CSPs, proving \cref{thm:intro-sym}. Finally, in \cref{sec:nontrivial}, we prove \cref{thm:intro-poly}, and then use it as a key building block to prove \cref{thm:intro-non-trivial}. In \cref{sec:3-ary-classification}, we then use \cref{thm:intro-non-trivial} to prove \cref{thm:3-ary-classufication-intro}.

\section{A Detailed Overview of Techniques}\label{sec:overview}

We start by reviewing, in \cref{ssec:KPS-review} the work of Khanna, Putterman and Sudan~\cite{KPS24} to highlight the challenges of getting efficient algorithms, and working over Abelian groups as opposed to finite fields. We then describe our efficient algorithm implementation of their strategy in \cref{ssec:efficient}. Then, in \cref{ssec:Zq} we describe our approach to extending their work to general (finite) cyclic groups. In~\cref{ssec:all-abelian} we explain we extend this work to all finite Abelian groups and then to all Abelian (even infinite) groups. While the CSP applications don't need it, we also remove the logarithmic dependence on $m$, the number of constraints, since $m$ can be exponentially large in general linear systems we look at. \cref{ssec:all-abelian} also includes an overview of the steps needed to achieve this.  We then show to get our classification of all Boolean symmetric CSPs in \cref{ssec:symmetric}. Finally, in \cref{ssec:higher-deg} 
show how to extend the entire approach to sparsification of higher degree polynomial constraints and how to use this to get a classification all non-trivially sparsifiable Boolean predicates.

\ifanon\subsection{The work of Khanna, Putterman, and Sudan}
\else \subsection{Previous work of {\protect \cite{KPS24}}}
\fi 
\label{ssec:KPS-review} 

Recall that the work of \cite{KPS24} showed that for any linear subspace (i.e. a code) $\calC \subseteq \F_q^m$ of dimension $n$, there exists a weighted subset $S \subseteq [m], w:S \ra \R^{+}$ of size $\widetilde{O}(n \log(q) / \eps^2)$ such that for any codeword $c \in \calC$, 
\[
 (1-\eps) \wt(v) \leq \wt_w(v|_S) \leq (1+\eps) \wt(v).
\]
In this context, $\wt_w(x)$ is meant to be the \say{weighted hamming weight}, i.e., $\wt_w(x) = \sum_{i=1}^m w(i) \mathbb{1}_{x_i \ne 0}$. 

Their sparsification relies on a counting bound they prove that asserts that for every code $\calC \subseteq \F_q^m$ of dimension $n$ and every integer $d$ either (1) $\calC$ has the property that for every positive integer $\alpha$ it has at most $n^\alpha\cdot q^{\alpha}$ codewords of weight at most $\alpha d$, or (2) there is a subset of coordinates $S \subseteq [m]$ and positive integer $t$  such that $|S| \leq d\cdot t$ and $\calC$ has more than $t$ independent codewords supported on $S$. If condition (1) holds for a judicious choice of $d$, then uniformly sampling coordinates in $[m]$ at rate roughly $1/d$ sparsifies the code $\calc$. But if condition 1 does not hold, then for $S$ being the maximal set with property (2), they show that sampling coordinates $i$ of $[m]\setminus S$ at rate $1/d$ and weighting them with weight $w(i) = d$ and retaining all coordinates of $S$ with weight $1$ leads to a good sparsifier. (The full sparsification to nearly-linear size in $n$ applies this idea recursively.)

The key thus is this counting bound, which is in turn proved by a ``contraction'' argument. This argument defines a randomized procedure that 
outputs a random codeword. The claim is that this if no set $S$ satisfies condition (2) then any fixed codeword $c$ of weight at most $\alpha d$ is output with probability at least $n^{-\alpha}\cdot q^{-\alpha}$. This immediately implies the counting lemma above.

The randomized procedure maintains a code $\calc'$ that is obtained from $\calc$ after a sequence of contractions as follows: (a) Sample a random coordinate $j$ in the support $\supp(\calc')$ of $\calc'$ (where $\supp(\calc')$ is the set of coordinates where there exists a codeword of $\calc'$ that is non-zero). (b) Contract $\calc'$ on coordinate $j$, i.e., remove all codewords with $j$ in their support from $\calc'$. This procedure is repeated until 
$\dim(\calc') \leq \alpha$ at which point we output a random codeword of $\calc'$. To see that this satisfies the claim, one assumes that no set $S$ satisfies condition (2) and fixes a random codeword $c \in \calc$ of weight at most $\alpha d$ and show that each step of contraction (as in (a) above) preserves membership of $c \in \calc'$ with probability at least 
$\alpha d/|\supp(\calc')|$ and the product of this quantity over the iterations of contraction telescopes to $n^{-\alpha}$. 
When the algorithm terminates $\calc'$ has dimension at most $\alpha$ and so has at most $q^\alpha$ codewords. Thus if $c$ is still a codeword of $\calc'$ then it will be output with probability $q^{-\alpha}$. 

This concludes our summary of the work of \cite{KPS24}. Before turning to our work we highlight the challenges towards making their analysis algorithmic, and to extending it to settings beyond linear codes over $\F_q$. 

\paragraph{Algorithmic challenge:} The description above including that of contraction is needlessly wasteful --- in order to maintain the code $\calc'$ one does not need to maintain the full set of codewords of $\calc'$. It suffices to maintain a basis and the contraction can be easily explained as a step of Gaussian elimination on this basis. (Indeed this is already done in \cite{KPS24}.) The key challenge to making this argument algorithmic is that of finding a set $S$ satisfying condition (2) when it exists. For instance if $\calc$ restricted to $S$ has dimension $t=1$, finding $S$ corresponds to finding a codeword of minimum weight in $\calc$, a well-known NP-hard task~\cite{Vardy} even to approximate~\cite{DMS03}. This is the key algorithmic barrier that we need to overcome in this work. 

\paragraph{Beyond Fields:} Turning to extensions beyond $\F_q$, we first note that such an extension is necessary for us to get a classification. For instance the predicate $P(x_1,\ldots,x_6) = 1 \Leftrightarrow \sum_i x_i \not\in \{0,6\}$ needs to works with linear structures (``modules'') over $\Z_6$. The entire analysis above involving notions like dimension and basis is linear algebraic, and Gaussian elimination only works in this setting. It is well-known that elements of linear algebra sometimes completely break down beyond the setting of fields, leading to obstacles such as inability to prove strong lower bounds for ACC circuits~\cite{BeigelBarrington}, while opening up paths for surprising designs~\cite{Grolmusz} and codes~\cite{Efremenko}. A priori it is not clear if there is a way to extend the analysis to, say, additive sets over $\calc \subseteq \Z_6^m$.  

\paragraph{Beyond linear-size sparsifications:} Finally while the methods of \cite{KPS24} are great when it comes to proving the existence of linear-sized sparsifications, they fail to distinguish between different settings where the best sparsification is not trivial (of size $n^r$ for an $r$-ary predicate $P$) but super-linear. They either offer sparsifications of (nearly) linear size or revert to trivial sparsifications. A challenge before our work is to develop techniques to identify settings with sparsification complexity in between the linear and the trivial regime.

\subsection{Towards an Efficient Algorithm}\label{ssec:efficient}

As explained above finding a small set $S$ of coordinates that supports many codewords is not known to be algorithmically feasible (and quite possibly is NP-hard). The key starting point of our work is the 
observation that we do not have to \emph{exactly} recover the set $S$ of rows to remove. Instead, as long as we can recover some \emph{superset} $T$ of these rows $S$ which is not too large, the rest of the sparsification algorithm will work unhindered. A similar insight is also seen in \cite{BK96} when creating their efficient sparsification routine for graphs. In the graph setting, \cite{BK96} define a notion of strength of an edge and show that simply storing $2d\log(n)$ spanning forests suffices for capturing all edges of strength $\leq d$. After this, one can then show that sampling the rest of the edges in a graph at rate roughly $\log(n) / d$ will preserve all cuts in the graph to a factor of $(1 \pm \eps)$. At a high-level, our approach is in a similar spirit, though we need a different combinatorial object.  

Instead of considering spanning forests, we introduce the notion of \say{maximum spanning subsets}. For a matrix $G$ generating the code $\calC$, our goal is to select a subset of rows $T_1$ such that the number of distinct codewords in $\text{Span}(G|_{T_1})$ is equal to the number of distinct codewords in $\text{Span}(G)$. We show that if one iteratively calculates and stores $2d\log n$ disjoint maximum spanning subsets (yielding a set $T$) we efficiently compute a set $T$ that is not too large and satisfies $S \subseteq T$. Note that our notion of largeness is much weaker. $S$ was promised to have size at most $d\cdot t$ (where $t$ is the dimension of $S$) whereas our $T$ has size at most $2dn\log n$. But this turns out to be good enough for efficient sparsification! 

The key analysis step is showing that such a set $T$ contains $S$: This is done by looking at how much of $S$ has been collected after each block of removal of $cd$ disjoint maximum spanning subsets. Suppose the initial dimension of $S$ is $t$. We note that if the dimension of $S$ is still at least $t/c$ after this block of removals, then each iteration within the block must have recovered at least $t/c$ new coordinates from $S$, and this is a contradiction to the claim that $S$ only had dimension $t$ to start with. We conclude the dimension of $S$ drops by a constant factor in each block and so $\log n$ blocks of removal of maximum disjoint spanning subsets completely covers all of $S$. 
This allows us to make the counting lemma of \cite{KPS24} algorithmic and thus get efficient algorithms for all their settings.

\subsection{Sparsifying additive sets over $\Z_q$ for general $q$} 
\label{ssec:Zq}

We now discuss sparsifying more general linear spaces, and start by identifying three specific properties of the contraction procedure from \cite{KPS24} that we would like to emulate in our setting. 
\begin{enumerate}
    \item[] Property 1: If we contract a code $\calc'$ on coordinate $j$, then a codeword $c$ remains in the contracted code if and only if $c_j = 0$. 
    \item[] Property 2: There is a bounded number of contractions one can do before $\calc'$ has dimension at most $\alpha$. 
    \item[] Property 3: Each contraction preserves the linear structure; i.e., $\calc'$ is always a linear code. Furthermore $\calc'$ is specified by at most $n$ vectors in $\F_q^m$ and this specification can be maintained under contraction in polynomial time.
\end{enumerate}

Roughly, the first two items are necessary to ensure that we can lower-bound the probability that a codeword $c$ remains in the span of the generating matrix after repeated contractions. The final item is necessary to ensure that the contraction is well-behaved and can be manipulated in polynomial time, in the sense that one can contract on an already contracted code in polynomial time.

The natural analog of linear codes in $\F_q^m$ is additive sets over $\calc \subseteq \Z_q^m$. There is no immediate notion of a basis, though one can start with a generating set of elements of $\calc$ given by a matrix $G \in \Z_q^{m \times n}$ such that $\calc = \{G.x | x \in \Z^n\}$. But this generating set, or even its size, is not unique --- for example the set of vectors\footnote{For lack of a better word, we abuse terminology here in referring to elements of $\Z_q^m$ as vectors even though they don't form a vector space.} generated by $(3,0)$ and $(0,2)$ in $\Z_6^2$ is the same as the set generated by $(3,2)$. Nevertheless, our goal is to sparsify such a matrix to one of size $\widetilde{O}(n)$. 

A natural choice for contraction would be to pick a random coordinate $j$ where some element of $\calc'$ is non-zero and then to remove from $\calc'$ all elements that are non-zero on the $j$th coordinate, and this is what we do. This immediately yields Property 1. It also preserves the additive structure, so the first part of Property 3 follows. To make the representation explicit we give an algorithm based on GCD computation (reminiscent of the Hermite Normal form computation). (See~\cite{Mic14a}) The tricky part is to argue Property 2. Here we can no longer count on reduction in dimension since we no longer have a vector space to work with. However group theory comes to our rescue --- we notice that $\calc$ is a subgroup of $\Z_q^m$ of size at most $q^n$ and after each contraction it shrinks to a strict subgroup. Since subgroups have size at most half of any group containing them, it follows that there can be at most $n \log q$ iterations of contraction.

Armed with this contraction procedure we are able to establish a type of codeword-counting bound as in \cite{Kar93, KPS24}. Specifically, we prove the following result, which plays a key role in our later sparsification method.

\begin{theorem}[Informal version of \cref{thm:karger}]\label{thm:kargerIntro}
    For every additive set  $\calC \subseteq \Z_q^m$ of size at most  $q^n$s, and for every integer $d \geq 1$, there exists a set $S \subseteq [m]$ of size at most $n \log (q) \cdot d$, such that upon removing these rows, for any integer $\alpha \geq 1$, the resulting additive set has at most $\binom{n \log(q)}{\alpha} \cdot q^{\alpha+1}$ distinct elements of weight $\leq \alpha d$.
\end{theorem}

The formal proof of this statement appears in \cref{sec:decomposition}.
As an immediate consequence of this result, we are able to directly take advantage of some of the tools used in \cite{KPS24} to conclude the existence of $(\eps, \widetilde{O}(n \log^2(m) \log^2(q) / \eps^2))$-sparsifiers for \emph{unweighted} codes $\calC \subseteq \Z_q^m$, with associated generating matrix $G \in \Z_q^{m \times n}$. This is formally stated and proved as \cref{thm:codeSparsifyGeneralLength}.

However, as in \cite{KPS24}, this result is imperfect for several reasons. First, the dependence on $\log(m)$ in the sparsifier size is unaffordable, as $m$ can be exponentially large in $n$, resulting in losses of polynomial factors of $n$ in the sparsifier size. (We can't afford to simply re-invoke the sparsification result as is because the sparsified code now has associated weights.) Second, we need an analog to the efficient procedure to find a small superset $T$ of the set $S$ alluded to in \cref{thm:kargerIntro} to get an efficient algorithm. The latter procedure turns out to be not too different than in the $\F_q$ case though leads to additional $\log q$ factor losses in the bound. Finally, we still only have a result for cyclic groups and not general Abelian groups (including infinite ones). We address these issues briefly next before moving on to the extensions to higher degree constraints.

\subsection{Getting near-linear sized sparsifiers over all Abelian groups}
\label{ssec:all-abelian}

To address the dependence on $m$, we create analogs of some of the key pieces used in the framework of \cite{KPS24}, and also introduce new techniques. To recap, the framework from \cite{KPS24} first uses a simple algorithm to make quadratic size code-sparsifiers for codes of originally \emph{unbounded} length. This algorithm operates by sampling each coordinate $i$ of the code at rate $p_i = \frac{n}{\eps^2 \min_{c \in \calC: c_i \neq 0} \wt(c)}$. That is, the sampling rate for coordinate $i$ is inversely proportional to the minimum weight codeword which is non-zero in its $i$th coordinate. Unfortunately, this quantity is in fact NP-hard to calculate (even to approximate), and further, the analysis used to show that these sampling rates preserve codeword weights is tuned specifically for the case of fields. To bypass this, we take advantage of the fact that we do not actually require \emph{exactly} quadratic size sparsifiers, and in particular, any polynomially-bounded (in $n$) size sparsifier suffices. Thus, we can simply invoke the result discussed above for creating $(\eps, \widetilde{O}(n \log^2(m) \log^2(q) / \eps^2))$ sparsifiers, which will yield sparsifiers of polynomial size. As we will see later, this procedure can be made to run in polynomial time, and thus we can bypass the NP-hardness of this first \say{quadratic-size sparsifier} step used in \cite{KPS24}.

The second component of this algorithm is the so-called \say{weight-decomposition} step from \cite{KPS24}. Roughly, the previous step returns a polynomially-bounded size sparsifier. But, each coordinate in this sparsifier can have weight which is potentially unbounded in $n$, and the original algorithm we defined \emph{only works on unweighted codes}. If we naively try to take this weighted code and turn it into an unweighted code by repeating each coordinate a number of times proportional to its weight, we will end up back where we started, with a code of possibly exponential length.

Thus, the objective is to break the code up into several parts, each of which corresponds to coordinates with weights in a given range $[\alpha^i, \alpha^{i+1}]$, for some value $\alpha$. As in \cite{KPS24}, the key insight comes from the fact that if a given codeword is non-zero in coordinates with large weight, it suffices to simply approximately preserve the weight of the codeword on these large weight coordinates, and ignore the lower weight coordinates. On the large weight coordinates, the ratio of the weights of the coordinates is bounded, so we can afford to simply repeat each coordinate a number of times proportional to its weight, and then sparsify this unweighted code. 

However, the tricky part comes in showing how to cleanly restructure the code such that once we have preserved the codewords on coordinates of large weight, they are removed from the code, and instead it is only the codewords which were all $0$ on these large weight coordinates that remain to be sparsified. Our key insight here is that it turns out that this corresponds \emph{exactly} with performing a contraction on the coordinates of the code with large weight. This returns a new code where the only surviving codewords are those which were all $0$ on the large weight coordinates. In particular, the number of remaining distinct codewords has sufficiently decreased such that we can take the union of all of the sparsifiers of all the different weight levels, while still not exceeding near-linear size. This leads to the proof of \cref{thm:mainRestated} in the general, non-field setting.

\paragraph{Affine Abelian Contraction Algorithm}

After establishing the result for $\Z_q$, the difficulty now comes in extending the algorithm to general Abelian groups $A$. We do this in \cref{sec:abelianConstraints}, where we show a generalization of the contraction argument to Abelian groups, and then we are immediately able to conclude the existence of \say{code} sparsifiers in this regime, and by equivalence, CSP sparsifiers for affine, Abelian predicates by using the same framework established for $\Z_q$; we discuss the generalization of this contraction argument below.

In the more general Abelian group setting, our first difficulty comes in associating a \say{code} with the Abelian group $A$. In the setting of $\Z_q$, such an association is natural, as we simply place the coefficients of our affine equation as the entries in a row of the generating matrix $G$. In the Abelian setting, we instead take advantage of the Fundamental Theorem of Finite Abelian groups \cite{Pin2010}. Roughly speaking, this theorem states that for any Abelian group $A$, we can say that $A$ is isomorphic to a group of the form
\[
\Z_{q_1} \times \Z_{q_2} \times \dots \times \Z_{q_u}.
\]

In particular, for any element $a \in A$, we can create a (linear) bijection sending $a$ to a tuple $(d^{(a)}_{1}, \dots d^{(a)}_u) \in \Z_{q_1} \times \Z_{q_2} \times \dots \times \Z_{q_u}$. Then, our result proceeds by creating a generating matrix $G$ over these tuples, that is, each entry of $G$ is an entry in $\Z_{q_1} \times \Z_{q_2} \times \dots \times \Z_{q_u}$. The benefit now comes from the fact that arithmetic in \emph{each individual entry of a tuple} operates the same as arithmetic over $\Z_q$. In particular, the contraction algorithm we defined for arbitrary $\Z_q$ extends to $\Z_{q_1} \times \Z_{q_2} \times \dots \times \Z_{q_u}$ by running the contraction algorithm once for each $\Z_{q_i}$ for $i \in [u]$ (once for each entry in the tuple). The correctness of this contraction algorithm then (largely) follows from the correctness of the algorithm over $\Z_q$, and so we can conclude the existence of a similar decomposition theorem / counting bound. This then allows us to conclude \cref{thm:affine-abelian}.

\subsection{Sparsifying Symmetric CSPs}
\label{ssec:symmetric} 

As a direct consequence of our sparsifiability result for affine predicates over $\Z_q$, we are able to precisely identify the symmetric, Boolean predicates $P$ for which $\CSP(P)$ is sparsifiable to near-linear size, formally stated as \cref{thm:intro-sym}.

The formal proof of this statement appears in \cref{sec:symmetric}. To prove this theorem, we show that for any symmetric predicate $P: \zo^r \ra \zo$, if the zeros of $P$ are periodic, then one can write a simple affine equation expressing the zeros of $P$. As before, we take \emph{periodic} to mean that if $P_0(x) = 0, P_0(y) = 0$ for $x, y \in \{ 0, \dots r \}$, then for $z = 2x - y$, or $z = 2y - x$, we have $P_0(z) = 0$. If this is the case, we show there will be an equation of the form 
\[
P(x) = \mathbf{1}[\sum_{i = 1}^r a_i x_i + b \neq 0 \mod q],
\]
where $q$ is some integer $\leq r+1$. In this context, $q$ will be the \emph{period} of $P$, meaning that the levels of the predicate $P$ which evaluate to $0$ will be exactly $q$ apart. It is inherent in this formulation that the period of $P$ may be a composite number, hence requiring the generalization of the contraction method to non-fields. We can consider for instance the predicate $P: \zo^6 \ra \zo$, where $P(x) = 0$ if and only if $|x| = 1, 5$. It follows then that we can express 
\[
P(x) = \mathbf{1}[\sum_{i = 1}^6 x_i + 3 \neq 0] \mod 4,
\]
and one can check there is no similar way to express the predicate as such an affine equation modulo a prime. 

We complement the result above by showing that if the zeros of a predicate $P$ are \emph{not} periodic, then there must exist a projection to an AND of arity at least $2$. Indeed, any single witness to non-periodicity, i.e. a triple of the form $c_1, c_2, c_3 = 2c_2 - c_1$, where $P$ is $0$ on strings with $c_1$ or $c_2$ $1$'s, but is $1$ on $c_3$ $1$'s is enough to conclude the existence of a projection to an AND of arity $2$. Thus, periodicity of the zeros of a symmetric predicate is a necessary and sufficient condition for being sparsifiable to near-linear size. 

\subsection{Non-trivial sparsification for almost all predicates}
\label{ssec:higher-deg} 

Finally, we prove a broad generalization of a result from Filtser and Krauthgamer \cite{FK17}. \cite{FK17} showed that when dealing with predicates of the form $P: \zo^2 \ra \zo$, all CSP instances with predicate $P$ are sparsifiable to near-linear size in the number of variables if and only if $|P^{-1}(1)| \neq 1$, i.e. the predicate $P$ does not have only one satisfying assignment. 

We extend this result, and show that for any $r$ (i.e. not only $r=2$ as in \cite{FK17}), if a predicate $P: \zo^r \ra \zo$ satisfies $|P^{-1}(1)| \neq 1$, then we can sparsify CSP instances with predicate $P$ to size $\widetilde{O}(n^{r-1})$. In fact, our result is slightly stronger than this, as it even applies to a CSP with {\em multiple} predicates. We show that for any CSP with predicates $P_1, \dots P_m$ such that each $P_i$ satisfies $|P_i^{-1}(1)| \neq 1$, the entire system can be sparsified to size $\widetilde{O}(n^{r-1})$. Our result is formally stated as \cref{thm:intro-non-trivial}.

This result is tight as there exist predicates with at least $2$ satisfying assignments that require sparsifiers of size $\Omega(n^{r-1})$. Indeed, consider for instance the predicate $P: \zo^r \ra \zo$ such that $P(1, x_1, \dots x_{r-1}) = 1$, and $P(0, x_1, \dots x_{r-1}) = \text{AND}(x_1, \dots x_{r-1})$. This predicate will have $2^{r-1} + 1$ satisfying assignments, but also has a projection to an AND of arity $r-1$, and thus requires sparsifiers of size $\Omega(n^{r-1})$.

Deriving the sparsification result is non-trivial. Indeed, the sparsification procedures created in this paper and the prior work of \cite{KPS24} both rely on being able to represent a predicate as the non-zero assignments to an affine predicate. So, in order to generalize our sparsification method to a broader class of predicates, we show that for any predicate $P: \zo^r \ra \zo$, if there exists a degree $\ell$ polynomial $Q$ over $\F_q$ such that $\forall y \in \zo^r, P(y) = \mathbf{1}[Q(y) \neq 0]$, we can then sparsify CSPs with predicate $P$ to size $\widetilde{O}(n^{\ell} / \eps^2)$. To see why this result follows intuitively, for a degree $\ell$ polynomial on $n$ variables, there are $O(n^{\ell})$ possible terms in this polynomial. Naturally, the polynomial is already \emph{linear} over all of these terms, so if we make a new variable for each term, we can represent the polynomial as a linear equation over all of the new variables (but this leads to a degradation in sparsifier size as we are now sparsifying over a larger universe of variables).

Beyond this point, the difficulty comes in showing that if a predicate has at least two satisfying assignments, there exists a polynomial of degree $r-1$ which exactly captures this predicate. Unfortunately, we are unable to show this fact exactly: instead, we show that for any two assignments $y_1, y_2 \in \zo^r$, one can create a polynomial $P_{y_1, y_2}$ over $\Z_2$ of degree $r-1$ which is $1$ only on $y_1, y_2$. It follows then that for a general predicate $P: \zo^r \ra \zo$ with satisfying assignments $y_1, \dots y_s$, one can write 
\[
P = P_{y_1, y_2} \vee P_{y_1, y_3} \vee \dots \vee P_{y_1, y_s}.
\]

Thus, all that remains is to show that one can simulate the OR of several predicates without blowing up the size of the sparsifier. To do this, we take advantage of our result on the sparsifiability of Abelian groups. Indeed, for an example such as the one above, we instead operate over the Abelian group $(\Z_{2})^s$. The elements of this group look like tuples
\[
(x_0, x_1, \dots x_{s-1}).
\]

Ultimately, we then create the predicate over $(\Z_{2})^s$ which looks like 
\[
P(x) = (P_{y_1, y_2}(x), P_{y_1, y_3}(x), \dots, P_{y_1, y_s}(x)) = (P_{y_1, y_2}, P_{y_1, y_3}, \dots, P_{y_1, y_s})(x).
\]
Note that because each polynomial $P_{y_1, y_i}(x)$ only takes values from $\Z_2$, the above polynomial is zero (i.e. the zero tuple) if and only if every polynomial $P_{y_1, y_i}$ evaluates to $0$. Otherwise, if any individual polynomial evaluates to $1$ on $x$, the output is non-zero. Finally, we can conclude by invoking our result on the sparsifiability affine Abelian groups. 

One consequence of this result is that it gives us a \emph{complete} characterization of the sparsifiability of CSPs with predicates on $3$ variables. Indeed, in the work of \cite{KPS24}, it was shown that any predicate $P$ of arity $3$ which does not have a projection to AND is sparsifiable to near-linear size (in $n$). Using the above result, we know that any predicate $P$ of arity $3$ which has at least $2$ satisfying assignments is sparsifiable to quadratic size. Thus, the only predicate which is not sparsifiable to quadratic size is the predicate with exactly one satisfying assignment, namely, $\text{AND}(x_1, x_2, x_3)$ up to negations. For this predicate, one can clearly see a sparsification lower bound of size $\Omega(n^3)$. Thus, piecing this all together, we give a \emph{complete} characterization of the sparsifiability of predicates of arity $3$. That is,

\aryclassificationintro*

\section{Preliminaries}\label{sec:preliminaries}

In this work, when we use $\F_q$, this refers to the field with $q$ elements (and as such $q$ must necessarily be prime or a prime power). When we say $\Z_q$, this is simply the group of integers $\mod q$, and as such we do not require that $q$ is prime or a prime power (and in fact, the focus will be on the case where $q$ is composite). 

\begin{definition}
    We say that a \emph{code} $\calC \subseteq \Z_q^m$ is simply a subspace of $\Z_q^m$ closed under linear combinations. Thus, we can associate with any such code $\calC$ a \emph{generating matrix} $G \in \Z_q^{m \times n}$ such that $\calC = \text{Span}(G)$ (specifically, $\calC = \{ Gx: x \in \Z_q^n \}$).
    
    The \emph{coordinates} of the code are $[m]$. When we refer to the \emph{number of coordinates}, this is interchangeable with the \emph{length} of the code, which is exactly $m$.
\end{definition}

Note that typically code are generated over fields, but in this work we do not restrict ourself to this setting.

In this work, we will be concerned with code sparsifiers as defined below.

\begin{definition}
    For a code $\calC \subseteq \Z_q^m$ with associated generating matrix $G \subseteq \Z_q^{m \times n}$, a $(1 \pm \eps)$-sparsifier for $\calC$ is a subset $S \subseteq m$, along with a set of weights $w_S: S \ra \mathbb{R}^{+}$ such that for any $x \in \Z_q^n$
    \[
    (1 - \eps) \wt(Gx) \leq \wt_S(G|_S x) \leq (1 + \eps) \wt(Gx).
    \]

    Here, $\wt_S$ is meant to imply that if the codeword is non-zero in its coordinate corresponding to an element $i \in S$, then it contributes $w_S(i)$ to the weight. We will often denote $G|_S$ with the corresponding weights as $\widetilde{G}$.
\end{definition}

We next present a few simple results for code sparsification that we will use frequently.

\begin{claim}\label{clm:verticalDecomp}
    For a vertical decomposition of a generating matrix, \[
    G = \begin{bmatrix}
        G_1 \\
        G_2 \\
        \vdots \\
        G_k
    \end{bmatrix},
    \]
    if we have a $(1 \pm \eps)$ sparsifier to codeword weights in each $G_i$, then their union is a $(1 \pm \eps)$ sparsifier for $G$.
\end{claim}
\begin{proof}
    Consider any codeword $c \in \text{Span}(G)$. Let $c_i$ denote the restriction to each $G_i$ in the vertical decomposition. It follows that if in the sparsifier $\wt(\hat{c}_i) \in (1 \pm \eps) \wt(c_i)$, then $\wt(\hat{c}) = \sum_i\wt(\hat{c}_i) \in (1 \pm \eps) \sum_i \wt(c_i)  = (1 \pm \eps) \wt(c)$. 
\end{proof}

\begin{claim}\label{clm:composingApproximations}
Suppose $\calC'$ is $(1 \pm \delta)$ sparsifier of $\calC$, and $\calC''$ is a $(1 \pm \eps)$ sparsifier of $\calC'$, then $\calC''$ is a $(1 - \eps)(1-\delta), (1 + \eps)(1 + \delta)$ sparsifier to $\calC$ (i.e. preserves the weight of any codeword to a factor $(1 - \eps)(1 - \delta)$ below and $(1 + \eps)(1 + \delta)$ above).
\end{claim}

\begin{proof}
    Consider any codeword $\calC x$. We know that $(1 - \eps)\wt(\calC x) \leq \wt(\calC'x) \leq (1 + \eps)\wt(\calC x)$. Additionally, $(1 - \delta)\wt(\calC'x) \leq \wt(\calC''x) \leq (1 + \delta)\wt(\calC'x)$. Composing these two facts, we get our claim. 
\end{proof}

\begin{claim}\label{clm:concentrationBound}{\rm (\cite{FHH11})}
    Let $X_1, \dots X_{\ell}$ be random variables such that $X_i$ takes on value $1 / p_i$ with probability $p_i$, and is $0$ otherwise. Also, suppose that $\min_i p_i \geq p$. Then, with probability at least $1 - 2e^{-0.38 \eps^2 \ell p}$,
    \[
    \sum_i X_i \in (1 \pm \eps) \ell.
    \]
\end{claim}

We will use the following result from the work of Khanna, Putterman, and Sudan \cite{KPS24}:

\begin{theorem}\label{thm:AND}
    For a predicate $P: \zo^r \ra \zo$, if there exists $\pi: \{x_1, \dots x_r\} \ra \{0, 1, a, \neg a, b, \neg b\}$ such that $P(\pi(x_1), \dots \pi(x_r)) = \text{AND}(a,b)$, then there exist CSPs with predicate $P$ that require sparsifiers of size $\Omega(n^2 / r^2)$.
\end{theorem}

While the result \cite{KPS24} is specifically only stated for ANDs of arity 2, note that their argument generalizes to ANDs of arity $r$. I.e., if there exists a projection to ANDs of arity $\ell$, then there exist instances which require sparsifiers of size $\Omega(n^{\ell})$ for any constant $0< \eps < 1$.

We will frequently make use of the following facts about the Euclidean algorithm:

\begin{fact}[Euclidean Algorithm](See among many others, \cite{Wei24})\label{fact:GCDAlgo}
    There exists an algorithm which upon being given a list of integers $y_1, \dots y_r$:
    \begin{enumerate}
    \item Sets $x_1 = y_1, \dots x_r = y_r$.
        \item Only performs operations of the form $x_i \leftarrow x_i + \ell \cdot x_j$ (for $\ell \in \Z$).
        \item Terminates with $x_1 = \text{GCD}(y_1, \dots y_r)$.
    \end{enumerate}
\end{fact}

Note that while many algorithms are only stated for computing the GCD of any two numbers, one can simply calculate the GCD of more numbers iteratively, adding one new numbers to the computation in each iteration. I.e., in the case of 3 numbers, one can compute $\text{GCD}(x_1, \text{GCD}(x_2, x_3))$.

\section{A Decomposition Theorem for Codes over $\Z_q$}\label{sec:decomposition}
 
We start by working towards a proof of \cref{thm:affine-abelian} for the special case of finite cyclic groups, i.e., groups of the form $\Z_q$ for some (potentially composite) integer $q$. 
In this section, we will prove a decomposition theorem (\cref{thm:karger}) for general linear spaces over $\Z_q$ and follow this with an efficient algorithm for computing this decomposition (\cref{thm:efficientKarger}). This theorem is analogous to Theorem 2.1 in \cite{KPS24}, with the key difference being our decomposition is also algorithmic. This forms the basis of all of our algorithmic sparsification results. We will use this efficient decomposition in \cref{sec:sparsificationAlgorithm} to prove \cref{thm:affine-abelian} over $\Z_q$. 

To do this, we first define a contraction algorithm on the generating matrix of this linear space. We show that with high probability a codeword of low-weight will survive this contraction procedure, and remain in the span of the generating matrix, provided the \say{support} of the generating matrix never becomes too small. Using this, we are able to prove a decomposition theorem about linear spaces over $\Z_q$. Either there exists a linear subspace with small \say{support} that contains many distinct codewords, or the contraction procedure shows that there can not be too many codewords of light weight.

\begin{definition}
    For a linear space $\calC \subseteq \Z_q^{m}$, we say that the \textbf{support} of $\calC$ is the set of coordinates that are not always $0$. That is,
    \[
    \Supp(\calC) = \{i \in [m]: \exists c \in \calC: c_i \neq 0 \}  .
    \]
\end{definition}

\begin{definition}
    Using the definition of support, we can likewise define the \textbf{density} of a linear space. For $\calC \subseteq \F_q^m$,
    \[
    \Density(\calC) = \frac{\log_2 |\calC|}{|\Supp(\calC)|}.
    \]
\end{definition}

\begin{definition}
    For a linear subspace $\calC \subseteq \Z_q^m$, a \textbf{subcode} of $\calC$ is any subspace $\calC' \subseteq \calC$ which is closed under linear combinations.
\end{definition}

We are now ready to define our contraction algorithm.

\begin{algorithm}[H]
    \caption{$\Contract(G, j)$}\label{alg:contract}
    Run the Euclidean GCD algorithm (\cref{fact:GCDAlgo}) on the $j$th row of generating matrix $G$ (treating the entries as being over $\Z$), using column operations to get the GCD of the $j$th row into a column, say this column is $v$. Swap $v$ with the first column.\\
    Add multiples of $v$ to cancel out the $j$th entry of every other column in the $j$th coordinate. I.e., for $i \in \{2, \dots n\}$, $G_i \leftarrow G_i - ((G_i)_j / v_j) \cdot v$. \\
    Replace column $v$ with $\eta \cdot v$, where $\eta = \min \{ \ell \in [1, \dots, q]: v_j \cdot \ell = 0 \}$. \\
    \Return{G}
\end{algorithm}

\begin{algorithm}[H]
\caption{Repeated Contractions$(G, \alpha, q)$}\label{alg:manyContract}
    Input generating matrix $G \in \Z_q^{m \times n}$. \\
    \While{more than $q^{\alpha + 1}$ distinct codewords in the span of $G$}
    {
    Choose a random non-zero row $j$ of $G$. \\
    Set $G$ = Contract$(G, j)$.
    }
\end{algorithm}

\begin{claim}\label{clm:stillInSpan}
    Consider a codeword $c \in \calC$ where $\calC$ is the span of $G$. If we run $Contract(G, j)$ on a coordinate $j$ such that $c_j = 0$, then $c$ will still be in the span of $G$ after the contraction.
\end{claim}

\begin{proof}
    First, we show that after line $1$ in the \cref{alg:contract}, $c$ is still in the span. Indeed, the Euclidean GCD algorithm only ever makes column operations of the form $v_1 + r \cdot v_2 \rightarrow v_1$ (see \cref{fact:GCDAlgo}). This means that every step of the algorithm is invertible, so in particular, after every step of the Euclidean GCD algorithm, the span of the matrix $G$ will not have changed. 

    Now, in line $2$ of \cref{alg:contract}, this again does not change the span of $G$. Indeed, we can simply undo this step by adding back multiples of $v$ to undo the subtraction. Since the span of $G$ hasn't changed after this step, $c$ must still be in the span. Further, note that this step is possible \emph{because} $v_j$ is the GCD of the $j$th row of the generating matrix. I.e., there must be some multiple of $v_j$ which cancels out every other entry in the $j$th row.

    Finally, in line $3$, the span of $G$ actually changes. However, since $c$ was zero in its $j$th coordinate, it could only contain an integer multiple of $\eta$ times the column $v$ (again where $\eta$ is defined as $\eta = \min \{ \ell \in [q]: v_j \cdot \ell = 0 \}$), as otherwise it would be non-zero in this coordinate. Hence $c$ will still be in the span of the code after this step, as it is only codewords that are non-zero in coordinate $j$ that are being removed.
\end{proof}

\begin{claim}
    After calling $\Contract(G, j)$ on a non-zero coordinate $j$, the number of distinct codewords in the span of $G$ decreases by at least a factor of $2$.
\end{claim}

\begin{proof}
    Consider the matrix $G$ after line $2$ of \cref{alg:contract}. By the previous proof, we know that the span of $G$ has not changed by line $2$. Now, after scaling up $v$ by $\eta$, we completely zero out row $j$ of $G$. Previously, for any codeword in the code which was zero in coordinate $j$, we could correspondingly add a single multiple of column $v$ to it to create a new codeword that was non-zero in its $j$th coordinate. Thus, the number of codewords which were non-zero in coordinate $j$ was at least as large as the number which were zero. After scaling up column $v$, codewords which were non-zero in this coordinate are no longer in the span, so the number of distinct codewords has gone down by at least a factor of $2$.
\end{proof}

\begin{claim}\label{clm:lowdensity}
    Let $\calC$ be the span of $G \subseteq (\Z_q)^{m \times n}$, and let $d$ be an integer. Suppose every subcode $\calC' \subseteq \calC$ satisfies $\Density(\calC') < \frac{1}{d}$. Then for any $\alpha \in \Z^+$, the number of distinct codewords of weight $\leq \alpha d$ is at most $\binom{n\log(q)}{\alpha } \cdot q^{\alpha + 1}$.
\end{claim}

\begin{proof}
    Consider an arbitrary codeword $c$ in the span of $G$ of weight $\leq \alpha d$. Let us consider what happens when we run \cref{alg:manyContract}. In the worst case, we remove only a factor of $2$ of the codewords in each iteration. Thus, suppose that this is indeed the case. Suppose that the code starts with $q^n$ distinct codewords. Now, after $\ell$ iterations, this means in the worst case there are $2^{n \log (q) - \ell}$ distinct codewords remaining. Note that after running these contractions, the resulting generating matrix defines a subcode $\calC'$ of our original space. Then, by our assumption on the density of every subcode, at this stage, the number of non-zero rows in the generating matrix must be at least $(n \log(q) - \ell)d$. Thus, the probability that $c$ remains in the span of the generating matrix after the next contraction is at least
    \[
    \Pr[c \text{ survives iteration}] \geq 1 - \frac{\alpha d}{(n \log(q) - \ell)d} = 1 - \frac{\alpha}{n \log(q) - \ell},
    \]
    where this is simply the probability that we sample a coordinate in which $c_j = 0$.

    Now, we can consider the probability that $c$ survives all iterations. Indeed,
    \begin{align*}
        \Pr[c \text{ survives all iterations}] & \geq \frac{n \log(q) - \alpha}{n\log (q)} \cdot \frac{n \log(q) - \alpha - 1}{n \log (q) - 1} \cdot \dots \cdot \frac{\alpha + 1 - \alpha}{\alpha + 1} \\
        & = \binom{n\log(q)}{\alpha }^{-1}.
    \end{align*}
    At this point, the number of remaining codewords is at most $q^{\alpha + 1}$. Hence, the total number of codewords of weight $\leq \alpha d$ is at most $\binom{n\log(q)}{\alpha } \cdot q^{\alpha + 1}$.
\end{proof}

\begin{theorem}\label{thm:karger}
    For a code $\calC \subseteq \Z_q^{n}$ with at most $q^{n}$ distinct codewords, and any arbitrary integer $d \geq 1$, at least one of the following is true:
    \begin{enumerate}
        \item There exists $\calC' \subseteq \calC$ such that $\Density(\calC') \geq \frac{1}{d}$.
        \item For any integer $\alpha \geq 1$, the number of distinct codewords of weight $\leq \alpha d$ is at most $\binom{n\log(q)}{\alpha } \cdot q^{\alpha + 1}$.
    \end{enumerate}
\end{theorem}

\begin{proof}
    Suppose condition $1$ does not hold, and invoke \cref{clm:lowdensity}. 
\end{proof}

\begin{corollary}\label{cor:moduleKarger}
    For any linear code $\calC$ over $\Z_q^m$, with $\leq q^n$ distinct codewords, and for any integer $d \geq 1$, there exists a set of at most $n \log (q) \cdot d$ rows, such that upon removing these rows, for any integer $\alpha \geq 1$, the resulting code has at most $\binom{n\log(q)}{\alpha } \cdot q^{\alpha+1}$ distinct codewords of weight $\leq \alpha d$.
\end{corollary}

\begin{proof}
    We use \cref{thm:karger}. Suppose for the code $\calC$ that condition 2 does not hold. Then, condition 1 holds, which means there exists a set $n'd$ rows, with at least $2^{n'}$ distinct codewords that are completely contained on these rows. Now, by removing these rows, this yields a new code where the number of distinct codewords has decreased by a factor of $2^{n'}$. Thus, if removing these rows yields a new code $\calC'$, this new code has at most $q^n / 2^{n'}$ distinct codewords. Now, if condition 2 holds, for $\calC'$, we are done. Otherwise, we can again apply this decomposition, removing another $n''d$ rows to yield a new code with at most $2^{n \log(q) - n' - n''}$ distinct codewords. In total, as long as we remove rows in accordance with condition 1, we can only ever remove $n \log(q) d$ rows total, before there are no more distinct codewords remaining, and condition 2 must hold.

    Thus, there exists a set of at most $n \log(q) \cdot d$ rows, such that upon their removal, the new code satisfies condition 2.
\end{proof}

\subsection{Efficient Algorithms for Computing Decomposition}

In this subsection, we will show how to \emph{efficiently} find (a superset of) the rows necessary for the decomposition guaranteed to exist. In particular, while we may not be able to exactly identify the set of rows $S$ of the generating matrix which need to be removed, we will be able to find a set $T$ of $\leq 2n \log(q)\log(n) \cdot d $ rows, such that $S \subseteq T$. It follows then that if we remove the set of rows $T$, then uniform sampling should suffice for preserving the weights of codewords.

We summarize this in the following theorem:

\begin{theorem}\label{thm:efficientKarger}
    For any linear code $\calC$ over $\Z_q^m$, with $\leq q^n$ distinct codewords, and for any integer $d \geq 1$, there exists a set $S$ of at most $n \log (q) \cdot d$ rows, such that upon removing these rows, for any integer $\alpha \geq 1$, the resulting code has at most $\binom{n\log(q)}{\alpha } \cdot q^{\alpha+1}$ distinct codewords of weight $\leq \alpha d$. Further, there is an efficient algorithm for recovering a set $ T \subseteq [m]$ such that $S \subseteq T$, and $|T| \leq nd\log(q)(\log(n) + \log(q))$.
\end{theorem}

Roughly speaking, the method we use in this section should be thought of as a linear algebraic analog to storing spanning forests of a graph. \cite{BK96} showed that if one stores $2 \log(n) \cdot d$ spanning forests of a graph, then one can sample the rest of the graph at rate roughly $\frac{1}{d}$ while still preserving the sizes of all cuts in the graph. In our case, the natural analog of a spanning forest will be a subset $T_1$ of the rows of the generating matrix $G$, such that the number of distinct codewords in the span of $G|_{T_1}$ is the same as in $G$. We define this notion more formally below:

\begin{definition}
    For a generating matrix $G \in \Z_q^{m \times n}$, we say that a \textbf{maximum spanning subset} $T \subseteq [m]$ is a subset of the rows of the generating matrix $G$ such that the number of distinct codewords in the span of $G|_T$ is the same as in the span of $G$.
\end{definition}

\begin{claim}
    In order to efficiently construct a maximum spanning subset of $G \in \Z_q^{m \times n}$ of size $\leq n \log(q)$, it suffices to be able to \emph{efficiently} compute the number of distinct codewords in the span of an arbitrary generating matrix $H \in \Z_q^{m' \times n'}$.
\end{claim}

\begin{proof}
    Consider the following simple algorithm: 
    
    \begin{algorithm}[H]
    \caption{BuildMaxSpanningSubset$(H)$}
        Initialize $T = \emptyset$. \\
        Let $k = 0$. \\
        \For{$i \in [m]$}{
        If the number of distinct codewords in the span of $H|_{T \cup \{i \}}$ is $\geq k$, then set $T = T \cup \{i \}$, and $k \leftarrow $ the number of distinct codewords in the span of $H|_{T \cup \{i \}}$.
}
\Return{$T$}
    \end{algorithm}

    Clearly, this is efficient if the procedure for checking the number of distinct codewords in the span of $G|_T$ is efficient. Further, it is clear to see that this yields a spanning subset, as any row which is not kept does not increase the number of codewords. So, it remains to prove the size bound. For this, we simply remark that if adding a row to $G|_T$ increases the number of distinct codewords, then it \emph{at least} doubles the number of distinct codewords. To see why, suppose we have a set $T$ and a set $T \cup \{ i\}$ such that $G|_{T \cup \{i \}}$ has more distinct codewords in its span than $G|_T$. Consider any two messages $x_1, x_2 \in \Z_q^n$ such that $G|_Tx_1 = G|_Tx_2$, yet $G|_{T \cup \{i \}} x_1 \neq G|_{T \cup \{i \}} x_2$. It must be the case then that $ G|_{T \cup \{i \}} (x_1-x_2) $ is non-zero only in the last coordinate (corresponding to row $i$), as the two vectors are equal for all the rows in $T$. To conclude, we can then see that for any codeword $z = G|_T x$, there are at least two \emph{distinct} corresponding codewords in the span of $G|_{T \cup \{i\}}$, i.e., $G|_{T \cup \{i\}} x$ and $G|_{T \cup \{i\}} (x + x_1 - x_2)$.

    Thus, each row which increases the number of codewords increases it by at least a factor of $2$. Because there can be at most $q^n$ distinct codewords, this means the size of the spanning subset will be at most $n \log(q)$.
\end{proof}

\begin{claim}\label{clm:runtimeContraction}
    There is an efficient algorithm for exactly calculating the number of distinct codewords in the span of a generating matrix $G \in \Z_q^{m \times n}$.
\end{claim}

\begin{proof}
    First, note that from the analysis of the contraction algorithm, we can assume without loss of generality that $G$ is given to us in a form where the first row has a single non-zero entry in the first column (as if $G$ is not, then we can efficiently get $G$ into such a form in time $O(mn \log(q))$). Now, let this non-zero entry be denoted by $a \in \Z_q$, and let $c$ be the smallest non-negative integer such that $c \cdot a = 0$ over $\Z_q$. Let $G'$ denote the resulting matrix when we replace the first column $c_1$ of $G$ with $c \cdot c_1$. In particular, this forces the entire first row of $G'$ to be $0$. 
    
    Then, we will show that the number of distinct codewords in the span of $G$ is exactly $c$ times greater than the number of distinct codewords in the span of $G'$.

    To see this, we will show that there is an equivalence class in the codewords of $G$, such that there are exactly $c$ codewords in the span of $G$ that map to each codeword in the span of $G'$. In particular, for a codeword $z \in G'$, we denote the corresponding codewords in the span of $G$ by $z + \alpha \cdot c_1$, where $\alpha \in \{0, 1, \dots c-1 \}$. Clearly, this yields a map from elements in the image of $G'$ to \emph{every} codeword in the span of $G$. Indeed, for any codeword in the span of $G$, it can be written as 
    \[
    y = \sum_{i = 1}^n \alpha_i c_i = (\alpha_1 \mod c) \cdot c_1 + ((\alpha_1 - (\alpha_i \mod c)) / c) \cdot c \cdot c_1 + \sum_{i = 2}^n \alpha_i c_i .
    \]

It remains to show that this map is unique. Indeed, suppose that for a codewords $y$ in the image of $G$ we can write it as $y = z_1 + \beta_1 c_1 = z_2 + \beta_2 c_1$. Now, observe that it must be the case that $\beta_1 = \beta_2$ because $(z_1)_1 = 0$ and $(z_2)_1 = 0$. This follows because the first row of the matrix $G'$ is all zeros, so any codeword generated by $G'$ must be zero in its first coordinate. Further, because $c$ is the smallest value such that $c \cdot (c_1)_1 = 0$, for any values $0 \leq \beta < c$, the values $\beta \cdot (c_1)_1$ must be distinct. Thus, we see that $y = z_1 + \beta_1 c_1 = z_2 + \beta_1 c_1$, and so it follows that $z_1 = z_2$. 

Thus, the natural algorithm for computing the number of distinct codewords is as follows: given the generating matrix $G$, contract on the first row to get a generating matrix $G'$, and keep track of the value $c$ that was used to scale the first column. Then, inductively, the number of distinct codewords in the span of $G$ is $c$ times the number of distinct codewords in the span of $G'$.

The running time for this algorithm is $O(m n^2 \log^2(q))$, as there can be at most $n \log(q)$ rows we contract on, each of which takes time $O(m n \log(q))$.
\end{proof}

Next, we will show that storing disjoint maximum spanning subsets suffices for recovering the set $S$ guaranteed by \cref{cor:moduleKarger}. First however, let us introduce some notation regarding this set $S$ of bad rows in $G$. 

Recall that we are given a parameter $d$, and we wish to remove a set of $\leq n \log(q) d$ bad rows, such that the codeword counting bound will hold with parameter $d$ after removing these rows. As mentioned, $S$ is really the support of the union of several subspaces which are \emph{particularly} dense. That is, if we let $C$ denote the span of $G$, and we analyze the following expression
\[
\max_{C' \subseteq C} \frac{\log |C'|}{|\Supp(C')|},
\]
if the expression is $\geq 1/d$, then the subspace is too dense for our counting bound to hold. When this occurs, we let $C'_1$ denote the corresponding maximizing subspace, we let $S_1$ denote the support of $C'_1$, and we let $k_1$ denote the log of the number of distinct codewords removed by taking away the rows in $S_1$. Now, after removing this subspace $C'_1$ (i.e., removing the support of $C'_1$ from $C$), it is still possible that the counting bound is violated. Thus, we continue to analyze the expression 
\[
\max_{C' \subseteq C|_{\bar{S_1}}} \frac{\log |C'|}{|\Supp(C')|},
\]
and if the expression is $\geq 1/d$, then again we let $C'_2$ denote the maximizing subspace, $S_2$ denote the support of $C'_2$, and we let $k_2 = \log|C_{\bar{S_1}}| - \log|C_{\bar{S_1} \cap \bar{S_{2}}}|$. We continue doing this iteratively until finally 
\[
\max_{C' \subseteq C|_{\bar{S_1} \cap \dots \cap \bar{S_{\ell}} }} \frac{\log |C'|}{|\Supp(C')|} < 1/d.
\]
We let the sequence of subspaces that are recovered be denoted by $C'_1, \dots C'_{\ell}$, let $S_i = \Supp(C'_i)$, and let $k_i = \log|C_{\bar{S_1} \cap \dots \bar{S_{i-1}}}| - \log|C_{\bar{S_1} \cap \dots \bar{S_{i-1}} \cap \bar{S_{i}}}|$. Under these definitions, note that $S = S_1 \cup \dots \cup S_{\ell}, k = \sum_{i = 1}^{\ell} k_i$, and therefore $|S| \leq k d$.

Now, let $T_1, \dots T_{2 d (\log(n) + \log(q))}$ denote \emph{disjoint} maximum spanning subsets of a generating matrix $G \in \Z_q^{m \times n}$. Note that computing such a set of disjoint maximum spanning subsets can be done efficiently, by iterating through the rows of $G$ to first create $T_1$, and then removing the rows from $T_1$ from $G$, re-iterating through $G$ and creating a maximum spanning subset $T_2$, etc.

That is, we can use the following algorithm:

\begin{algorithm}[H]
    \caption{ConstructSpanningSubsets($H, t$)} \label{alg:constructSpanningSubsets}
    \For{$i \in [t]$}{
    Let $T_i = \mathrm{BuildMaxSpanningSubset}(H|_{\bar{T_1} \cap \dots \bar{T_{i-1}}})$.
    }
    \Return{$T_i: i \in [t]$}.
\end{algorithm}

Clearly, the spanning subsets are disjoint, and the run-time is bounded by $O(t \cdot mn^2\log^2(q))$.

Thus, it remains to show the following claim:

\begin{claim}\label{clm:containsBadSet}
    For a generating matrix $G \in \Z_q^{m \times n}$, any choice of $d$ and the set $S$ of bad rows (i.e. those guaranteed by \cref{cor:moduleKarger}) for $G$ and parameter $d$, for any disjoint maximum spanning subsets $T_1, \dots T_{2 d\log(q) (\log(n) + \log(q))}$, we have that $S \subseteq T_1 \cup T_2 \dots \cup T_{2 d\log(q) (\log(n) + \log(q))}$.
\end{claim}

\begin{proof}

First, we will adopt the notation for $S_i, k_i, C'_i, T_i$ for $i = 1, \dots \ell$ that was created in the preceding paragraph.

Let us consider a single maximum spanning subset $T_1$ of $G$ along with the set $S$. We let $k = \sum_{i = 1}^{\ell} k_i$ denote the sum of the logs of the number of distinct codewords that were removed. In particular then, we know that after removing the supports of $C'_1, \dots C'_{\ell}$ from $C$ (i.e., removing the set $S$), then $\log(|C_{\bar{S}}|) = \log(|C|) - k$. So, in order to get a \emph{maximum spanning subset} (where the log of the number of distinct codewords is $\log(|C|)$), it must be the case that $T_1$ includes at least $k / \log(q)$ rows from $S$, as each row we include increases the number of codewords by a factor of at most $q$ (and the log by at most $\log(q)$), and there is a factor of $2^k$ distinct codewords that must be captured by these rows. However, naively repeating this argument does not suffice, as it is possible that as we recover rows in the $T_i$'s, the remaining number of distinct codewords to be recovered from $C|_S$ decreases (i.e., perhaps $\log(|C_{\bar{T}}|) - \log(|C_{\bar{T} \cap \bar{S}}|) << k$). Instead, we argue in a manner reminiscent of \cite{BK96}. After the first $2 d\log(q)$ disjoint maximum spanning subsets are recovered (and denote the set of rows they recover by $T$), there are two cases:
    \begin{enumerate}
        \item One case is that 
        $\log(|C_{\bar{T}}|) - \log(|C_{\bar{T} \cap \bar{S}}|) \geq k/2$.
        However, this yields a contradiction, as we know that the set $S$ is of size at most $kd$. The above implies that for $2d\log(q)$ iterations, each maximum spanning subset must have recovered at least $k / (2\log(q))$ (distinct) rows from $S$. This is because any maximum spanning subset \emph{must} contain at least $k/(2\log(q))$ rows from the support of these subspaces (otherwise they are not \emph{maximally} spanning) in order to have the maximum possible number of distinct codewords. But, then we must have recovered at least $kd$ rows from $S$ total, which would have exhausted the entire set. 
        \item The other case is that $\log(|C_{\bar{T}}|) - \log(|C_{\bar{T} \cap \bar{S}}|) < k/2$. That is, among the first $2d\log(q)$ maximum spanning subsets that are recovered, these contained sufficiently many rows from $S$ such that the difference between the log of the remaining number of distinct codewords in $C_{\bar{T}}$ versus in $C_{\bar{T} \cap \bar{S}}$ decreased by more than half. We want to argue that this means there are at most $kd/2$ remaining rows that must be recovered. The key claim we will use is the following: \begin{claim}
            If $\log(|C_{\bar{S_1} \cap \dots \cap \bar{S_{i-1}}}|) - \log(|C_{\bar{S_1} \cap \dots \cap \bar{S_{i}}}|) = k_i$ with corresponding support $S_i$, and after removing a set $T$ $\log(|C_{\bar{S_1} \cap \dots \cap \bar{S_{i-1}} \cap \bar{T}}|) - \log(|C_{\bar{S_1} \cap \dots \cap \bar{S_{i}} \cap \bar{T}}|) < k_i - s_i$, then $|S_i / T| \leq (k_i - s_i)d $.
        \end{claim}
        \begin{proof}
            First, note that after removing $S_1, \dots S_{i-1}$, $C'_i$ was the maximizer for the expression 
            \[
            \max_{C' \subseteq C|_{\bar{S_1} \cap \dots \bar{S_{i-1}}}} \frac{\log(|C'|)}{|\Supp(C')|},
            \]
            and in particular, $C'_i$ achieved some value $1 / d_i \geq 1 / d$ for this expression, and had $\log(|C_{\bar{S_1} \cap \dots \cap \bar{S_{i-1}}}|) - \log(|C_{\bar{S_1} \cap \dots \cap \bar{S_{i}}}|) = k_i$. Now, we claim that if we removed some rows $T$ from the support of $C'_i$ such that now $\log(|C_{\bar{S_1} \cap \dots \cap \bar{S_{i-1}} \cap \bar{T}}|) - \log(|C_{\bar{S_1} \cap \dots \cap \bar{S_{i}} \cap \bar{T}}|)$ is $\leq k_i - s_i$, then $T$ must have recovered at least $s_i \cdot d_i$ rows from the support of $C'_i$. This follows because instead of removing the entirety of $S_i$, we instead only remove $T \cap S_i$, yet still find a reduction in log of the number of distinct codewords by $k_i$. In particular then, the subspace defined on $T \cap S_i$ contains $\geq s_i$ of the log of the number of distinct codewords contributed by $C|_{S_i}$.
            
            So, if $T$ removed fewer than $s_i \cdot d_i$ rows, yet still decreased the log of the number of distinct codewords by $s_i$, this means there must have been a \emph{more optimal} subspace $C''_i$, defined on a support of size $< s_id_i$ with at least $2^{s_i}$ distinct codewords, therefore contradicting the optimality of $C'_i$. Thus, we get the stated claim. 
        \end{proof}
        Now, note that originally, the spaces $C'_1, \dots C'_{\ell}$ had logs of the number of distinct codewords they contributed totaling $k_1, \dots k_{\ell}$. We are in the case where after removing the rows from $T$, $\log(|C_{\bar{T}}|) - \log(|C_{\bar{T} \cap \bar{S}}|) < k/2$. If we let $s_i$ denote 
        $(\log(|C_{\bar{S_1} \cap \dots \cap \bar{S_{i-1}}}|) - \log(|C_{\bar{S_1} \cap \dots \cap \bar{S_{i}}}|) ) - (\log(|C_{\bar{S_1} \cap \dots \cap \bar{S_{i-1}} \cap \bar{T}}|) - \log(|C_{\bar{S_1} \cap \dots \cap \bar{S_{i}} \cap \bar{T}}|))$ 
        then it must be the case that $|S_i / T| \leq (k_i - s_i)d$. 
        
        In particular, 
        \[
        |S / T| = \sum_{i = 1}^{\ell} |S_i / T| \leq \sum_{i = 1}^{\ell} (k_i - s_i)d = kd - d \sum_{i = 1}^{\ell} s_i \leq kd/2,
        \]
        because $\sum_{i = 1}^{\ell} s_i \geq k/2$ (half of the log of the number of distinct codewords contributed by these spaces has been recovered). Thus, the number of remaining rows that have to be recovered is bounded by $kd/2$.
        \end{enumerate}

        In particular, we now repeat this $\log(n) + \log(q)$ times. In the $i$th iteration, we are guaranteed that one of the following happens:
        \begin{enumerate}
            \item The sum of the log of the number of distinct codewords in the spaces in the subspaces $C'_1, \dots C'_{\ell}$ is at least $k / 2^i$. If this is the case, then we will recover all of the rows in the support. This is because by induction, after the $i-1$st iteration, there will be at most $(kd / 2^{i-1})$ rows remaining in $S$ to be recovered. If at the end of the $i$th iteration, the sum of the logs of the number of distinct codewords of $C'_1, \dots C'_{\ell}$ is at least $k / 2^i$, then this means we will have recovered $2 k d / 2^i = kd / 2^{i-1}$ new rows from $S$ in the $i$th iteration, which will be the entirety of the remaining rows.
            \item The remaining the sum of the logs of the number of distinct codewords in the spaces $C'_1, \dots C'_{\ell}$ is $< k / 2^i$. In this case, by the same argument as above, the remaining number of rows that must be recovered from $S$ is bounded by $kd / 2^i$.
        \end{enumerate}
        After $\log(n) + \log(q)$ iterations, the remaining number of distinct codewords contributed by rows (not already recovered) in $S$ goes to $0$, and therefore we must have recovered all of $S$. 

        This yields the desired claim. 
\end{proof}

\begin{proof}[Proof of \cref{thm:efficientKarger}]
    The set $S$ exists because of \cref{cor:moduleKarger}. The efficient algorithm for recovering the set $T$ of the given size follows from \cref{clm:containsBadSet}.
\end{proof}

\section{Sparsifying Codes Over $\Z_q$}\label{sec:sparsificationAlgorithm}

In this section we use the decomposition theorem of the previous section to derive a proof of \cref{thm:affine-abelian} for the special case of the group $\Z_q$ and a dependence on $\log^2(q)$.
Specifically, given an arbitrary (possibly weighted) generating matrix $G \in \Z_q^{m \times n}$, our goal is to return a sparsifier for this generating matrix. Note that our codeword counting bounds as stated hold only for \emph{unweighted} codes. Thus, our first stepn is generalizing the weight-class decomposition technique from \cite{KPS24} to the non-field setting. 

\subsection{Weighted Decomposition}

We give an algorithm to show that given a weighted code, decomposes the code cleanly into weight classes. To this end, we suggest the following procedure upon being given a code of length $m$ and at most $q^n$ distinct codewords in Algorithm \ref{alg:WeightClassDecomposition}. 

\begin{algorithm}
	\caption{WeightClassDecomposition$(\calC, \eps, \alpha)$}\label{alg:WeightClassDecomposition}
	Let $E_i$ be all coordinates of $\calC'$ that have weight between $[\alpha^{i-1}, \alpha^i]$. \\
	Let $\Dodd = E_1 \cup E_3 \cup E_5 \cup \dots$, and let $\Deven = E_2 \cup E_4 \cup E_6 \cup \dots$. \\
	\Return{$\Dodd, \Deven$}.
\end{algorithm}

Next, we prove some facts about this algorithm.

\begin{lemma}\label{clm:PreserveApproxWeightDecomp}
	Consider a code $\calC$ with at most $q^n$ distinct codewords and length $m$. Let 
	\[
	\Dodd, \Deven = \mathrm{WeightClassDecomposition}(\calC, \eps, (m / \eps)^3).
	\] To get a $(1 \pm \eps)$-sparsifier for $\calC$, it suffices to get a $(1 \pm \eps)$ sparsifier to each of $\Dodd, \Deven$.
\end{lemma}

\begin{proof}
	First, note that the creation of $\Dodd, \Deven$ forms a \emph{vertical} decomposition of the code $\calC'$. Thus, by \cref{clm:verticalDecomp}, if we have a $(1\pm \eps)$ sparsifier for each of $\Dodd, \Deven$, we have a $(1\pm \eps)$ sparsifier to $\calC$.
\end{proof}

Because of the previous claim, it is now our goal to create sparsifiers for $\Dodd, \Deven$. Without loss of generality, we will focus our attention only on $\Deven$, as the procedure for $\Dodd$ is exactly the same (and the proofs will be the same as well). At a high level, we will take advantage of the fact that 
\[
\Deven = E_2 \cup E_4 \cup \dots ,
\]
where each $E_i$ contains coordinates (i.e. rows of the generating matrix) with weights in the range $[\alpha^{i-1}, \alpha^i]$ where we have now set $\alpha =(m / \eps)^3$. Because the code $\calC$ is of length $m$ we know that the length of $\Deven$ is also $\leq m$. Thus, whenever a codeword $c \in \Deven$ is non-zero in a coordinate in $E_i$, we can effectively ignore all coordinates of lighter weights $E_{i-2}, E_{i-4}, \dots$. This is because any coordinate in $E_{\leq i-2}$ has weight which is at most a $\frac{\eps^3}{m^3}$ fraction of any single coordinate in $E_{i}$. Because there are at most $m$ coordinates in $\Deven$, it follows that the total possible weight of all rows in $E_{\leq i-2}$ is still at most a $O(\eps / m)$ fraction of the weight of a single coordinate in $E_{i}$. Thus, we will argue that when we are creating a sparsifier for codewords that are non-zero in a row in $E_i$, we will be able to effectively ignore all rows corresponding to $E_{\leq i-2}$. Thus, our decomposition is quite simple: we first restrict our attention to $E_i$ and create a $(1 \pm \eps)$ sparsifier for these rows. Then, we transform the remaining code such that only codewords which are all zeros on $E_i$ remain. We present this transformation below:

\begin{algorithm}
	\caption{SingleSpanDecomposition$(\Deven, \alpha, i)$}\label{alg:SingleSpanDecomposition}
	Let $E_i$ be all rows of $\Deven$ with weights between $\alpha^{i-1}$ and $\alpha^i$. \\
	Let $G$ be a generating matrix for $\Deven$. \\
	Store $G|_{E_i}$. \\
	Let $G' = G$. \\
	\While{$G'|_{E_i}$ is not all zero}
	{
		Find the first non-zero coordinate of $G'|_{E_i}$, call this $j$. \\
		Set $G' = \Contract(G', j)$. \\
}
	
	\Return{$G|_{E_i}$, $G'|_{\bar{E_i}}$}
\end{algorithm}

\begin{claim}\label{clm:conserveCodewords}
	If the span of $G$ originally had $2^{n'}$ distinct codewords, and the span of $G|_{E_i}$ has $2^{n''}$ distinct codewords, then after \cref{alg:SingleSpanDecomposition}, the span of  $G'|_{\bar{E_i}}$ has $2^{n' - n''}$ distinct codewords.
\end{claim}

\begin{proof}
	After running the above algorithm, $G'$ is entirely $0$ on the rows corresponding to $E_i$, hence it follows that after running the algorithm, $G'$ and $G'|_{\bar{E_i}}$ have the same number of distinct codewords. Now, we will argue that the span of $G'$ has at most $2^{n' - n''}$ distinct codewords. Indeed, for any codeword $c$ in the span of $G'$, $c$ is also in the span of the original $G$. However, for this same $c$, in the original $G$ we could add any of the $2^{n'}$ distinct vectors which are non-zero on the rows corresponding to $E_i$. Thus, the span of $G$ must have at least $2^{n'}$ times as many distinct codewords as $G'$. This concludes the claim. 
\end{proof}

\begin{claim}
	For any codeword $c$ in the span of $G$, if $c$ is zero in the coordinates corresponding to $E_i$, then $c$ is still in the span of $G'$ after the contractions of \cref{alg:SingleSpanDecomposition}.
\end{claim}

\begin{proof}
	This follows from \cref{clm:stillInSpan}. If a codeword is $0$ in a coordinate which we contract on, then it remains in the span. Hence, if we denote by $c$ a codeword which is zero in all of the coordinates of $E_i$, then $c$ is still in the span after contracting on the coordinates in $E_i$.
\end{proof}

\begin{claim}\label{clm:recursivebreakdown}
	In order to get a $(1 \pm \eps)$ approximation to $\Deven$, it suffices to combine a $(1 \pm \eps / 2)$ approximation to $G|_{E_i}$ and a $(1 \pm \eps)$ approximation to $G'|_{\bar{E_i}}$. 
\end{claim}

\begin{proof}
	For any codeword $c \in \Deven$ which is non-zero on the coordinates $E_i$, it suffices to get a $(1 \pm \eps/2)$ approximation to their weight on $G|_{E_i}$, as this makes up at least a $(1 \pm \eps / m)$ fraction of the overall weight of the codeword.
	
	For any codeword $c \in \Deven$ which is zero on rows $E_i$, then $c$ is still in the span of $G'$, and in particular, its weight when generated by $G$ is exactly the same as its weight in $G'_{\bar{E_i}}$ (as it is zero in the coordinates corresponding to $E_i$, we can ignore these coordinates). Hence, it suffices to get a $(1 \pm \eps)$ approximation to the weight of $c$ on $G'_{\bar{E_i}}$. Taking the union of these two sparsifiers will then yield a sparsifier for every codeword in the span of $\Deven$.
\end{proof}

This then yields the decomposition we will use to construct sparsifiers:

\begin{algorithm}
    \caption{SpanDecomposition$(\Deven, \alpha)$}\label{alg:SpanDecomposition}
    Let $\Deven' = \Deven = E_2 \cup E_4 \cup E_6 \dots$. \\
    Let $S = \{ \}$.\\
    \While{$\Deven'$ is not empty}{
    Let $i$ be the largest integer such that $E_i$ is non-empty in $\Deven'$. \\
    Let $G|_{E_i}, G'|_{\bar{E_i}} =$ SingleSpanDecomposition$(\Deven', \alpha, i)$. \\
    Let $\Deven'$ be the span of $G'|_{\bar{E_i}}$, and let $H_i = G|_{E_i}$. \\
    Add $i$ to $S$.
    }
    \Return{$S, H_i$ for every $i \in S$}
\end{algorithm}

\begin{claim}\label{clm:conserveRank}
    Let $S, H_i$ be as returned by \cref{alg:SpanDecomposition}. Then, $\sum_{i \in S} \log(|\text{Span}(H_i) |) =  \log(|\text{Span}(\Deven)|)$.
\end{claim}

\begin{proof}
    This follows because from line 5 of \cref{alg:SpanDecomposition}. In each iteration, we store $G|_{E_i}$, and iterate on $G'|_{E_i}$. From \cref{clm:conserveCodewords}, we know that
    \[
    (\text{number of distinct codewords in } G|_{E_i}) \cdot (\text{number of distinct codewords in } G'|_{\bar{E_i}})\]\[ = (\text{number of distinct codewords in } G),\]
    thus taking the log of both sides, we can see that the sum of the logs of the number of distinct codewords is preserved.
\end{proof}

\begin{lemma}\label{clm:OnlyApproxHi}
    Suppose we have a code of the form $\Deven$ created by \cref{alg:WeightClassDecomposition}. Then, for $S, (H_i)_{i \in S} = \mathrm{SpanDecomposition}(\Deven)$, it suffices to get a $(1 \pm \eps/2)$ sparsifier for each of the $H_i$ in order to get a $(1 \pm \eps)$ sparsifier for $\Deven$.
\end{lemma}

\begin{proof}
This follows by inductively applying \cref{clm:recursivebreakdown}. Let our inductive hypothesis be that getting a $(1 \pm \eps/2)$ sparsifier to each of codes returned of \cref{alg:SpanDecomposition} suffices to get a $(1 \pm \eps)$ sparsifier to the code overall. We will induct on the number of recursive levels that \cref{alg:SpanDecomposition} undergoes (i.e., the number of distinct codes returned by the algorithm). In the base case, we assume that there is only one level of recursion, and that \cref{alg:SpanDecomposition} simply returns a single code. Clearly then, getting a $(1 \pm \eps/2)$ sparsifier to this code suffices to sparsify the code overall. 

Now, we prove the claim inductively. Assume the algorithm returns $\ell$ codes. After the first iteration, we decompose $\Deven$ into $H_i = G|_{E_i}$ and $G'|_{\bar{E_i}}$. By \cref{clm:recursivebreakdown}, it suffices to get a $(1 \pm \eps/2)$ sparsifier to $H_i$, while maintaining a $(1 \pm \eps)$ sparsifier to $G'|_{\bar{E_i}}$. By invoking our inductive claim, it then suffices to get a $(1 \pm \eps/2)$ sparsifier for the $\ell -1$ codes returned by the algorithm on $G'|_{\bar{E_i}}$. Thus, we have proved our claim.
\end{proof}

\subsection{Dealing with Bounded Weights}

From the previous section, we know that for a code of length $m$, we can decompose the code into disjoint sections, where each section has weights bounded in the range $[\alpha^i, \alpha^{i+1}]$. In this section, we will show how we can sparsify these codes with weights in a bounded range.

Let us consider any $H_i$ that is returned by \cref{alg:SpanDecomposition}, when called with parameter $\alpha$. By construction, $H_i$ will contain weights only in the range $[\alpha^{i-1}, \alpha^{i}]$ and will have at most $O(m)$ coordinates. In this subsection, we will show how we can turn $H_i$ into an unweighted code with at most $O(m \cdot \alpha / \eps)$ coordinates by essentially repeating each coordinate $j$ about $w(j)$ times (where $w(j)$ is the weight of the corresponding coordinate in $H_i$). First, note however, that we can simply pull out a factor of $\alpha^{i-1}$, and treat the remaining graph as having weights in the range of $[1, \alpha]$. Because multiplicative approximation does not change under multiplication by a constant, this can be done without loss of generality. Formally, consider the following algorithm:

\begin{algorithm}
    \caption{MakeUnweighted$(\calC, \alpha, i, \eps)$}\label{alg:MakeUnweighted}
    Divide all edge weights in $\calC$ by $\alpha^{i-1}$. \\
    Make a new unweighted code $\calC'$ by duplicating every coordinate $j$ of $\calC$ $\lfloor 10 w(j) / \eps \rfloor$ times. \\
    \Return{$\calC', \alpha^{i-1} \cdot \frac{\eps}{10}$}
\end{algorithm}

\begin{lemma}\label{clm:PreserveApproxUnweighted}
Consider a code $\calC$ with weights bounded in the range $[\alpha^{i-1}, \alpha^i]$. To get a $(1 \pm \eps)$ sparsifier for $\calC$ it suffices to return a $(1 \pm \eps / 10)$ sparsifier for $\calC' =$ MakeUnweighted$(\calC, \alpha, i, \eps)$ weighted by $\alpha^{i-1} \cdot \frac{\eps}{10}$.
\end{lemma}

\begin{proof}
    It suffices to show that $\calC'$ is $(1 \pm \eps/10)$ sparsifier for $\calC$, as our current claim will then follow by Claim \ref{clm:composingApproximations} (composing approximations). Now, to show that $\calC'$ is $(1 \pm \eps/10)$ sparsifier for $\calC$, we will use Claim \ref{clm:verticalDecomp} (vertical decomposition of a code), and show that in fact the weight contributed by every coordinate in $\calC$ is approximately preserved by the copies of the coordinate introduced in $\calC'$.
    
    Without loss of generality, let us assume that $i = 1$, as otherwise pulling out the factor of $\alpha^{i-1}$ in the weights clearly preserves the weights of the codewords. Indeed, for every coordinate $j$ in $\calC$, let $w(j)$ be the corresponding weight on this coordinate, and consider the corresponding $\lfloor 10 w(j) / \eps \rfloor$ coordinates in $\calC'$. We will show that the contribution from these coordinates in $\calC'$, when weighted by $\eps/10$, is a $(1 \pm \eps/10)$ approximation to the contribution from $j$. 

    So, consider an arbitrary coordinate $j$. Then,
    \[
    \frac{10w}{\eps} - 1 \leq \lfloor 10 w(j) / \eps \rfloor \leq \frac{10w}{\eps}.
    \]

    When we normalize by $\frac{\eps}{10}$, we get that the combined weight of the new coordinates $w'$ satisfies 
    \[
    w(j) - \eps/10 \leq w'(j) \leq w(j).
    \]

    Because $w \geq 1$, it follows that this yields a $(1 \pm \eps/10)$ sparsifier, and we can conclude our statement.
\end{proof}

\begin{claim}\label{clm:boundUnweightedSupport}
    Suppose a code $\calC$ of length $m$ has weight ratio bounded by $\alpha$, and minimum weight $\alpha^{i-1}$. Then, MakeUnweighted$(\calC, \alpha, i, \eps)$ yields a new unweighted code of length $O(m \alpha / \eps)$.
\end{claim}

\begin{proof}
    Each coordinate is repeated at most $O(\alpha / \eps)$ times.
\end{proof}

\subsection{Sparsifiers for Codes of Polynomial Length}

In this section, we introduce an efficient algorithm for sparsifying codes. We will take advantage of the decomposition proved in \cref{cor:moduleKarger} in conjunction with the following claim:

\begin{claim}\label{clm:preserveWeightSubsample}
    Suppose $\calC$ is a code with at most $q^{n}$ distinct codewords over $\Z_q$, and let $b \geq 1$ be an integer such that for any integer $\alpha \geq 1$, the number of codewords of weight $\leq \alpha b$ is at most $\binom{n\log(q)}{\alpha } \cdot q^{\alpha + 1} \leq (qn)^{2\alpha}$. Suppose further that the minimum distance of the code $\calC$ is $b$. Then, sampling the $i$th coordinate of $\calC$ at rate $p_i \geq \frac{\log(n) \log(q) \eta}{b \eps^2}$ with weights $1/p_i$ yields a $(1 \pm \eps)$ sparsifier with probability $1 - 2^{-(0.19\eta - 110) \log n} \cdot n^{-101}$.
\end{claim}

\begin{proof}

    Consider any codeword $c$ of weight $[\alpha b / 2, \alpha b]$ in $\calC$. We know that there are at most $(qn)^{\alpha}$ codewords that have weight in this range. The probability that our sampling procedure fails to preserve the weight of $c$ up to a $(1 \pm \eps)$ fraction can be bounded by Claim \ref{clm:concentrationBound}. Indeed,
    \[
    \Pr[\text{fail to preserve weight of } c] \leq 2e^{-0.38 \cdot \eps^2 \cdot \frac{\alpha b}{2} \cdot \frac{\eta \log (n) \log(q)}{\eps^2 b}} = 2e^{-0.19 \alpha \eta \log (n)\log(q)}.
    \]
    Now, let us take a union bound over the at most $(qn)^{2\alpha}$ codewords of weight between $[\alpha b / 2, \alpha]$. Indeed,
    \begin{align*}
    \Pr[\text{fail to preserve any } c \text{ of weight } [\alpha b / 2, \alpha b]] &\leq 2^{2\alpha \log (qn)} \cdot 2e^{-0.19 \alpha \eta \log (n)\log(q)} \\
     & \leq 2^{\alpha \cdot (-0.19\eta + 2) \log (n)\log(q)} \\
     & \leq 2^{\alpha \cdot (-0.19\eta + 2) \log (n)} \\
     & \leq 2^{-(0.19\eta - 110) \alpha \log n} \cdot 2^{-108 \alpha \log n} \\
     & \leq 2^{-(0.19\eta - 110) \log n} \cdot n^{-108 \alpha},
    \end{align*}
    where we have chosen $\eta$ to be sufficiently large. Now, by integrating over $\alpha \geq 1$, we can bound the failure probability for any integer choice of $\alpha$ by $2^{-(0.19\eta - 110) \log n} \cdot n^{-101}$.
\end{proof}

Next, we consider \cref{alg:CodeDecomposition}:

\begin{algorithm}[H]
\caption{CodeDecomposition$(\calC, d)$}\label{alg:CodeDecomposition}
Let $T$ be $\cup_i T_i$ for $T_i$ the sets of coordinates returned by ConstructSpanningSubsets$(\calC, 2d(\log(n) + \log(q)))$. \\
Let $\calC'$ be the code $\calC$ after removing the set of coordinates $T$. \\
\Return{$T, \calC'$}
\end{algorithm}

Intuitively, the set $T$ returned by \cref{alg:CodeDecomposition} contains all of the \say{bad} rows which were causing the violation of the codeword counting bound. We know that if we removed \emph{only} the true set of bad rows, denoted by $S$, then we could afford to simply sample the rest of the code at rate roughly $1/d$ while preserving the weights of all codewords. Thus, it remains to show that when we remove $T$ (a superset of $S$) that this property still holds. More specifically, we will consider the following algorithm:

\begin{algorithm}
    \caption{CodeSparsify$(\calC \subseteq \Z_q^{m}, n, \eps, \eta)$}\label{alg:CodeSparsify}
    Let $m$ be the length of $\calC$. \\
    \If{$m \leq 100 \cdot n \cdot \eta \log^2(n)\log^2(q) / \eps^2$}{\Return{$\calC$}}
    Let $d = \frac{m \eps^2}{2\eta \cdot n \log^2 (n) \log^2(q)}$. \\
    Let $T, \calC' = \text{CodeDecomposition}(\calC, \sqrt{d} \cdot \eta \cdot \log(n) \log(q) / \eps^2)$.
    Let $\calC_1 = \calC|_T$.
    Let $\calC_2$ be the result of sampling every coordinate of $\calC'$ at rate $1 / \sqrt{d}$. \\
    \Return{$\mathrm{CodeSparsify}(\calC_1, n, \eps, \eta) \cup \sqrt{d} \cdot \mathrm{CodeSparsify}(\calC_2, n, \eps, \eta)$ }
\end{algorithm}

\begin{lemma}\label{clm:overallSpace}
    In Algorithm \ref{alg:CodeSparsify}, starting with a code $\calC$ of size $2d n \log^2 (n) \log^2(q) / \eps^2$, after $i$ levels of recursion, with probability $1 - 2^{i} \cdot 2^{-\eta n}$, the code being sparsified at level $i$, $\calC^{(i)}$ has at most 
    \[
    2(1 + 1 / 2 \log \log (n))^{i} \cdot d^{1/2^i} \cdot \eta \cdot n \log^2(n)\log^2(q) / \eps^2
    \]
    surviving coordinates.
\end{lemma}

\begin{proof}
    Let us prove the claim inductively. For the base case, note that in the $0$th level of recursion the number of surviving coordinates in $\calC^{(0)} = \calC$ is $d \cdot 2n \log^2 (n) \log^2(q) / \eps^2$, so the claim is satisfied trivially.

    Now, suppose the claim holds inductively. Let $\calC^{(i)}$ denote a code that we encounter in the $i$th level of recursion, and suppose that it has at most \[
    2(1 + 1 / 2 \log \log (n))^{i} \cdot d^{1/2^i} \cdot \eta \cdot n \log^2(n) \log^2(q) / \eps^2
    \]
    coordinates. Denote this number of coordinates by $\ell$. Now, if this number is smaller than $100 n \eta \log^2(n) \log^2(q) / \eps^2$, we will simply return this code, and there will be no more levels of recursion, so our claim holds vacuously. Instead, suppose that this number is larger than $100 n \eta \log^2 (n) \log^2(q) / \eps^2$, and let $d' = \frac{\ell \eps^2}{2\eta n \log^2(n) \log^2(q)} \leq (1 + 1 / 2 \log \log (n))^{i} \cdot d^{1/2^i}$.

    Then, we decompose $\calC^{(i)}$ into two codes, $\calC_1$ and $\calC_2$. $\calC_1$ is the restriction of $\calC$ to the set of disjoint maximum spanning subsets. By construction, we know that $T$ is constructed by calling ConstructSpanningSubsets with parameter $\sqrt{d'}\eta \log(n) \log(q) / \eps^2$, and therefore
    \[
    |T| \leq 2 \sqrt{d'}n \eta \log^2(n) \log^2(q) / \eps^2 \leq 2 (1 + 1 / 2 \log \log (n))^{i} n \eta d^{1/2^{i+1}}\log^2(n) \log^2(q) / \eps^2 ,
    \]
    satisfying the inductive claim. 

    For $\calC_2$, we define random variables $X_1 \dots X_{\ell}$ for each coordinate in the support of $\calC_2$. $X_i$ will take value $1$ if we sample coordinate $i$, and it will take $0$ otherwise. Let $X= \sum_{i = 1}^{\ell} X_i$, and let $\mu = \E[X]$. Note that 
    \[
    \frac{\mu^2}{\ell} = \left ( \frac{\ell}{\sqrt{d'}} \right )^2 / {\ell} = \frac{\ell}{d'} \geq \eta \cdot n \cdot \log^2(n) \log^2(q) / \eps^2.
    \]

    Now, using Chernoff, \[
    \Pr[X \geq (1 + 1 / 2 \log \log (n)) \mu] \leq e^{\frac{-2}{4 \log^2 \log(n)} \cdot \eta \cdot n \cdot \log(n) \log(q) / \eps^2} \leq 2^{-\eta n}, 
    \]
    as we desire. Since $\mu = \ell / \sqrt{d'} \leq (1 + 1 / 2 \log \log (n))^{i} \cdot d^{1/2^{i+1}} \cdot \eta \cdot n\log^2(n) \log^2(q) / \eps^2$, we conclude our result. 
    
    Now, to get our probability bound, we also operate inductively. Suppose that up to recursive level $i-1$, all sub-codes have been successfully sparsified to their desired size. At the $i$th level of recursion, there are at most $2^{i-1}$ codes which are being probabilistically sparsified. Each of these does not exceed its expected size by more than the prescribed amount with probability at most $2^{-\eta n}$. Hence, the probability all codes will be successfully sparsified up to and including the $i$th level of recursion is at least $1 - 2^{i-1} 2^{-\eta n} - 2^{i-1}2^{-\eta n} = 1 - 2^i 2^{-\eta n}$. 
\end{proof}

\begin{lemma}\label{clm:goodApprox1Iter}
For any iteration of Algorithm \ref{alg:CodeSparsify} called on a code $\calC$, $\calC_1 \cup \sqrt{d} \cdot \calC_2$ is a $(1 \pm \eps)$ sparsifier to $\calC$ with probability at least $1 - 2^{-(0.19\eta - 110) \log n} \cdot n^{-101}$.
\end{lemma}

\begin{proof}
First, let us note that the set $T$ returned from \cref{alg:CodeDecomposition} is a superset of the bad set $S$ specified in \cref{cor:moduleKarger} (this follows from \cref{clm:containsBadSet}). Thus, we can equivalently view the procedure as producing three codes: $\calC|_S, \calC|_{T / S}$ and $\calC_{\bar{T}} = \calC'$. For our analysis, we will view this procedure in a slightly different light: we will imagine that first the algorithm removes exactly the bad set of rows $S$, yielding $\calC|_S$ and $\calC_{\bar{S}}$. Now, for this second code, $\calC_{\bar{S}}$, we know the code-word counting bound will hold, and in particular, random sampling procedure will preserve codeword weights with high probability. However, our procedure is \emph{not} uniformly sampling the coordinates in $\calC_{\bar{S}}$, because some of these coordinates are in $T / S$, and thus are preserved exactly (i.e. with probability $1$). For this, we will take advantage of the fact that preserving coordinates with probability $1$ is \emph{strictly better} than sampling at any rate $<1$. Thus, we will still be able to argue that the ultimate result $\calC_1 \cup \sqrt{d} \cdot \calC_2$ is a $(1 \pm \eps)$ sparsifier to $\calC$ with high probability. 

    As mentioned above, we start by noting that $\calC', \calC|_S, \calC_{T / S}$ form a \emph{vertical} decomposition of $\calC$. $\calC|_S$ is preserved exactly, so we do not need to argue concentration of the codewords on these coordinates. Hence, it suffices to show that $\sqrt{d} \cdot \calC_1 \cup \calC_{T / S}$ is a $(1 \pm \eps)$-sparsifier to $\calC' \cup \calC_{T / S}$. 

    To see that $\sqrt{d} \cdot \calC_1 \cup \calC_{T / S}$ is a $(1 \pm \eps)$-sparsifier to $\calC' \cup \calC_{T / S}$, first note that every codeword in $\calC' \cup \calC_{T / S}$ is of weight at least $\sqrt{d} \cdot \eta \cdot \log(n) \log(q) / \eps^2$. This is because if there were a codeword of weight smaller than this, there would exist a subcode of $\calC' \cup \calC_{T / S}$ with $2$ distinct codewords, and support bounded by $\sqrt{d} \cdot \eta \cdot \log(n) \log(q) / \eps^2$. But, because we have removed the set $S$ of bad rows, we know that there can be no such sub-code remaining in $\calC' \cup \calC_{T / S}$. Thus, every codeword in $\calC' \cup \calC_{T / S}$ is of weight at least $\sqrt{d} \cdot \eta \cdot \log(n) \log(q) / \eps^2$. 

    Now, we can invoke Claim \ref{clm:preserveWeightSubsample} with $b = \sqrt{d} \eta \log(n) \log(q) / \eps^2$. Note that the hypothesis of Claim \ref{clm:preserveWeightSubsample} is satisfied by virtue of our code decomposition. Indeed, we removed coordinates of the code such that in the resulting $\calC' \cup \calC_{T / S}$, for any $\alpha \geq 1$, there are at most $(qn)^{2\alpha}$ codewords of weight $\leq \alpha \sqrt{d} \eta \log(n) \log(q) / \eps^2$. Using the concentration bound of Claim \ref{clm:preserveWeightSubsample} yields that with probability at least $1 - 2^{-(0.19\eta - 110) \log n} \cdot n^{-101}$, when samplin every coordinate at rate $\geq 1/ \sqrt{d}$ the resulting sparsifier for $\calC' \cup \calC_{T / S}$ is a $(1 \pm \eps)$ sparsifier, as we desire. Note that we are using the fact that every coordinate is sampled with probability $\geq 1/ \sqrt{d}$ (in particular, those in $T -S$ are sampled with probability $1$).
\end{proof}

\begin{corollary}\label{clm:overallCodeAccuracy}
    If Algorithm \ref{alg:CodeSparsify} achieves maximum recursion depth $\ell$ when called on a code $\calC$, and $\eta > 600$, then the result of the algorithm is a $(1 \pm \eps)^{\ell}$ sparsifier to $\calC$ with probability $\geq 1 - (2^{\ell}-1) \cdot2^{-(0.19\eta - 110) \log n} \cdot n^{-101}$
\end{corollary}

\begin{proof}
    We prove the claim inductively. Clearly, if the maximum recursion depth reached by the algorithm is $0$, then we have simply returned the code itself. This is by definition a $(1 \pm \eps)^{0}$ sparsifier to itself.

    Now, suppose the claim holds for maximum recursion depth $i - 1$. We will show it holds for maximum recursion depth $i$. Let the code we are sparsifying be $\calC$. We break this into $\calC_1$, $\calC'$, and sparsify these. By our inductive claim, with probability $1 - (2^{i-1}-1) \cdot 2^{-(0.19\eta - 110) \log n} \cdot n^{-101}$ each of the sparsifiers for $\calC_1, \calC'$ are $(1 \pm \eps)^{i-1}$ sparsifiers. Now, by Lemma \ref{clm:goodApprox1Iter} and our value of $\eta$, $\calC_1, \calC'$ themselves together form a $(1 \pm \eps)$ sparsifier for $\calC$ with probability $1 - 2^{-(0.19\eta - 110) \log n}\cdot n^{-101}$. So, by using Claim \ref{clm:composingApproximations}, we can conclude that with probability $1 - (2^i -1) \cdot 2^{-(0.19\eta - 110) \log n} \cdot n^{-101}$, the result of sparsifying $\calC_1, \calC'$ forms a $(1 \pm \eps)^i$ approximation to $\calC$, as we desire.
\end{proof}

We can then state the main theorem from this section:

\begin{theorem}\label{thm:codeSparsifyGeneralLength}
    For a code $\calC$ over $\Z_q$ with at most $q^n$ distinct codewords, and length $m$, Algorithm \ref{alg:CodeSparsify} creates a $(1 \pm \eps)$ sparsifier for $\calC$ with probability $1 - \log (m) \cdot 2^{-(0.19\eta - 110) \log n} \cdot n^{-100}$ with at most \[
    O(n \eta \log(n) \log^2(q) \log^2(m) (\log \log (m))^2 / \eps^2)
    \]
    coordinates.
\end{theorem}

\begin{proof}
    For a code of with $q^n$ distinct codewords, and length $m$, this means that our value of $d$ as specified in the first call to Algorithm \ref{alg:CodeSparsify} is at most $m$ as well. As a result, after only $\log \log m$ iterations, $d = m^{1 / 2^{\log \log m}} = m^{1 / \log m} = O(1)$. So, by Corollary \ref{clm:overallCodeAccuracy}, because the maximum recursion depth is only $\log \log m$, it follows that with probability at least $1 - (2^{\log \log m} -1) \cdot 2^{-(0.19\eta - 110) \log n} \cdot n^{-101}$, the returned result from Algorithm \ref{alg:CodeSparsify} is a $(1 \pm \eps)^{\log \log m}$ sparsifier for $\calC$.

    Now, by Lemma $\ref{clm:overallSpace}$, with probability $\geq 1 - 2^{\log \log m} \cdot 2^{-\eta n} \geq 1 - \log (m) \cdot 2^{-(0.19\eta - 110) \log n} \cdot  2^{-n}$, every code at recursive depth $\log \log m$ has at most 
    \[
    (1 + 1 / 2 \log \log (n))^{\log\log m} \cdot m^{1/\log m} \cdot \eta \cdot n \log(n)\log(q) / \eps^2 = O(n \eta \log (n) \log^2(q) \cdot e^{\frac{\log \log m}{\log \log n}} / \eps^2)
    \]
    coordinates. Because the ultimate result from calling our sparsification procedure is the \emph{union} of all of the leaves of the recursive tree, the returned result has size at most 
    \[
    \log (m)  \cdot e^{\frac{\log \log m}{\log \log n}} \cdot O(n \eta \log (n)\log^2(q) / \eps^2) = O(n \eta \log(n) \log^2(q) \log^2(m) / \eps^2),
    \]
    with probability at least $1 - \log (m) \cdot 2^{-(0.19\eta - 110) \log n} \cdot n^{-101}$. 

    Finally, note that we can replace $\eps$ with a value $\eps' = \eps / 2 \log \log m$. Thus, the resulting sparsifier will be a $(1 \pm \eps')^{\log \log m} \leq (1 \pm \eps)$ sparsifier, with the same high probability.

    Taking the union bound of our errors, we can conclude that with probability $1 - \log (m) \cdot 2^{-(0.19\eta - 110) \log n} \cdot n^{-100}$, Algorithm \ref{alg:CodeSparsify} returns a $(1 \pm \eps)$ sparsifier for $\calC$ that has at most 
    \[
    O(n \eta \log(n)\log^2(q) \log^2(m) (\log \log (m))^2 / \eps^2)
    \]
    coordinates.
\end{proof}

However, as we will address in the next subsection, this result is not perfect: 
\begin{enumerate}
    \item For large enough $m$, there is no guarantee that this probability is $\geq 0$ unless $\eta$ depends on $m$.
    \item For large enough $m$, $\log^2(m)$ may even be larger than $n$.
\end{enumerate}

\subsection{Final Algorithm}

Finally, we state our final algorithm in Algorithm \ref{alg:FinalCodeSparsify}, which will create a $(1 \pm \eps)$ sparsifier for any code $\calC \subseteq \Z_q^m$ with $\leq q^n$ distinct codewords preserving only $\widetilde{O}(n \log^2(q) / \eps^2)$ coordinates. Roughly speaking, we start with a weighted code of arbitrary length, use the weight class decomposition technique, sparsify the decomposed pieces of the code, and then repeat this procedure now that the code will have a polynomial length. Ultimately, this will lead to the near-linear size complexity that we desire. We write a single iteration of this procedure below:

\begin{algorithm}
    \caption{FinalCodeSparsify$(\calC, \eps)$}\label{alg:FinalCodeSparsify}
    Let $n$ be $\log_q(|\calC|)$. \\
    Let $m$ be the length of the code. \\
    Let $\alpha = (m/\eps)^3$, and $\Dodd, \Deven = $WeightClassDecomposition$(\calC, \eps, \alpha)$. \\
    Let $S_{\text{even}}, \{ H_{\text{even}, i} \} = $SpanDecomposition$(\Deven, \alpha)$. \\
    Let $S_{\text{odd}}, \{ H_{\text{odd}, i} \} = $SpanDecomposition$(\Dodd, \alpha)$. \\
    \For{$i \in S_{\text{even}}$}{
    Let $\widehat{H}_{\text{even}, i}, w_{\text{even}, i}=$ MakeUnweighted$({H}_{\text{even}, i}, \alpha, i, \eps/8)$. \\
    Let $\widetilde{H}_{\text{even}, i} = $CodeSparsify$(\widehat{H}_{\text{even}, i}, \log_q(|\mathrm{Span}(\widehat{H}_{\text{even}, i})|), \eps/80, 100 (\log(m / \eps) \log \log(q))^2)$.
    }
    \For{$i \in S_{\text{odd}}$}{
    Let $\widehat{H}_{\text{odd}, i}, w_{\text{odd}, i}=$ MakeUnweighted$({H}_{\text{odd}, i}, \alpha, i, \eps/8)$. \\
    Let $\widetilde{H}_{\text{odd}, i} = $CodeSparsify$(\widehat{H}_{\text{odd}, i}, \log_q(|\mathrm{Span}(\widehat{H}_{\text{odd}, i})|), \eps/80, 100 (\log(m/\eps) \log \log(q))^2)$.
    }
    \Return{$\bigcup_{i \in S_{\text{even}}} \left ( w_{\text{even}, i} \cdot \widetilde{H}_{\text{even}, i} \right ) \cup \bigcup_{i \in S_{\text{odd}}} \left ( w_{\text{odd}, i} \cdot \widetilde{H}_{\text{odd}, i} \right )$}
\end{algorithm}

First, we analyze the space complexity. WLOG we will prove statements only with respect to $\Deven$, as the proofs will be identical for $\Dodd$.

\begin{claim}\label{clm:singleCallSparse}
    Suppose we are calling Algorithm \ref{alg:FinalCodeSparsify} on a code $\calC$ with $q^n$ distinct codewords. Let $\teveni = \log_q(\text{Span}((\widehat{H}_{\text{even}, i}))|)$ from each call to the for loop in line 5. 
    
    For each call $\widetilde{H}_{\text{even}, i} = $CodeSparsify$(\widehat{H}_{\text{even}, i}, \log_q(|\mathrm{Span}(\widehat{H}_{\text{even}, i})|), \eps/10, 100 (\log(n/\eps) \log \log(q))^2)$ in Algorithm \ref{alg:FinalCodeSparsify}, the resulting sparsifier has
    \[
    O\left ( \teveni \log(\teveni) \log^4(m/\eps) \log^2(q) (\log \log (m/\eps) \log \log(q))^2 / \eps^2 \right )
    \]
    coordinates with probability at least $1 - \log(m/ \eps) \cdot 2^{-\Omega(\log^2(m / \eps) (\log \log (q))^2)}$.
\end{claim}

\begin{proof}
    We use several facts. First, we use Theorem \ref{thm:codeSparsifyGeneralLength}. Note that the $m$ in the statement of Theorem \ref{thm:codeSparsifyGeneralLength} is actually a $\text{poly}(m / \eps)$ because $\alpha = m^3 / \eps^3$, and we started with a weighted code of length $O(m)$. So, it follows that after using Algorithm \ref{alg:MakeUnweighted}, the support size is bounded by $O(m^4 / \eps^3)$. We've also added the fact that $\eta$ is no longer a constant, and instead carries $O((\log(m/\eps) \log \log(q))^2)$, and carried this through to the probability bound. 
\end{proof}

\begin{lemma}\label{clm:overallDevenSize}
    In total, the combined number of coordinates over $i \in S_{\text{even}}$ of all of the $\widetilde{H}_{\text{even}, i}$ is at most $\widetilde{O}(n \log^4(m)\log^2(q) / \eps^2)$ with probability at least $1 - \log(m\log(q) / \eps) \cdot 2^{-\Omega(\log^2(m/ \eps) (\log \log (q))^2)}$.
\end{lemma}

\begin{proof}
    First, we use \cref{clm:conserveRank} to see that 
    \[
    \sum_{i \in S_{\text{even}}} \log_q( | \text{Span}(\widehat{H}_{\text{even}, i}) | ) \leq n.
    \]
    Thus, in total, the combined length (total number of coordinates preserved) of all the $\widetilde{H}_{\text{even}, i}$ is 
    \begin{align*}
        & \sum_{i \in S_{\text{even}}} \text{number of coordinates in }\widehat{H}_{\text{even}, i} \\
        &\leq \sum_{i \in S_{\text{even}}} O\left ( \teveni \log(\teveni) \log^4(m/\eps) \log^2(q) (\log \log (m/\eps) \log \log(q))^2 / \eps^2 \right ) \\
        & \leq \sum_{i \in S_{\text{even}}} (\teveni) \cdot \widetilde{O} \left (\log^4(m)  \log^2(q) / \eps^2 \right ) \\
        & = n \cdot \widetilde{O} \left (\log^4(m)  \log^2(q) / \eps^2 \right ) \\
        & = \widetilde{O}(n \log^4(m)\log^2(q) / \eps^2).
    \end{align*}

    To see the probability bound, we simply take the union bound over all at most $n$ distinct $\widetilde{H}_{\text{even}, i}$, and invoke Claim \ref{clm:singleCallSparse}.
\end{proof}

Now, we will prove that we also get a $(1 \pm \eps)$ sparsifier for $\Deven$ when we run Algorithm \ref{alg:FinalCodeSparsify}.

\begin{lemma}\label{clm:DevenCorrectness}
    After combining the $\widehat{H}_{\text{even}, i}$ from Lines 5-8 in Algorithm \ref{alg:FinalCodeSparsify}, the result is a $(1 \pm \eps/4)$-sparsifier for $\Deven$ with probability at least $1 - \log(m\log(q) / \eps) \cdot 2^{-\Omega(\log^2(m/ \eps) (\log \log (q))^2)}$.
\end{lemma}

\begin{proof}
    
    We use Lemma \ref{clm:OnlyApproxHi}, which states that to sparsify $\Deven$ to a factor $(1 \pm \eps/4)$, it suffices to sparsify each of the $H_{\text{even}, i}$ to a factor $(1 \pm \eps / 8)$, and then combine the results.

    Then, we use Lemma \ref{clm:PreserveApproxUnweighted}, which states that to sparsify any $H_{\text{even}, i}$ to a factor $(1 \pm \eps / 8)$, it suffices to sparsify $\widehat{H}_{\text{even}, i}$ to a factor $(1 \pm \eps / 80)$, where again, $\widehat{H}_{\text{even}, i}$ is the result of calling Algorithm \ref{alg:MakeUnweighted}. Then, we must multiply $\widehat{H}_{\text{even}, i}$ by a factor $\alpha^{i-1} \cdot \eps / 10$.    
    
    Finally, the resulting code $\widehat{H}_{\text{even}, i}$ is now an unweighted code, whose length is bounded by $O(m^4 /\eps^3)$, with at most $q^\teveni$ distinct codewords. The accuracy of the sparsifier then follows from Theorem \ref{thm:codeSparsifyGeneralLength} called with parameter $\eps / 80$.

    The failure probability follows from noting that we take the union bound over at most $n\log(q)$ $H_{\text{even}, i}$. By Theorem \ref{thm:codeSparsifyGeneralLength}, our choice of $\eta$, and the bound on the length of the support being $O(m^4 / \eps^3)$, the probability bound follows.
\end{proof}

For Theorem \ref{thm:codeSparsifyGeneralLength}, the failure probability is characterized in terms of the number of distinct codewords of the code that is being sparsified. However, when we call Algorithm \ref{alg:CodeSparsify} as a sub-routine in Algorithm \ref{alg:FinalCodeSparsify}, we have no guarantee that the number of distinct codewords is $\omega(q)$. Indeed, it is certainly possible that the decomposition in $H_i$ creates $n$ different codes, each with $q$ distinct codewords in their span. Then, choosing $\eta$ to only be a constant, as stated in Theorem \ref{thm:codeSparsifyGeneralLength}, the failure probability could be constant, and taking the union bound over $n$ choices, we might not get anything meaningful. To amend this, instead of treating $\eta$ as a constant in Algorithm \ref{alg:CodeSparsify}, we set $\eta = 100 (\log(m/\eps) \log \log(q))^2$, where now $m$ is the length of the original code $\calC$, \emph{not} in the current code that is being sparsified $H_i$. With this modification, we can then attain our desired probability bounds.

\begin{theorem}\label{thm:mainRestated}
For any code $\calC$ with $q^n$ distinct codewords and length $m$ over $\Z_q$, Algorithm \ref{alg:FinalCodeSparsify} returns a $(1 \pm \eps)$ sparsifier to $\calC$ with $\widetilde{O}(n \log^4(m)\log^2(q) / \eps^2)$ coordinates with probability $\geq 1 - 2^{-\Omega((\log(m/\eps) \log \log(q))^2)}$.
\end{theorem}

\begin{proof}
First, we use Lemma \ref{clm:PreserveApproxWeightDecomp}. This Lemma states that in order to get a $(1 \pm \eps)$ sparsifier to a code $\calC$, it suffices to get a $(1 \pm \eps/4)$ sparsifier to each of $\Deven, \Dodd$, and then combine the results.

    Then, we invoke Lemma \ref{clm:DevenCorrectness} to conclude that with probability $\geq 1 - 2^{-\Omega((\log(m/\eps) \log \log(q))^2)}$, Algorithm \ref{alg:FinalCodeSparsify} will produce $(1 \pm \eps/4)$ sparsifiers for $\Deven, \Dodd$.

    Further, to argue the sparsity of the algorithm, we use Lemma \ref{clm:overallDevenSize}. This states that with probability $\geq 1 - 2^{-\Omega( (\log(m/\eps) \log \log(q))^2)}$, Algorithm \ref{alg:FinalCodeSparsify} will produce code sparsifiers of size $\widetilde{O}(n \log^4(m) \log^2(q) / \eps^2)$ for $\Deven, \Dodd$. 

    Thus, in total, the failure probability is at most $2^{-\Omega((\log(m/\eps) \log \log(q))^2)}$, the total size of the returned code sparsifier is at most $\widetilde{O}(n \log^4(m)\log^2(q) / \eps^2)$, and the returned code is indeed a $(1 \pm \eps)$ sparsifier for $\calC$, as we desire.

    Note that the returned sparsifier may have some duplicate coordinates because of Algorithm \ref{alg:MakeUnweighted}. Even when counting duplicates of the same coordinate separately, the size of the sparsifier will be at most $\widetilde{O}(n \log^4(m) \log^2(q) / \eps^2)$. We can remove duplicates of coordinates by adding their weights to a single copy of the coordinate.
\end{proof}

\begin{claim}\label{clm:finalSparsifyRuntime}
    Running \cref{alg:FinalCodeSparsify} on a code of length $m$ with parameter $\eps$ takes time $\text{poly}(mn\log(q) / \eps)$.
\end{claim}

\begin{proof}
    Let us consider the constituent algorithms that are invoked during the execution of \cref{alg:FinalCodeSparsify}. First, we consider weight class decomposition. This groups rows together by weight (which takes time $\widetilde{O}(\log(m))$). Next, we invoke SpanDecomposition, which then contracts on the rows in the largest weight class. Note that in the worst case, we perform $O(n \log(q))$ contractions, as each contraction reduces the number of codewords by a factor of $\geq 2$. Further, each contraction takes time $O(mn\log(q))$ as the total number of rows is bounded by $m$, and there are at most $n\log(q)$ columns in the generating matrix. Thus, the total runtime of this step is $\widetilde{O}(m n^2 \log^2(q))$.

    The next step is to invoke the algorithm MakeUnweighted. Because the value of $\alpha$ is $m^3 / \eps^3$, this takes time at most $O(m^4 / \eps^3)$ to create the new code with this many rows. 
    
    Finally, we invoke CodeSparsify on a code of length $\leq m^4 / \eps^3$ and with at most $q^{\teveni}$ distinct codewords. Note that there are $\polylog(m)$ nodes in the recursive tree that is built by CodeSparsify. Each such node requires removing the set $T$ which is a set of $\leq \widetilde{O}(\sqrt{m^4 / \eps^3}) = \widetilde{O}(m^2 / \eps^{1.5})$ maximum spanning subsets. Constructing each such subset (by \cref{clm:runtimeContraction}) takes time at most $O((m^4 \eps^3)^2 n \log(q))$. After this step, the subsequent random sampling is efficiently doable. Thus, the total runtime is bounded by $\text{poly}(mn\log(q) / \eps)$, as we desire.

\end{proof}

Note that creating codes of linear-size now simply requires invoking \cref{alg:FinalCodeSparsify} two times (each with parameter $\eps/2$). Indeed, because the length of the code to begin with is $\leq q^n$, this means that after the first invocation, the resulting $(1 \pm \eps/2)$ sparsifier $\calC'$ that we get maintains $\leq \widetilde{O}(n^5 \log^6(q) / \eps^2)$ coordinates. In the second invocation, we get a $(1 \pm \eps/2)$-sparsifier $\calC''$ for $\calC'$, whose length is bounded by $\widetilde{O}(n \log^4(n^5 \log^6(q) / \eps^2) \log^2(q) / \eps^2) = \widetilde{O}(n \log^2(q) / \eps^2)$, as we desire.

Formally, this algorithm can be written as:
\begin{algorithm}
    \caption{Sparsify($\calC, \eps$)}\label{alg:Sparsify}
    $\calC' = $FinalCodeSparsify$(\calC, \eps/2)$. \\
    \Return{FinalCodeSparsify($\calC', \eps/2$)}
\end{algorithm}

\begin{theorem}\label{thm:efficientSparsifyZq}
    \cref{alg:Sparsify} returns a $(1 \pm \eps)$-sparsifier to $\calC$ of size $\widetilde{O}(n \log^2(q) / \eps^2)$ with probability $1 - 2^{-\Omega((\log(n/\eps) \log \log(q))^2)}$ in time $\text{poly}(mn\log(q) / \eps)$.
\end{theorem}

\begin{proof}
    Indeed, because the length of the code to begin with is $\leq q^n$, this means that after the first invocation, the resulting $(1 \pm \eps/2)$ sparsifier $\calC'$ that we get maintains $\leq \widetilde{O}(n^5 \log^6(q) / \eps^2)$ coordinates. In the second invocation, we get a $(1 \pm \eps/2)$-sparsifier $\calC''$ for $\calC'$, whose length is bounded by $\widetilde{O}(n \log^4(n^5 \log^6(q) / \eps^2) \log^2(q) / \eps^2) = \widetilde{O}(n \log^2(q) / \eps^2)$, as we desire. To see the probability bounds, note that $m \geq n$, and thus both processes invocations of FinalCodeSparsify succeed with probability $1 - 2^{-\Omega((\log(n/\eps) \log \log(q))^2)}$.

    Finally, to see that the algorithm is efficient, we simply invoke \cref{clm:finalSparsifyRuntime} for each time we fun the algorithm. Thus, we get our desired bound. 
\end{proof}

\section{Sparsifying Affine Predicates over Abelian Groups}\label{sec:abelianConstraints}

In this section, we will generalize the result from the previous section. Specifically, whereas the previous section showed that one can sparsify CSP systems where each constraint is of the form $\sum_{i = 1}^r a_i b_i \neq a_0 \mod q$, for $a_i \in \Z_q, b_i \in \zo$, we will show that in fact, for any Abelian group $A$, we can sparsify predicates of the form $\sum_{i = 1}^r a_i b_i \neq a_0 \mod q$, for $a_i \in A, b_i \in \zo$.

Concretely, we first define affine constraints.

\begin{definition}
    We say a constraint (or predicate) $P: \zo^r \ra \zo$ is \textbf{affine} if it can be written as
    \[
    P(b_1, \dots b_r) = \mathbf{1}[\sum_{i = 1}^r a_i b_i \neq a_0],
    \]
    where $a_i \in A, b_i \in \zo$, and the addition is done over an Abelian group $A$. 
\end{definition}

Going forward, for an Abelian group $A$, we will let $q$ be the number of elements in $A$. We will make use of the following fact:

\begin{fact}\label{fact:finiteAbelian}\cite{Pin2010}
    Any finite Abelian group $A$ is isomorphic to a direct product of the form 
    \[
    \Z_{q_1} \times \dots \times \Z_{q_u},
    \]
    where each $q_j$ is a prime power. We will let $q = \prod_{i = 1}^u q_i$.
\end{fact}

Going forward, we will only work in accordance with this isomorphism. That is, instead of considering constraints of the form \[ \mathbf{1}[\sum_{i = 1}^r a_i b_i \neq a_0],
    \]
    where $a_i \in A, b_i \in \zo$, we will simply consider constraints of the same form only with $a_i \in \Z_{q_1} \times \dots \times \Z_{q_u}$. We will view these elements as tuples $(d_1, \dots d_u)$, where $d_j \in \Z_{q_j}$.

Now, we can introduce the analog of the contraction algorithm in the Abelian case. We will let $G \in \left ( \Z_{q_1} \times \dots \times \Z_{q_u} \right )^{m \times n}$ be a generating matrix, which generates a code $\calC \subseteq \left ( \Z_{q_1} \times \dots \times \Z_{q_u} \right )^{m}$. In particular, each entry in the matrix $G$ will be from $\left ( \Z_{q_1} \times \dots \times \Z_{q_u} \right )$. Being the generating matrix for $\calC$ implies that $\calC = \{Gx : x \in \Z^n \} = \{Gx : x \in \Z_q^n \}$ (i.e., all linear combinations of columns of $G$), where $q = q_1 \cdot \dots \cdot q_u$. 

For a codeword $c \in \calC$, we will still say that $\wt(c) = |\{j \in [n]: c_j \neq 0 \}|$, where $c_j = 0$ if the corresponding coordinate is the tuple $(0, 0, 0, \dots 0) \in \Z_{q_1} \times \dots \times \Z_{q_u}$. As before, we say 
\[
\Density(\calC) = \frac{\log_2(|\calC|)}{|\Supp(\calC)|}.
\]

Going forward, we will refer to the $j$th row of the generating matrix $G$. This is a row vector in $\left ( \Z_{q_1} \times \dots \times \Z_{q_u} \right )^{n}$. Roughly speaking, our new contraction algorithm, when contracting on a specific coordinate $j$, will perform the contraction algorithm on each index of the tuples in row $j$. We will use the notation $G_j$ to refer to the $j$th row of the generating matrix $G$, and we will use the notation $G_j^{(p)}$ to refer to the component of $G_j$ corresponding to $\Z_{q_p}$. This will be a row vector in $\Z_{q_p}^{n}$. Note that we will be consistent in referring to the coordinate of a codeword $c$ as the entry in $[m]$, while the the index of the $j$th coordinate will refer to a specific entry in the tuples of that coordinate.

\begin{algorithm}
    \caption{$\ContractAbelian(G, j)$}\label{alg:contractAbelian}
    Let $G_j$ be the $j$th row of the generating matrix $G$.\\
    \For{ $p \in [u]$ }{
    \If{$G_j^{(p)}$ is non-zero}{
    Run the Euclidean GCD algorithm on $G_j^{(p)}$, using column operations to get the GCD of the $G_j^{(p)}$ into the first entry of $G_j^{(p)}$. Call this first column $v$.\\
    Add multiples of $v$ to cancel out every other non-zero entry in $G_j^{(p)}$. \\
    Replace column $v$ with $c \cdot v$, where $c = \min \{ \ell \in [q_p]: v_j \cdot \ell = 0 \}$. \\
    }
    }
    \Return{$G$}
\end{algorithm}

\begin{claim}\label{clm:stillInSpanAbelian}
    Consider a codeword $c \in \calC$ where $\calC$ is the span of $G \in \left ( \Z_{q_1} \times \dots \times \Z_{q_u} \right )^{m \times n}$. If we run $ContractAbelian(G, j)$ on a coordinate $j$ such that $c_j = 0$, then $c$ will still be in the span of $G$ after the contraction.
\end{claim}

\begin{proof}
    First, we show that after line $4$ in the \cref{alg:contractAbelian}, $c$ is still in the span. Indeed, the Euclidean GCD algorithm only ever makes column operations of the form $v_1 + c \cdot v_2 \rightarrow v_1$. This means that every step of the algorithm is invertible, so in particular, after every step of the Euclidean GCD algorithm, the span of the matrix $G$ will not have changed.

    Now, in line $5$ of \cref{alg:contractAbelian}, this again does not change the span of $G$. Indeed, we can simply undo this step by adding back multiples of $v$ to undo the subtraction. Since the span of $G$ hasn't changed after this step, $c$ must still be in the span.

    Finally, in line $6$, the span of $G$ actually changes. However, since $c$ was zero in its $j$th coordinate, it must have been $0$ in the $p$th index of the $j$th coordinate as well. This means it could only contain an integer multiple of $c$ times the column $v$ (again where $c$ is defined as $c = \min \{ \ell \in [q]: v_j \cdot \ell = 0 \}$), as otherwise it would be non-zero in this index. Hence $c$ will still be in the span of the code after this step, as it is only codewords that are non-zero in the $p$th index of coordinate $j$ that are being removed.
\end{proof}

\begin{claim}
    After calling $\ContractAbelian(G, j)$ on a non-zero coordinate $j$, the number of distinct codewords in the span of $G$ decreases by at least a factor of $2$.
\end{claim}

\begin{proof}
    Consider the matrix $G$ after line $5$ of \cref{alg:contractAbelian}. By the previous proof, we know that the span of $G$ has not changed by line $2$. Now, after scaling up $v$ by $c$, we completely zero out the $p$th index of row $j$ of $G$. Previously, for any codeword in the code which was zero in coordinate $j$, we could correspondingly add a single multiple of column $v$ to it to create a new codeword that was non-zero in the $p$th index of its $j$th coordinate. Thus, the number of codewords which were non-zero in coordinate $j$ was at least as large as the number which were zero. After scaling up column $v$, codewords which were non-zero in this coordinate are no longer in the span, so the number of distinct codewords has gone down by at least a factor of $2$.
\end{proof}

\begin{algorithm}
\caption{Repeated Contractions$(G, \alpha, q)$}\label{alg:manyContractAbelian}
    Input generating matrix $G \in \Z_q^{m \times n}$. \\
    \While{more than $q^{\alpha + 1}$ distinct codewords in the span of $G$}
    {
    Choose a random non-zero row $G_j$ of $G$. \\
    Set $G$ = ContractAbelian$(G, j)$.
    }
\end{algorithm}

\begin{claim}\label{clm:lowdensityAbelian}
    Let $\calC$ be the span of $G \in \left ( \Z_{q_1} \times \dots \times \Z_{q_u} \right )^{m \times n}$, and let $d$ be an integer. Suppose every subcode $\calC' \subseteq \calC$ satisfies $\Density(\calC') < \frac{1}{d}$. Then for any $\alpha \in \Z$, the number of distinct codewords of weight $\leq \alpha d$ is at most $\binom{n\log(q)}{\alpha } \cdot q^{\alpha + 1}$, where $q = \prod_{i = 1}^u q_i$.
\end{claim}

\begin{proof}
    Consider an arbitrary codeword $c$ in the span of $G$ of weight $\leq \alpha d$. Let us consider what happens when we run \cref{alg:manyContractAbelian}. In the worst case, we remove only a factor of $2$ of the codewords in each iteration. Thus, suppose that this is indeed the case and that the code starts with $q^n$ distinct codewords. Now, after $\ell$ iterations, this means in the worst case there are $2^{n \log (q) - \ell}$ distinct codewords remaining. Note that after running these contractions, the resulting generating matrix defines a subcode $\calC'$ of our original space. Then, by our assumption on the density of every subcode, the number of non-zero rows in the generating matrix must be at least $(n \log(q) - \ell)d$. Thus, the probability that $c$ remains in the span of the generating matrix after the next contraction is at least
    \[
    \Pr[c \text{ survives iteration}] \geq 1 - \frac{\alpha d}{(n \log(q) - \ell)d} = 1 - \frac{\alpha}{n \log(q) - \ell}.
    \]

    Now, we can consider the probability that $c$ survives all iterations. Indeed,
    \begin{align*}
        \Pr[c \text{ survives all iterations}] & \geq \frac{n \log(q) - \alpha}{n \log (q)} \cdot \frac{n \log(q) - \alpha - 1}{n \log (q) - 1} \cdot \dots \cdot \frac{\alpha + 1 - \alpha}{\alpha + 1} \\
        & = \binom{n\log(q)}{\alpha }^{-1}.
    \end{align*}
    At this point, the number of remaining codewords is at most $q^{\alpha + 1}$. Hence, the total number of codewords of weight $\leq \alpha d$ is at most $\binom{n\log(q)}{\alpha } \cdot q^{\alpha + 1}$.
\end{proof}

\begin{corollary}\label{cor:moduleKargerAbelian}
    For any linear code $\calC$ over $\left ( \Z_{q_1} \times \dots \times \Z_{q_u} \right )^m$, with $\leq q^n$ distinct codewords, and for any integer $d \geq 1$, there exists a set $S$ of at most $n \log (q) \cdot d$ rows, such that upon removing these rows, for any integer $\alpha \geq 1$, the resulting code has at most $\binom{n\log(q)}{\alpha } \cdot q^{\alpha+1}$ distinct codewords of weight $\leq \alpha d$.
\end{corollary}

\begin{proof}
    Suppose for the code $\calC$ that there is a subcode of density $\geq 1/d$. Then, this means there exists a set $n'd$ rows (coordinates), with at least $2^{n'}$ distinct codewords that are completely contained on these rows. Now, by removing these rows, this yields a new code where the number of distinct codewords has decreased by a factor of $2^{n'}$. Thus, if removing these rows yields a new code $\calC'$, this new code has at most $q^n / 2^{n'}$ distinct codewords. Now, if there is no subcode of density $\geq 1/d$ in $\calC'$, we are by the preceding claim (i.e. we will satisfy the counting bound). Otherwise, we can again apply this decomposition, removing another $n''d$ rows to yield a new code with at most $2^{n \log(q) - n' - n''}$ distinct codewords. In total, as long as we remove rows in accordance with condition 1, we can only ever remove $n \log(q) d$ rows total, before there are no more distinct codewords remaining, at which point the counting bound must hold.

    Thus, there exists a set of at most $n \log(q) \cdot d$ rows, such that upon their removal, we satisfy the aforementioned counting bound.
\end{proof}

\begin{theorem}\label{thm:startabelian}
For any code $\calC \subseteq \left ( \Z_{q_1} \times \dots \times \Z_{q_u} \right )^{m}$ with $q^n$ distinct codewords, there is an efficient algorithm for creating $(1 \pm \eps)$ sparsifiers to $\calC$ with $\widetilde{O}(n \log^2(q) / \eps^2)$ weighted coordinates.
\end{theorem}

\begin{proof}
    Note that we can simply replace \cref{alg:contract} with \cref{alg:contractAbelian} in every instance in which it appears in \cref{alg:FinalCodeSparsify}. Because this contraction algorithm satisfies all the same properties as the original, the theorems follow immediately. Note that we set $q = q_1 \cdot q_2 \cdot \dots \cdot q_u$, and hence do not get any improvement by working over $\Z_{p_1} \times \Z_{p_2}$ as opposed to $\Z_{p_1 \cdot p_2}$. For completeness, we provide a more extensive proof in the appendix (see \cref{sec:abelianComplete}).
\end{proof}

With this, we are now able to prove the efficient sparsification for any CSP using an affine predicate $P$ over an Abelian group.

\begin{theorem}\label{thm:mainRestatedAbelian}
    Let $P$ be an affine predicate over an Abelian group $A$. Then, for any CSP instance $C$ with predicate $P$ on a universe of $n$ vaiables, we can efficiently $(1 \pm \eps)$-sparsify $C$ to $\widetilde{O}(n \log^2(|A|) / \eps^2)$ constraints.
\end{theorem}

\begin{proof}
    Indeed, let the predicate $P$ be given. Then, we can write $P$ as a constraint of the form 
    \[ 
    P(b_1, \dots b_r) = \mathbf{1}[\sum_{i = 1}^r a_i b_i \neq a_0],
    \]
    (invoking \cref{fact:finiteAbelian})
    where $a_i \in \Z_{q_1} \times \dots \times \Z_{q_u}$ and $b_i \in \zo$. Now, we will show that we can create a generating matrix $G$ such that for any assignment $x \in \zo^n$, there is a corresponding codeword in the code which is non-zero in coordinate $j$ if and only if constraint $j$ is satisfied. Indeed, let us initialize a generating matrix $G \in (\Z_{q_1} \times \dots \times \Z_{q_u})^{m \times (n+1)}$ where $m$ is the number of constraints in $C$, and for each variable $x_1, \dots x_n$ in the universe, there is a single corresponding column among the first $n$ columns. Now, suppose that the $j$th constraint of $C$ is of the form 
    \[
    \mathbf{1}[\sum_{i = 1}^r a_i x_{j_i} \neq a_0].
    \]
    Then, in the $j$th row of the generating matrix, let us place the value $a_i$ in the $j_i$th column. Finally, in the last column ($n+1$st column), let us place $-a_0$. 

    With this generating matrix, it follows that for any assignment $x \in \zo^n$, we can define a corresponding message $x' = [x_1, \dots x_n, 1]$ such that $(Gx')_j$ is non-zero if and only if $C_j(x)$ (the $j$th constraint) is satisfies (equal to $1$) on this assignment. Indeed,
    $C_j(x) = 1$ if and only if 
    \[
    \sum_{i = 1}^r a_i x_{j_i} \neq a_0,
    \]
    or equivalently if 
        \[
    \sum_{i = 1}^r a_i x_{j_i} -a_0 \neq 0.
    \]
    Correspondingly, \[
    (Gx')_j = \sum_{i = 1}^r a_i x_{j_i} -a_0,
    \]
    and the weight of the codeword $Gx'$ is 
    \[
    \wt(Gx') = \left | \{j \in [m]: \sum_{i = 1}^r a_i x_{j_i} -a_0 \neq 0 \} \right |,
    \]
    which exactly equals the value of the CSP on assignment $x$.
    Thus because there is an exact equivalence between the weight contributed by the coordinates in the codeword corresponding to $x'$ and the weight contributed by the constraints with assignment $x$, returning a $(1 \pm \eps)$ sparsifier for the code generated by $G$ also creates a sparsifier for the CSP $C$. Because this former task can be done efficiently (\cref{thm:startabelian}), this yields our desired result. 
\end{proof}

\subsection{Infinite Abelian Groups}

The previous section has the disadvantage of the sparsifier size having a dependence on the size of the Abelian group. In this section we show that this is not necessary, and instead we can replace this with a polynomial dependence on the arity of the predicates. First, we make the following observation regarding affine Abelian predicates:

\begin{claim}\label{clm:closedLattice}
    Let $P: \zo^r \ra \zo$ be an affine Abelian predicate such that $P(0^r) = 0$. Then, the space $P^{-1}(0)$ is closed under integer linear combinations over $\Z$.
\end{claim}

\begin{proof}
    Let $y_1, \dots y_{\ell} \in P^{-1}(0)$, and $\alpha_1, \dots \alpha_{\ell}$ be such that $\sum_{j = 1}^{\ell} \alpha_j y_j \in \zo^r$ (when addition is done over $\Z$). We claim then that $P(\sum_{j = 1}^{\ell} \alpha_j y_j) = 0$ also. 

    This follows simply from $P$ being an affine Abelian predicate. It must be the case that $P(b_1, \dots b_r) = \mathbf{1}[\sum_i a_i b_i \neq 0]$, for some $a_i \in A$, where $A$ is an Abelian group. Then, 
    \[
    P(\sum_{j = 1}^{\ell} \alpha_j y_j) = \mathbf{1}[\sum_i a_i (\sum_{j = 1}^{\ell} \alpha_j y_j)_i \neq 0] = \mathbf{1}[\sum_{j = 1}^{\ell} \alpha_j \sum_i a_i (y_j)_i \neq 0].
    \]
    But, because each $y_j \in P^{-1}(0)$, it must be the case that $\sum_i a_i (y_j)_i = 0$. Thus, the entire sum must be $0$, we can conclude that $P(\sum_{j = 1}^{\ell} \alpha_j y_j) = 0$, as we desire.
\end{proof}

Now, we will show that any predicate whose unsatisfying assignments form a closed space under integer linear combinations will be representable over a finite Abelian group. An immediate consequence of this is that affine Abelian predicates over infinite Abelian groups are sparsifiable to finite size. We explain this more in depth below. 

\begin{definition}
    For a matrix $B \in \Z^{d \times \ell}$, the lattice generated by $B$ (often denoted by $\calL(B)$), is the set $\{Bx: x \in \Z^{\ell} \}$.
\end{definition}

From the preceding claim, we know that for an affine Abelian predicate $P$, the set of unsatisfying assignments is closed as a lattice. Thus, our goal will be to show that for any lattice, as long as the coefficients of the generating matrix are bounded, then we can characterize exactly which elements are in the lattice with a finite set of constraints over a finite alphabet. It follows then that we will be able to express this as an affine predicate over a finite Abelian group. 

\begin{lemma}\label{lem:smallCoeff}
    Suppose $B$ is a $d \times \ell$ matrix with entries in $\Z$, such that the magnitude of the largest sub-determinant is bounded by $M$, and $\rank(B) = k$. Then, every element of the lattice generated by the columns of $B$ is given exactly by the solutions to $d-k$ linear equations and $k$ modular equations. All coefficients of the linear equations are bounded in magnitude by $M$, and all modular equations are written modulo a single $M' \leq M$.
\end{lemma}

\begin{proof}
    First, if $\rank(B) = k < d$, this means that there exist $k$ linearly independent rows such that the remaining $d-k$ rows are linear combinations of these rows. Let us remove these rows for now, and focus on $B' \in \Z^{k \times \ell}$ where now the matrix has full row-rank. 

    It follows that for this matrix $B'$ we can create a new matrix $\hat{B}$ such that the lattice generated by $B'$ (denoted by $\calL(B') = \calL(\hat{B})$) and $\hat{B}$ is in Hermite Normal Form (HNF) \cite{Mic14a}. In this form, $\hat{B}$ is lower triangular with the diagonal entries of $\hat{B}$ satisfying 
    \[
    \det(\hat{B}) = \prod_{i = 1}^k \hat{B}_{i,i} \leq \max_{k \times k \text{ subrectangle A}}\det(B'_A) \leq M.
    \]
Further, all columns beyond the $k$th column will be all zeros, so we can remove these from the matrix (because the lattice generated by the first $k$ columns will be the same as the lattice generated by \emph{all} columns).
    
    We can now define the \emph{dual} lattice to $\calL(B') = \calL(\hat{B}))$. For a lattice $\Lambda \subseteq \Z^k$, we say that 
    \[
    \dual(\Lambda) = \{x \in  \Q^k: \forall y \in \Lambda, \langle x, y \rangle \in Z\}.
    \]
    
    Here it is known that the dual is an exact characterization of the lattice $\Lambda$. I.e., any vector in $\Lambda$ will have integer-valued inner product with \emph{any} vector in the dual, while for any vector not in $\Lambda$, there exists a vector in the dual such that the inner-product is not integer valued \cite{Mic14b}. 

    Now, for our matrix $\hat{B}$, it is known that one can express the dual lattice to $\calL(\hat{B})$ with a generating matrix $\hat{D} = \hat{B} (\hat{B}^T \hat{B})^{-1}$ \cite{Mic14b}. As a result, it must be the case that $\calL(\hat{D}) \subseteq \Z^{k} / \det(\hat{B})$. If a vector $x$ of length $k$ is not in $\calL(\hat{B})$, it must be the case that there exists a column $y$ of $\hat{D}$ such that $\langle x, y \rangle \notin \Z$. Otherwise, if the inner-product with every column is in $\Z$, it follows that for any vector in the dual, the inner product would also be in $\Z$, as we can express any vector in the dual as an integer linear combination of columns in $\hat{D}$. Thus, it follows that membership of a vector $x$ in $\calL(\hat{B})$ can be tested exactly by the $k$ equations $\forall i \in [k]: \langle x, \hat{D}_i \rangle \in \Z$, where $\hat{D}_i$ is the $i$th column of $\hat{D}$.
    
    Now, because every entry of $\hat{D}$ has denominator dividing $\det(\hat{B})$, it follows that we can scale up the entire equation by $\det(\hat{B})$. Thus, an equivalent way to test if $x \in \calL(\hat{B})$ is by checking if $\forall i \in [k]: \langle x, \det(\hat{B}) \cdot \hat{D}_i \rangle = 0 \mod \det(\hat{B})$. Now, all the coefficients of these equation are integers, and we are testing whether the sum is $0$ modulo an integer. Thus, we can test membership of any $k$-dimensional integer vector in $\calL(\hat{B})$ with $k$ modular equations over $\det{\hat{B}} \leq M$.

    The above argument gives a precise way to characterize when the restriction of a $d$ dimensional vector to a set of coordinates corresponding with linearly independent rows in $B$ is contained in the lattice generated by these same rows of $B$. It remains to show that we can also characterize when the dependent coordinates (i.e. coordinates corresponding to the rows that are linearly dependent on these rows) are contained in the lattice. Roughly speaking, the difficulty here arises from the fact that we are operating with a non-full dimensional lattice. I.e., there exist directions that one can continue to travel in $\Z^n$ without ever seeing another lattice point. In this case, we do not expect to be able to represent membership in the lattice with a modular linear equation, as these modular linear equations rely on periodicity of the lattice. 
    
    Instead, here we rely on the fact that for any of the rows of $B$ that are linearly dependent, we know that there is a way to express it as a linear combination of the set of linearly independent rows. WLOG, we will assume the first $k$ rows $r_1, \dots r_k$ are linearly independent, and we are interested in finding $c_i$ such that $\sum_{i = 1}^k c_i r_i = r_{k+1}$. Now, consider any subset $A$ of $k$ linearly independent columns amongst these $k$ rows. We denote the corresponding restriction of the rows to these columns by $r_i^{(A)} \in \zo^k$. It follows that if we want a linear combination of these rows such that $\sum_{i = 1}^k c_i r_i^{(A)} = r_{k+1}^{(A)}$, we can express this a constraint of the form $M^{(A)} c = (r_{k+1}^{(A)})^T$, where we view the $i$th column of $M$ as being the (transpose of) $r_i^{(A)}$ and $c$ as being the vector of values $c_1, \dots c_k$. Using Cramer's rule, we can calculate that $c_i = \det(M^{(A)}_i) / \det(M^{(A)})$, where $M^{(A)}_i$ is defined to be the matrix $M^{(A)}$ with the $i$th column replaced by $(r_{k+1}^{(A)})^T$. In particular, this means that we can express $r_{k+1}^{(A)} = \sum_{i = 1}^k \det(M^{(A)}_i) / \det(M^{(A)}) \cdot r_i^{(A)}$, and because $A$ corresponds to a set of linearly independent columns, it must also be the case that $r_{k+1} = \sum_{i = 1}^k \det(M^{(A)}_i) / \det(M^{(A)}) \cdot r_i$. We can re-write this as an integer linear equation by expressing $r_{k+1} \cdot \det(M^{(A)}) = \sum_{i = 1}^k \det(M^{(A)}_i) \cdot r_i$. This means that for any valid vector $x \in \Z^d$ expressable as a linear combination of the columns of $B$, it must be the case that $x_{k+1} \cdot \det(M^{(A)}) = \sum_{i = 1}^k \det(M^{(A)}_i) \cdot x_i$, as the $k+1$st coordinate is always a specific linear combination of the first $k$ coordinates.

    We can repeat the above argument for each of the $d-k$ linearly independent rows. Let $j$ denote the index of a row linearly dependent on the first $k$ rows. It follows that $x_j \cdot \det(M^{(A)}) = \sum_{i = 1}^k \det(M^{(A),j}_i) \cdot x_i$, where now $M^{(A),j}_i$ is the $d\times d$ matrix $M^{(A)}$ where the $i$th column has been replaced with $(r_j^{(A)})^T$. Note that every coefficient that appears in these equations above is of the form $\det(C)$ where $C$ is a $d \times d$ submatrix of $B$. It follows that each of these coefficients is bounded in magnitude by $M$, where $M$ is again defined to be the maximum magnitude of the determinant of any square sub-matrix. 

    To conclude, we argue that for any vector $x \in \calL(B)$, $x$ satisfies all of the above linear equations and modular equations. The first part of the proof showed that amongst the set of linearly independent rows $S$, the dual of the lattice exactly captures when $x_S$ is in the lattice generated by $B_S$. That is, if $x_S$ is generated by $B_S$, then $x_S$ satisfies the above modular linear equations, while if $x_S$ is not in the span of $B_S$, then $x_S$ does not satisfy the modular linear equations. Now, if $x_S$ does not satisfy the modular linear equations, this is already a witness to the fact that $x$ is not in the $\calL(B)$. But, if $x_S$ is in the span of $B_S$, then if $x$ is in the span of $B$, it must also be the case that the coordinates of $x_{\bar{S}}$ satisfy the exact same linear dependence on the $x_S$ that $B_{\bar{S}}$ has on $B_S$. This is captured by our second set of linear equations.
\end{proof}

\begin{theorem}
     Let $P: \zo^r \ra \zo$ be a predicate of arity $r$. Let $S = P^{-1}(0) \subseteq \zo^r$ denote the unsatisfying assignments of $P$. If $S$ is closed under integer valued linear combinations, then CSPs with predicate $P$ on $n$ variables are efficiently-sparsifiable to size $\widetilde{O}(n\cdot r^4 / \eps^2)$.
\end{theorem}

\begin{proof}
Let us create a matrix $B \in \zo^{r \times |S|}$ where the $i$th column of $B$ is the $i$th element of $S$. Let $k$ be the rank of $B$. It follows that for any assignment $x \in \zo^{r}$, we can exactly express the membership of $x$ in $\calL(B)$ with $k$ modular linear equations, and $r-k$ linear equations. I.e., $x$ is in $\calL(B)$ if and only if all of these equations are satisfied. Note that the $d$ modular linear equations are all over modulus $M \leq \max_{k \in [r], k \times k \text{ subrectangle } A} \det(B'_A) \leq (r)^{r}$. Likewise, the integer linear equations also all have coefficients $\leq (r)^{r}$ (by \cref{lem:smallCoeff}). It follows that because $x \in \zo^{r}$, we can choose a prime $p$ such that $p \geq 2 \cdot r \cdot (r)^{r}$. Now, for any of the integer linear equations of the form $c_1 x_1 + \dots c_k x_k - c_{k+1} x_{k+1}$, it will be the case that for $x \in \zo^{r}$,
\[
c_1 x_1 + \dots c_k x_k - c_{k+1} x_{k+1} = 0 \iff c_1 x_1 + \dots c_k x_k - c_{k+1} x_{k+1} = 0 \mod p.
\]

This is because the expression on the left can never be as large as $p$ or $-p$ since we chose $p$ to be sufficiently large.

Thus, we can express $x \in \zo^{r}$ as being in the lattice $\calL(B)$ if and only if all $r$ modular equations are $0$. This is then the OR of $r$ modular equations, which can be expressed over the Abelian group $A = \Z_M \times \Z_m \dots \times Z_m \times Z_p \dots \times \Z_p$, where there are $k$ copies of $Z_m$, and $r$ copies of $\Z_p$. It follows that $|A| \leq \left ( 2 \cdot r \cdot (r)^{r} \right )^{r} = (2r)^{r} \cdot (r)^{^2 r^2}$. Thus, for any CSP on predicate $P$ of the above form on $n$ variables, we can efficiently $(1 \pm \eps)$ sparsify $P$ to size $\widetilde{O}(n\cdot  r^4 / \eps^2)$ via \cref{thm:mainRestatedAbelian}.

\end{proof}

\affineabelian*

\begin{proof}
    This follows because we can assume WLOG that $P(0^r) = 0$. Then, it must be the case that $P^{-1}(0)$ is closed under integer linear combinations (by \cref{clm:closedLattice}), and we can invoke the preceding theorem. 
\end{proof}

Note that here we are dealing with boolean predicates (i.e., operating on $\zo^r$). We can extend this lattice characterization to larger alphabets, and do so in \cref{sec:largerAlphabetAffine}.

\section{Impossibility of Sparsifying Affine CSPs Over Non-Abelian Groups}\label{sec:impossibility}

In this section, we will complement the result of the previous section and show that in fact, if an affine constraint is written as 
\[
P(b_1, \dots b_r) \sum_{i = 1}^r a_i b_i \neq a_0
\]
for $a_i \in H$, $b_i \in \zo$ and $H$ being a non-Abelian group, then there exist CSP instances with predicate $P$ that require sparsifiers of quadratic size. To show this, we will make use of \cref{thm:AND} from the work of Khanna, Putterman, and Sudan \cite{KPS24}. This theorem shows that it suffices to show that a predicate has a projection to an AND of arity $2$ in order to conclude that there exist CSP instances with this predicate that require sparsifiers of quadratic size.

\intrononabelian*

\begin{proof}
    Indeed, let $H$ be a non-Abelian group, and let $a, b$ be two elements which do not commute. That is, let us assume that $a +b \neq b + a$. Then, consider the following predicate $P$ of arity $4$:
    \[
    P(x_1, x_2, x_3, x_4) = \mathbf{1}[ax_1 + bx_2 -a x_3 - b x_4 \neq 0].
    \]

    Note that $P(0, 0, 0, 0) = 0$, $P(1, 0, 1, 0) =  \mathbf{1}[a-a \neq 0] = 0$, $P(0, 1, 0, 1) =  \mathbf{1}[b-b \neq 0] = 0$, but 
    \[
    P(1, 1, 1, 1) =  \mathbf{1}[a + b - a - b \neq 0] = 1,
    \]
    as if $a + b - a - b = 0$, then $a, b$ will commute with one another, which violates our original assumption. This means that $0000, 1010, 0101$ are unsatisfying assignments, while $1111$ is a satisfying assignment. In particular, we can consider the following restriction 
    \[
    \pi(x_1) = c, \pi(x_2) = d, \pi(x_3) = c, \pi(x_4) = d.
    \]
    It follows then that $P(\pi(x_1), \pi(x_2), \pi(x_3), \pi(x_4)) = \text{AND}(c, d)$. Thus, there exists a projection of $P$ to an AND of arity $2$, so we can invoke \cref{thm:AND} and conclude that there exist CSP instances on predicate $P$ which require sparsifiers with $\Omega(n^2)$ surviving constraints.
\end{proof}

In this sense, the preceding result we proved about sparsifying affine constraints over Abelian groups is in fact the strongest possible result we could hope to hold over affine constraints, as even the most simple non-Abelian groups will have instances which are not sparsifiable. 
    
\section{Sparsifying Symmetric CSPs}\label{sec:symmetric}

In this section, we will characterize the sparsifiability of symmetric CSPs. Namely, for a predicate $P: \zo^r \ra \zo$ we say that $P$ is symmetric if for any permutation $\pi: [r] \ra [r]$, $P(x) = P(\pi \circ x)$. This is equivalent to saying that $P(x)$ is uniquely determined by the number of non-zero entries in $x$. First, we will introduce the definition of a symmetric predicate being periodic.

\begin{definition}
    A symmetric predicate $P: \zo^r \ra \zo$ is \textbf{periodic} if for any $x,y \in \zo^r$ such that $P(x) = P(y) = 0$ and $|y| > |x|$, then for any $z \in \zo^r$ such that $|z| = 2|y| - |x|$ or $|z| = 2|x| - |y|$, $P(z) = 0$. In this context $|x|$ refers to the number of non-zero entries in $x$.
\end{definition}

With this, we are then able to state the main theorem of this section.

\begin{theorem}
    For a symmetric predicate $P: \zo^r \ra \zo$, all CSPs with predicate $P$ are sparsifiable to nearly-linear size if and only if $P$ is periodic.
\end{theorem}

Now, we are ready to prove the following:

\begin{lemma}\label{lem:notperiodic}
    Suppose a symmetric predicate $P: \zo^r \ra \zo$ is not periodic. Then there is a projection $\pi: \{x_1, \dots x_r\} \ra \{0, 1, a, \neg a, b, \neg b\}$ such that $P(\pi(x_1), \dots \pi(x_r)) = \text{AND}(a,b)$.
\end{lemma}

\begin{proof}
    Suppose that a symmetric CSP does not have this periodic property. This implies that there exists $a, b, c$ such that for $|x| = a$, $|x| = b$, $P(x) = 0$, but for $|x| = 2a - b$, $P(x) = 1$. WLOG we will assume $b>a$ but we can equivalently swap $a$ with $b$ in the following construction and assume the opposite. We will show that this contains an affine projection to AND, and therefore requires sparsifiers of quadratic size (see \cref{thm:AND}). Indeed, consider the following bit strings:
    \begin{enumerate}
        \item The bit string $y_1$ with $r-a$ $0$'s followed by $a$ $1$'s.
        \item The bit string $y_2$ with $r-b$ $0$'s followed by $b$ $1$'s.
        \item The bit string $y_3$ with $r - 2b +a$ $0$'s, followed by $b-a$ $1$'s followed by $b-a$ $0$'s, followed by $a$ $1$'s.
        \item The bit string $y_4$ with $r - 2b +a$ $0$'s, followed by all $1$'s.
    \end{enumerate}

    By the above conditions, $P(y_1) = P(y_2) = P(y_3) = 0$, while $P(y_4) = 1$. Now, we consider the projection $\pi:\{ x_1, \dots x_r \} \ra \{ 0, 1, x, \neg x, y, \neg y\}$ such that $\pi(x_1) = \pi(x_2) = \dots = \pi(x_{r-2b+a}) = 0$, $\pi(x_{r-2b+a+1}) = \dots = \pi(x_{r-b}) = x$, $\pi(x_{r-b+1}) = \dots = \pi(x_{r-a}) = y$, and all remaining are sent to $1$. Once can verify that under this projection, $P_{\pi} =$ AND.
\end{proof}

We now show that symmetric, perioidic predicates can be written as a simple affine equation. 

\begin{lemma}\label{lem:affinePeriodic}
    Let $P: \zo^r \ra \zo$ be a symmetric, periodic predicate. Then, there exist $c, \ell \in [r+1]$ such that 
    \[
    P(b_1, \dots b_r) = \mathbf{1}[\sum_{i = 1}^r b_i \neq c \mod \ell].
    \]
\end{lemma}

\begin{proof}
    Clearly, if there exists $\ell$ such that $P(x_1, \dots x_r) = \mathbf{1}[\sum_i x_i \neq c \mod \ell]$, then the zero levels of $P$ are evenly spaced out, and clearly satisfy the condition of being periodic. 
    
    Now, let $P: \zo^r \ra \zo$ be a symmetric, periodic predicate, and let $x, y \in \zo^r$ be such that $P(x) = P(y) = 0$ and $\left | |x| - |y| \right |$ is as small as possible (without being $0$). Let $\left | |x| - |y| \right | = \ell$. Now, note that the predicate $P$ can be written as 
    \[
    P(b_1, \dots b_r) = \mathbf{1}[\sum_{i = 1}^r b_i \neq |x| \mod \ell].
    \]

    This follows because by the definition of periodicity, every string $z$ such that $|z| = |x| - \ell$ or $|z| = |y| + \ell$ must satisfy $P(z) = 0$. Inductively then, every string $z'$ which satisfies $|z'| = |x| - 2\ell$, $|z'| = |y| + 2\ell$ must also satisfy $P(z') = 0$, and so on. Further, note that by the minimality of $\ell$, no strings $x, y$ such that $P(x) = P(y) = 0, |x| \neq |y|$ can be closer than hamming distance $\ell$. Thus, $P(b_1, \dots b_r)$ can be written as $ \mathbf{1}[\sum_{i = 1}^r b_i \neq |x| \mod \ell]$. 
    
    Finally, if there is only one level of the predicate which is $0$, we can choose $\ell = r+1$.
\end{proof}

Next, we will show that for any symmetric, periodic predicate, we can indeed sparsify CSPs using this predicate to nearly-linear size. 

\begin{lemma}\label{lem:sparsifyPeriodic}
    Let $P: \zo^r \ra \zo$ be a symmetric, periodic predicate. Let $C$ be a CSP using predicate $P$ on a universe of $n$ variables. Then, we can efficiently create a $(1 \pm \eps)$ sparsifier for $C$ with only $\widetilde{O}(n \log^2(r) / \eps^2)$ surviving weighted constraints.
\end{lemma}

\begin{proof}
So, let such a predicate $P$ be given. Per \cref{lem:affinePeriodic} there exists $c, \ell \in [r+1]$ such that
\[
P(b_1, \dots b_r) = \mathbf{1}[\sum_{i = 1}^r b_i \neq c \mod \ell].
\]

Now, we will show how to create a code over $Z_{\ell}$ which exactly captures the CSP built on predicate $P$. First, we create a generating matrix $G \in \Z_{\ell}^{m \times n+1}$, where $m$ is the number of constraints in the CSP, and $n$ is the size of the universe of variables. We associate each of the first $n$ columns to each of the $n$ variables. Next, let $C_j$ refer to the $j$th constraint of the CSP. Suppose that $C_j$ acts on variables $x_{j_1}, \dots x_{j_r}$. Then, in the corresponding $j$th row of the generating matrix $G$, we place a $1$ in the columns corresponding to $x_{j_1}, \dots x_{j_r}$, and place $-c \mod \ell$ in the final $n+1$st column. Now, consider an assignment to the variables $x_1, \dots x_n$ in the original CSP $C$. Note that the $j$th constraint $C_j$ is satisfied if and only if 
\[
\sum_{i = 1}^r x_{j_i} \neq c \mod \ell.
\]

Correspondingly, consider the code generated by the generating matrix $G$. For the message $x' = (x_1, \dots x_n, 1)$, note that the $j$th coordinate in the codeword $Gx'$ is 
\[
(Gx')_j = \sum_{i = 1}^r x_{j_i} -c \mod \ell.
\]
This means that the weight of the codeword $Gx'$ is exactly 
\[
\wt(Gx') = | \{j \in [m]: (Gx')_j \neq 0 \} | = | \{j \in [m]: \sum_{i = 1}^r x_{j_i} -c \neq 0 \mod \ell  \} | \] \[ = | \{j \in [m]: \sum_{i = 1}^r x_{j_i} \neq c \mod \ell \} | = | \{j \in [m]: C_j(x) = 1 \} |.
\]

Thus, any subset of the coordinates of $G$ which creates $(1 \pm \eps)$ sparsifier for the codewords in the code generated by $G$ must also yield a subset of the constraints of $C$ which yields a $(1 \pm \eps)$ sparsifier for the entire CSP. Because we can efficiently compute such sparsifiers for $G$ with only $\widetilde{O}(n \log^2(\ell) / \eps^2)$ surviving coordinates (by \cref{thm:efficientSparsifyZq}), we can thus efficiently find sparsifiers for $C$ with only $\widetilde{O}(n \log^2(\ell) / \eps^2)$ surviving constraints. Using that $\ell \leq r+1 \leq n+1$, we get our desired result. 
\end{proof}

Finally, we can now conclude the main result of this section:

\introsym*

\begin{proof}
    If $P$ is not periodic, then by \cref{lem:notperiodic} we know that $P$ has a projection to AND. Then, by invoking the work of Khanna, Putterman, and Sudan \cite{KPS24}, we know there exist CSP instances with predicate $P$ that require sparsifiers of size $\Omega_r(n^2)$.

    Otherwise, if $P$ is periodic, then by \cref{lem:sparsifyPeriodic} we can efficiently $(1 \pm \eps)$ sparsify any CSP instance on $P$ to only $\widetilde{O}(n  \log^2 (r) / \eps^2) = \widetilde{O}(n / \eps^2)$ constraints.
\end{proof}

\section{Non-trivial Sparsification for Almost All Predicates}\label{sec:nontrivial}

In this section, we show that almost all predicates can be non-trivially sketched. Indeed, given a predicate $P: \zo^r \ra \zo$, we show that as long as $P$ has $0$ or at least $2$ satisfying assignments, then any CSP instance with predicate $P$ on a universe of $n$ admits an $(\eps, \widetilde{O}(n^{r-1} / \eps^2))$-sparsifier. At a high level, we do this by extending our method for sparsifying systems of affine equations not being zero to sparsifying systems of polynomials not being equal to zero. Although this leads to a blowup in the size of our sparsifiers, it is distinctly more powerful, and allows us to model more general predicates. We recall \cref{thm:intro-poly}:

\intropoly*

Alternatively stated, the above implies the following:

\begin{corollary}
    Let $P: \zo^r \ra \zo$ be a predicate such that there exists a polynomial $Q$ with coefficients in an Abelian group $A$ of degree $\ell$ satisfying $\forall x \in \zo^r: \mathbf{1}[Q(x) \neq 0] = P(x)$. Then, all CSPs with predicate $P$ on a universe of $n$ variables efficiently admit $(\eps, \widetilde{O}(n^{\ell} \min(\log^2 (|A|), r^{4\ell}) / \eps^2))$-sparsifiers.
\end{corollary}

\begin{proof}
We will prove this via a reduction to code sparsification. We will prove the following lemma along the way:

\begin{lemma}\label{lem:polynomialrep}
    For any polynomials $P_1, \dots P_m$ of degree $\leq \ell$ over $A$, there exists generating matrix $G \in A^{m \times O(n^{\ell})}$ such that for any $x \in \zo^n$, there exists a corresponding $x' \in \zo^{O(n^{\ell})}$ such that $\forall j \in [m]$, $(Gx')_j = P_j(x')$. 
\end{lemma}

\begin{proof}

Indeed, let us consider the $j$th polynomial $P_j$. We can write
\[
P_j(x) = \sum_{i_1 + i_2 + \dots + i_n \leq \ell} a_{j, i_1, \dots i_n} \prod_{u = 1}^n x_u^{i_u}.
\]

Now, in our generating matrix, we associate each column of the generating matrix with a single term of each polynomial. In the $j$th row (corresponding to the $j$th polynomial), in the column corresponding to 
    the term $\prod_{u = 1}^n x_u^{i_u}$, we place $a_{j, i_1, \dots i_n}$. Note that there can be at most $O(n^{\ell})$ terms of degree $\leq \ell$, so it follows the size of the generating matrix is at $m \times O(n^{\ell})$.

Now consider any assignment $x \in \zo^n$. We will show that there is a corresponding assignment $x' \in \zo^{O(n^\ell)}$ such that $Gx'$ is non-zero in coordinate $j$ if and only if $P_j(x)$ was non-zero. From there, it follows that sparsifying the code generated by $G$ will yield a sparsifier for the system of polynomials.

So, let such an assignment $x \in \zo^n$ be given. Let $x' \in \zo^{O(\ell)}$ be such that the $p$th entry of $x'$ is the evaluation of the term corresponding to the $p$th column of $G$. That is, $x'_p = \prod_{u = 1}^n x_u^{i_u} |_x$, where $\prod_{u = 1}^n x_u^{i_u}$ is the term corresponding to the $p$th column of $G$.

Then, it follows that $(Gx')_j = \sum_{i_1 + i_2 + \dots + i_n \leq \ell} a_{j, i_1, \dots i_n} \prod_{u = 1}^n x_u^{i_u} = P_j(x)$.
\end{proof}

Now, by \cref{thm:mainRestatedAbelian}, we can efficiently sparsify the code generated by $G$ to only $\widetilde{O}(n^{\ell} \min(\log^2(|A|), r^{4 \ell}) / \eps^2)$ remaining coordinates, so it also follows that we can sparsify our set of polynomials to $\widetilde{O}(n^{\ell} \min(\log^2(|A|), r^{4 \ell}) / \eps^2)$ remaining weighted polynomials such that for any assignment $x \in \zo^n$, the number of non-zero polynomials is preserved to a $(1 \pm \eps)$ fraction. 
\end{proof}

We will also use the following trick for any predicate over any finite group, and in this context, $\Z_2$.

\begin{claim}\label{clm:ORaffine}
    Let $P_1, P_2, \dots P_{s}$ be such that each $P_i$ is an affine predicate over $\Z_2$. Then, one can write the predicate $P = P_1 \vee P_2 \vee \dots \vee P_s$ as single affine predicate over $(Z_{2})^{s}$, such that $P(x) \neq 0$ if and only if there exists some $i$ such that $P_i(x) \neq 0$.
\end{claim}

\begin{proof}
Indeed, each $P_i$ is affine so we can write $P_i(x) = \sum_{j = 1}^r a_{i,j} x_j - b_i$, where the addition is over $\Z_2$. Now, let the tuple $A^{(j)} = (a_{1,j}, \dots a_{s,j})$ and let $B = (b_1, \dots b_s)$. It follows that we can write 
\[
P(x) = \sum_{j = 1}^{r} x_j \cdot A^{(j)} -B.
\]
It follows that the $i$th entry of the tuple returned by $P(x)$ will be exactly $P_i(x) = \sum_{j = 1}^r a_{i,j} x_j - b_i$. Thus, $P(x)$ is zero if and only if each of the constituent $P_i(x)$'s also evaluates to zero. Otherwise, if at least one of the $P_i(x)$ is non-zero, $P(x)$ will also be non-zero. 
\end{proof}

\begin{claim}\label{clm:2sat}
    Let $P: \zo^r \ra \zo$ be a predicate with two satisfying assignments. Then, there exists a polynomial $L_{a,b}: \zo^r \ra \zo$ over $\Z_2$ of degree $r-1$ such that for any $y \in \zo^r$, $L_{a,b}(y) = 0$ if and only if $P(y) = 0$.
\end{claim}

\begin{proof}
    Let the two satisfying assignments be $a = a_1, \dots a_r$ and $b = b_1, \dots b_r$. WLOG let us assume that the first $t_1$ bits of $a$ and $b$ are $1$, the second $t_2$ bits of $a$ and $b$ are $0$, the next $t_3$ bits are $1$ for $a$ and $0$ for $b$, while the last $t_4$ bits of $a, b$ are $0$ for $a$ and $1$ for $b$. Consider then the following polynomial:
    \[
    L_{a, b}(y) = \prod_{i = 1}^{t_1} y_i \prod_{i = t_1 + 1}^{t_1 + t_2} (1 - y_1) \prod_{i = t_1 + t_2+2}^{t_1 + t_2 + t_3} (y_{t_1 + t_2 + 1} + y_{i} - 1) \prod_{i = t_1 + t_2+t_3}^{t_1 + t_2 + t_3 + t_4} (y_{t_1 + t_2 + 1} - y_{i}).
    \]
    Note that the degree of $L$ is $r-1$, and the only $y$'s for which the expression evaluates to something non-zero are $a$ and $b$ (and in this case it either evaluates to $1$ or $-1$). Over $\Z_2$, note that this is still the case, as $-1 = 1 \neq 0$ over $\Z_2$.
\end{proof}

\intronontrivial*

\begin{proof}
    Let such a CSP instance $C$ be given. Consider the $j$th constraint in this CSP, and call the corresponding predicate for the $j$th constraint $P_j: \zo^r \ra \zo$. Note that if $P_j$ is the constant $0$ predicate, we can simply remove it from $C$ without changing the value, so we will assume that $P_j$ has at least $2$ satisfying assignments. Let us write the satisfying assignments to $P_j$ as $a_1, \dots a_s$. It follows that we can define the polynomials $L_{a_1, a_2}, L_{a_1, a_3}, \dots L_{a_1, a_s}$ such that $L_{a_1, a_i}(y) \neq 0$ if and only if $y = a_i$ or $y = a_1$ in accordance with \cref{lem:polynomialrep}. Now, by \cref{lem:polynomialrep}, each of these polynomials can be written as an affine equation over $\Z_2$ over a universe of variables of size $O(n^{r-1})$. So, let $\hat{L}_{a_1, a_i}$ refer to the polynomial $L_{a_1, a_i}$ when instead viewed as a linear equation over $\F_2$ in a variable set of size $O(n^{r-1})$.

    Now, it follows that $P_j = L_{a_1, a_2} \vee L_{a_1, a_3} \vee  \dots \vee L_{a_1, a_s}$ and $s \leq 2^r$. Over the universe of variables of size $O(n^{r-1})$, it follows that we can write each polynomial as a linear equation by \cref{lem:polynomialrep}, and hence we can write $P_j =\hat{L}_{a_1, a_2} \vee \hat{L}_{a_1, a_3} \vee  \dots \vee \hat{L}_{a_1, a_s}$. Because $s \leq 2^r$, this means we can write each $P_j$ as a single affine constraint over $(\Z_{2})^{2^r}$, such that $P_j(x) \neq 0$ if and only if one of the $\hat{L}_{a_1, a_i}(x) \neq 0$. 

    Now, we can write the generating matrix $G$ for this linear space over $O(n^{r-1})$ variables on $(\Z_{2})^{2^r}$. It follows that for each original assignment $x \in \zo^n$, there exists a corresponding assignment $x' \in \zo^{O(n^{r-1})}$ such that for each $j \in [m]$, 
    \[
    (Gx')_j \neq 0 \iff \hat{L}_{a_1, a_2}(x') \vee \hat{L}_{a_1, a_3}(x') \vee  \dots \vee \hat{L}_{a_1, a_s}(x') \neq 0 \iff P_j(x) \neq 0.
    \]
    Hence, for any $x \in \zo^n$, there is a corresponding $x' \in \zo^{O(n^{r-1})}$ such that for ever $j \in [m]$, $(Gx')_j \neq 0 \iff P_j(x) \neq 0$. Therefore if a set of weighted indices is a sparsifier for the code generated by $G$, this same set of indices must be a sparsifier for the CSP over predicates $P_j$.

    Finally, we conclude by noting that we proved the existence of $(\eps, \widetilde{O}(n \log^2(q) / \eps^2))$ sparsifiers for affine Abelian predicates over any Abelian group of size $q$, in a universe of variables of size $n$. By applying this to our generating matrix $G$, this translates to a $(1 \pm \eps)$-sparsifier for $G$ of size $\widetilde{O}(n^{r-1} \cdot 4^r / \eps^2) = \widetilde{O}_r(n^{r-1}/ \eps^2)$.    
    \end{proof}

\begin{remark}
    Note that in many senses the aforementioned result is the best possible. Indeed, for predicates on $r$ variables which have 1 satisfying assignment, these predicates will have a projection to an AND of arity $r$, and thus require sparsifiers of size $\Omega(n^r)$ in the worst case. 

    Likewise, there exist predicates of arity $r$ with $2^{r-1}+1$ satisfying assignments which require sparsifiers of size $\Omega(n^{r-1})$ in the worst case. Consider for instance the predicate $P: \zo^r \ra \zo$, such that $P(1,...) = 1$, and $P(0,x_1,\dots x_{r-1}) = \text{AND}(x_1, \dots x_{r-1})$. This predicate has a projection to an AND of arity $r-1$, and thus requires sparsifiers of size $\Omega(n^{r-1})$. We are able to match this sparsifier size even when we are only told that the predicate has \emph{two} satisfying assignments, and further, our result holds for \emph{any} collection of predicates, provided each one has $0$, or at least $2$ satisfying assignments.
\end{remark}

\section{Classifying Predicates of Arity $3$}\label{sec:3-ary-classification}

In this section, we prove \cref{thm:3-ary-classufication-intro}. Recall this theorem:

\aryclassificationintro*

\begin{proof}
    Let any such predicate $P$ be given. First, the work of \cite{KPS24} shows that if $P$ does not have a projection to $\AND_2$, then $\CSP(P)$ can be written as an affine predicate (and in particular, can thus be efficiently sparsified to size $\widetilde{O}(n/\eps^2)$). Further, if $P$ has a projection to $\AND_2$, the beset possible size of any sparsifier is $\Omega(n^2)$. So, it remains only to distinguish between when $\CSP(P)$ is sparsifiable to near-quadratic size vs. cubic size. Now, suppose that $P$ has only $1$ satisfying assignment. Then $P$ (up to negation) is $\AND_3$, and requires sparsifiers of size $\Omega(n^3)$ (and trivially admits such sparsifiers). Finally, let us consider all other predicates, i.e., $P$ with projections to $\AND_2$, but not only $1$ satisfying assignment. These predicates require sparsifiers of size $\Omega(n^2)$ (by \cite{KPS24}), yet by \cref{thm:intro-non-trivial}, efficiently admit sparsifiers of size $\widetilde{O}(n^2 / \eps^2)$. This completes the proof.
\end{proof}

\section{Applications Beyond CSPs}

In this section, we discuss applications of our framework beyond just CSPs. In particular, we discuss efficient $\F_2$ Cayley-graph sparsification, hedge-graph sparsification, and sparsifying general hypergraphs with $\zo$-valued cardinality-based splitting functions. 

\subsection{Efficient Cayley-graph Sparsification over $\F_2$}\label{sec:cayleyGraph}

First, we recall the definition of a Cayley graph and the notion of a Cayley graph sparsifier.

\begin{definition}
    A Cayley graph $G$ is a graph with algebraic structure; its vertex set is defined to be a group, and the edges correspond to a set of generators $S$, along with weight $(w_i)_{i \in S}$. For every element in $s \in S$, and for every vertex $v$, there is an edge from $v$ to $v + s$ of weight $w_s$.
\end{definition}

\begin{definition}
    Given a Cayley graph $G$ with generating set $S$ over $\F_2^n$, we say that $\widetilde{G}$ with a re-weighted generating set $\widetilde{S} \subseteq \widetilde{S}$ is a $(1 \pm \eps)$ Cayley-graph sparsifier of $G$ if 
    \[
    (1 - \eps)L_G \preceq L_{\widetilde{G}} \preceq (1 + \eps) L_G,
    \]
    where $L_G$ here is used to denote the Laplacian of the graph $G$.
\end{definition}

\cite{KPS24} provided the first proof of the existence of $(1 \pm \eps)$ Cayley-graph sparsifiers over $\F_2^n$ where the resulting generating set retains only $\widetilde{O}(n / \eps^2)$ generators. At the core of their result is the following theorem:

\begin{fact}\label{clm:CayleytoCode}\cite{KPS24}
    Given a Cayley graph $G$ with generating set $S$ over $\F_2^n$, let $H$ be the matrix in $\F_2^{|S| \times n}$ where one places the generators as rows in the matrix (with their corresponding weights). If $\widetilde{H}$ is a $(1 \pm \eps)$ code-sparsifier of $H$, then the Cayley-graph $\widetilde{G}$ with weighted generators coming from the rows of $\widetilde{H}$ is a $(1 \pm \eps)$ Cayley-graph sparsifier of $G$.
\end{fact}

Using this, we can derive the following theorem:

\begin{theorem}
    Given a Cayley graph $G$ with generating set $S$ over $\F_2^n$ and a parameter $\eps \in (0,1)$, there is a polynomial time (in $|S|,n, 1 / \eps$), randomized algorithm which produces a $(1 \pm \eps)$ Cayley-graph sparsifier $\widetilde{G}$ with generating set $\widetilde{S} \subseteq S$ such that $|\widetilde{S}| = \widetilde{O}(n / \eps^2)$.
\end{theorem}

\begin{proof}
    Given the graph $G$ we create the generating matrix $H$ as per \cref{clm:CayleytoCode}. Next, we invoke \cref{thm:efficientSparsifyZq} to efficiently sparsify the generating matrix $H$. This yields a $(1 \pm \eps)$ code-sparsifier $\widetilde{H}$ of $H$ (in randomized polynomial time) such that $\widetilde{H}$ preserves only $\widetilde{O}(n / \eps^2)$ coordinates of $H$ with probability $1 - 1 / \mathrm{poly}(n)$. This yields the claim.
\end{proof}

\subsection{Cayley-graph Sparsifiers over $\Z_q^n$}\label{sec:cayleyGraphGeneral}

In this section, we show how to sparsify more general cayley-graphs over $\Z_q^n$ (where $q$ is an arbitrary composite number).

For this, we first recall the following characterization of the eigenvectors of cayley-graphs over $\Z_q^n$.

\begin{definition}
    Let $\Gamma$ be the group over $\Z_q^n$. Then we say that $\Gamma$ has $q^n$ \emph{characters}, one for each vector $r \in \Z_q^n$. We denote each character by $\chi_r: \Z_q^n \rightarrow \mathbb{C}$. For a vector $x \in \Z_q^n$, we have that 
    \[
    \chi_r(x) = e^{\frac{2 \pi i}{q} \cdot \langle r, x \rangle}.
    \]
    Further, for each character, we can define a vector $x_r \in \mathbb{C}^{\Gamma}$, where $(x_r)_a = \chi_r(a)$. 
    
    Then, if we let $G = \mathrm{Cay}(\Gamma, S)$, we have that $(x_r)_a $ is an eigenvalue of the adjacency matrix of $G$ with eigenvalue 
    \[
    \sum_{s \in S} \chi_r(s). 
    \]
    In general, Cayley graphs may have weights $w_s$ associated with each generator $s$. Then, the corresponding eigenvalues are simply the weighted sum.
\end{definition}

In particular, we can simplify the expression of the eigenvalues when we consider \emph{cyclically closed} Cayley graphs:

\begin{definition}
For a generator $s \in \Z_q^n$, we say the cycle induced by $s$ is $s, 2s, 3s, \dots (q-1)s$. We denote these cycles by $\mathrm{Cyc}(s)$, and we give each element in the cycle the same weight as the generator $s$.

We say a Cayley graph $G = \mathrm{Cay}(\Gamma, S)$ is cyclically closed if there exists a set $S'$ such that 
\[
\bigcup_{s \in S'} \mathrm{Cyc}(s) = S,
\]
where we use the convention that union operator adds the weights of generators. I.e., if an element $p$ appears in $\mathrm{Cyc}(s_1), \mathrm{Cyc}(s_2)$, then the weight of $p$ is $w_{s_1} + w_{s_2}$. Notationally, we say that $S = \mathrm{Cyc}(S')$.
\end{definition}

Now, observe that for a generator $s \in \Z_q^n$, and a character $\chi_r$, we have that
\[
\chi_r(s) + \chi_r(2s) + \dots + \chi_r((q-1)s) = \sum_{j = 1}^q \chi_r(js) - 1.
\]
In particular, when $\langle r, s \rangle = 0$, then we see that this expression evaluates to $q-1$. Otherwise, when $\langle r, s \rangle \neq 0$, it evaluates to $-1$, as we are summing together (all but one of) the powers of a root of unity. With this, we get the following characterization:

\begin{claim}\label{clm:eigenvalueCodewordWeights}
    Let $G = \mathrm{Cay}(\Z_q^n, \mathrm{Cyc}(S'))$. Then, the eigenvalue of $L_G$ corresponding to $\chi_r$ is 
    \[
 q \cdot \wt(Hr),
    \]
    where $H \in \Z_q^{n \times |S'|}$ is the generating matrix of a code over $\Z_q$ where there is a single row for each element $s \in S'$.
\end{claim}

\begin{proof}
    Recall that the eigenvalue of the adjacency matrix corresponding to $\chi_r$ is equal to
    \[
    \sum_{s \in \mathrm{Cyc}(S')} \chi_r(s) = \sum_{s \in S'}\left ( \sum_{j = 1}^q \chi_r(js) - 1 \right ) = \sum_{s \in S'} q \cdot \mathbf{1}[\langle r, s \rangle = 0] - 1.
    \]

    For the Laplacian, we simply subtract the corresponding eigenvalue of the adjacency matrix from the degree. For each $s \in S'$, because the Cayley graph is cyclically closed, the corresponding degree is $q-1$. Thus, the corresponding eigenvalue of $L_G$ is 
    \[
    \sum_{s \in S'} (q - 1 - (q \cdot \mathbf{1}[\langle r, s \rangle = 0] - 1)) = \sum_{s \in S'} w_s \cdot q \cdot \mathbf{1}[\langle r, s \rangle \neq 0].
    \]
    In particular, if we let $H \in \Z_q^{n \times |S'|}$ denote the generating matrix of a code over $\Z_q$ where there is a single row for each $s \in S'$, then we can observe that the eigenvalue of $L_G$ corresponding to $r \in \Z_q^n$ is exactly $q \cdot \wt(Hr)$, where we use the weighted notion of Hamming weight. 
\end{proof}

\begin{claim}\label{clm:codeSparsifierSuffices}
     Let $G = \mathrm{Cay}(\Z_q^n, \mathrm{Cyc}(S'))$, and let $H$ be its corresponding generating matrix over $\Z_q$. If $\widetilde{H}$ is a $(1 \pm \eps)$ code-sparsifier of $H$ with the rows of $\widetilde{H}$ being denoted by $\widetilde{S'}$, then $\widetilde{G} = \mathrm{Cay}(\Z_q^n, \mathrm{Cyc}(\widetilde{S'}))$ is a $(1 \pm \eps)$ spectral Cayley-graph sparsifier of $G$. 
\end{claim}

In the above claim, we are using the convention that the weight assigned to the coordinate corresponding to a generator $s \in \widetilde{S'}$ is the same as the weight assigned the same generator $s$ in the Cayley graph $\widetilde{H}$.

\begin{proof}
    First, recall that the eigenvectors of the Laplacian of an abelian Cayley graph are completely determined by the underlying group. Thus, $\widetilde{G}, G$ have the same eigenvectors. Thus, it remains only to show that their eigenvalues are within a factor of $(1 \pm \eps)$. For this, recall that by \cref{clm:eigenvalueCodewordWeights}, the eigenvalues of $G$ are exactly $q \cdot \wt(Hr)$ for each $r \in \Z_q^n$. Likewise, for the graph $\widetilde{G}$, the eigenvalues are exactly $q \cdot \wt(\widetilde{H}r)$, where in both instances the Hamming weight we use is the \emph{weighted} notion of Hamming weight. In particular, because $\widetilde{H}$ is a $(1 \pm \eps)$ code-sparsifier of $H$, we have that 
    \[
    q \cdot \wt(\widetilde{H}r) \in (1 \pm \eps) q \cdot \wt(Hr). 
    \]
    Hence, $\widetilde{G}$ is a $(1 \pm \eps)$ spectral-sparsifier of $G$. 
\end{proof}

With this, we are ready to conclude our main theorem:

\begin{theorem}
    Let $G = \mathrm{Cay}(\Z_q^n, \mathrm{Cyc}(S))$, and let $\eps \in (0,1)$. Then, there is a randomized, polynomial time algorithm (in $\log(q), \eps, n, |S|$) which returns (with high probability) a re-weighted sub-Cayley graph $\widetilde{G} = \mathrm{Cay}(\Z_q^n, \mathrm{Cyc}(\widetilde{S}))$, with $|\widetilde{S}| = \widetilde{O}(n \mathrm{polylog}(q) / \eps^2)$ such that $\widetilde{G}$ is a $(1 \pm \eps)$ spectral sparsifier of $G$. 
\end{theorem}

\begin{proof}
    By \cref{clm:codeSparsifierSuffices}, it suffices to simply compute a $(1 \pm \eps)$ code sparsifier of $H$, where $H$ is the generating matrix induced by $S$. By invoking \cref{thm:efficientSparsifyZq}, this can be done in the stated time, yielding the above theorem. 
\end{proof}

\subsection{Efficient Hedge-graph Sparsification}\label{sec:hedgegraphs}

In this section, we detail the procedure of creating efficient $(1 \pm \eps)$ hedge-graph cut sparsifiers. To start, we recap the definition of a hedge-graph.

\begin{definition}
    A hedge-graph $G = (V, E)$ is defined by a set of vertices $V$ and a set of hedges $E$. Each hedge $E$ is itself a collection of edges $\in \binom{V}{2}$. For a subset of vertices $S \subseteq V$, we say that a hedge $e$ is cut by $S$ if there exists an edge $f \in e$ such that $f$ is cut by $S$. We define cut-sizes globally by 
    \[
    \cut_G(S) = \sum_{e \in E}\mathbf{1}[e \text{ is cut by }S].
    \]
\end{definition}

As remarked in prior work (see \cite{GKP17}) hedge-graph cuts behave very differently from graph (or hypergraph) cuts. In particular, there exist hedge-graphs with respect to which the cut function is not submodular. In this same work \cite{GKP17} it was remarked that uniform random sampling at a rate roughly equal to the reciprocal of the minimum hedge-cut \emph{does not} preserve all the cuts in the sparsifier. Thus, an analysis mimicking \cite{BK96} for creating cut-sparsifiers of hedge-graphs. Nevertheless, we show here that the code sparsification framework is sufficiently general such that one can derive efficient sparsifications of many hedge-graphs. We formalize this below:

\begin{definition}
    Given a hedge-graph $G$, we say that a re-weighted sub-hedge-graph $\widetilde{G}$ is a $(1 \pm \eps)$ cut-sparsifier of $G$ if $\forall S \subseteq V$:
    \[
    (1 - \eps) \cut_G(S) \leq \cut_{\widetilde{G}}(S) \leq (1 + \eps)  \cut_G(S).
    \]
\end{definition}

We will also use the following observation in our discussion of hedge-graph sparsifiers:

\begin{fact}
    Each hedge $e \in E$ induces a set of connected components, which we denote $\mathcal{P}_e$, that partitions the vertex set $V$. The hedge $e$ is cut by a set $S \subseteq V$ \emph{if and only if} one of the constituent connected components $C_i \in \mathcal{P}_e$ is cut by the set $S$. Formally, $e$ is cut if and only if $\exists C_i \in \mathcal{P_e}: C_i \cap S \neq \emptyset \wedge C_i \cap \bar{S} \neq \emptyset$.
\end{fact}

Going forward, we also use $R_e$ to denote the number of connected components in $\mathcal{P}_e$ which are of size $\geq 2$. With this, we are able to state our main theorem:

\begin{theorem}
    Let $G$ be a hedge-graph on $n$ vertices, let $\eps \in (0,1)$, and let $R = \max_e R_e$. Then, there is a randomized polynomial time algorithm for computing a $(1 \pm \eps)$ cut-sparsifier $\widetilde{G}$ of $G$ (with high probability), such that $\widetilde{G}$ preserves only $\widetilde{O}(n R^2 / \eps^2)$ re-weighted hedges.
\end{theorem}

\begin{remark}
    Note that $R$ is separate from the size of the hedge. In particular, a hyperedge is a hedge where $R = 1$.
\end{remark}

\begin{proof}
    First, let us choose a prime $p \in [n, 2n]$. The proof proceeds by creating a code over $\F_{p^R}$. This means that the generating matrix $H$ we create is $\in \F_{p^R}^{m \times n}$. Now, for any hedge $e \in E$, let us use $C_1, \dots C_R$ to denote its connected components of size $\geq 2$. Our goal will be to write the cut-function of $G$ as the OR of a cut-condition of each of the individual components. Then, observe that we are exactly implementing the logic of a hedge-cut; namely, a hedge is cut if and only if some constituent component of the hedge is cut. 

    Next, for the field $\F_{p^R}$, recall that we create this field by extending the field $\F_p$. We let $\alpha_1, \dots \alpha_{R-1}$ denote the roots we use to extend the field. An equivalent way to view each element $x$ in $\F_{p^R}$ is as a linear combination of the $\alpha_i$'s:
    \[
    x = b_0 + b_1 \alpha_1 + \dots + b_{R-1}\alpha_{R-1},
    \]
    where $b_i \in \F_p$.

    In particular, an element $x$ is non-zero \emph{if and only if} $\exists b_i$ in the above representation such that $b_i \neq 0$. Thus gives us a natural way to express the cut function of each hedge, as the $\alpha_i$'s can essentially simulate an OR of each individual component being cut. So, before concluding, our final ingredient is an expression which evaluates to $0$ if and only if a certain component is not cut. Indeed let $C_i$ denote the component, and let $v^*$ denote the largest label of vertex in $C_i$. We can write:
    \[
    \mathbf{1}[C_i \text{ is cut}] = \mathbf{1}[(\sum_{v \in C_i} x_v) - |C_i|\cdot x_{v^*} \neq 0] =  \mathbf{1}[f_{C_i}(x) \neq 0] =.
    \]
One can verify that for $x = \mathbf{1}_S \in \zo^n$, the above expression exactly captures whether the cut $S$ splits the component $C_i$. To conclude then, for the hedge $e$, we simply include a row in the generating matrix of the form:
\[
\sum_{i = 0}^{\leq R-1} \alpha_i \cdot f_{C_{i-1}}(x). 
\]
By the above logic,
\[
\sum_{i = 0}^{\leq R-1} \alpha_i \cdot f_{C_{i-1}}(\mathbf{1}_S) = \mathbf{1}[e \text{ is cut}].
\]

In particular after creating the generating matrix $H$ in this manner, we can invoke \cref{thm:efficientSparsifyZq} to create a $(1 \pm \eps)$ code-sparsifier of $H$ with only $\widetilde{O}(n R^2 / \eps^2)$ re-weighted rows remaining. This sparsifier must preserve codeword weights for any vector $x \in \zo^n$, and hence the same selection of re-weighted hedges would constitute a $(1 \pm \eps)$ cut-sparsifier of the hedge-graph $G$. This yields the claim.  
\end{proof}

\subsection{Efficient Cardinality-based Splitting Function Sparsification}

First, let us recall the notion of generalized hypergraph sparsification:

\begin{definition}
    For a hypergraph $G = (V, E)$, a general hypergraph associates a splitting function $g_e: 2^e \rightarrow \R^+$ to each hyperedge $e \in E$. For a set $S \subseteq V$, we define 
    \[
    \cut_G(S) = \sum_{e \in E} g_e(S \cap e).
    \]
    We say a re-weighted sub-hypergraph of $G$ is a $(1 \pm \eps)$ cut-sparsifier of $\widetilde{G}$ if for all $S \subseteq V:$
    \[
    (1- \eps)\cut_G(S) \leq \sum_{e \in \widetilde{e}} \widetilde{w}_e \cdot g_e(S \cap e) \leq (1 + \eps)\cut_G(S).
    \]
\end{definition}

In the general hypergraph sparsification literature, one particularly important class of splitting functions are the so-called ``cardinality-based splitting function''.

\begin{definition}\label{def:cardinality}
    For a hyperedge $e \in E$, we say that $g_e: 2^e \rightarrow \R^+$ is a cardinality-based splitting function if there exists $f_e: \Z \rightarrow \R^+$ such that $\forall S \subseteq V$, $g_e(S \cap e) = f(|S \cap e|)$ (i.e., there is a function $f$ which depends only on the cardinality of the input set which dictates the value of the splitting function). 
\end{definition}

Beyond this, we call a splitting function a $\zo$-valued cardinality-based splitting function if $g_e$ is cardinality-based and maps to the range $\zo$.

In this regime, we derive the following theorem:

\begin{theorem}
    Let $G$ be a general hypergraph, and for every $e \in E$, suppose that $g_e$ is the same $\zo$-valued, cardinality-based splitting function. Then, $G$ is efficiently-sparsifiable to $\widetilde{O}(n /\eps^2)$ hyperedges if and only if $f_e$ is periodic.
\end{theorem}

Here, we are using $f_e$ in the same way as defined in \cref{def:cardinality}.

\begin{proof}
    Fix a hyperedge $e \in E$, and recall that for $S \subseteq V$, we have $g_e(S) = f_e(|e \cap S|) \in \zo$. In particular, because every hyperedge has the same splitting function, we can simply use $f = f_e$ for every $e \in E$. Note that this also necessarily means that every hyperedge is of the same arity, which we denote by $r$. 

    The key observation is that this general hypergraph is now equivalent to a CSP. Indeed, for any hyperedge $e \in E$, there is a corresponding constraint $f: \zo^r \rightarrow \zo$ operating on the variables $(y_1, \dots y_r)$ in $e$. For any cut $S \subseteq V$, and any assignment to the variables $x = \mathbf{1}_S$, we have that $g_e(S) = 1$ if and only if $f((\mathbf{1}_S)_{e_1}, \dots (\mathbf{1}_S)_{e_r}) = 1$.

    Now, we can simply invoke \cref{thm:intro-sym} to conclude the stated theorem. The efficiency follows from the same theorem.
\end{proof}

\ifanon 
\else 
\section{Acknowledgements}

We would like to thank Swastik Kopparty for supplying us with the proof of \cref{clm:symm-mod-6} and Srikanth Srinivasan for introducing us to the notion of polynomials over groups as used in \cref{thm:intro-poly}.
\fi

\bibliographystyle{alpha}
\bibliography{ref}

\appendix

\section{Detailed Proof of Sparsifiers for Abelian Codes}\label{sec:abelianComplete}

\subsection{Efficient Spanning Subsets for Abelian Codes}

Here, we re-produce the algorithms used to create spanning subsets for codes over $\Z_q$ in the new setting of $(\Z_{q_1} \times \dots \times \Z_{q_u})$. For a code $\calC \subseteq (\Z_{q_1} \times \dots \times \Z_{q_u})^{m}$, the following algorithms take as input a generating matrix $H \in (\Z_{q_1} \times \dots \times \Z_{q_u})^{m \times n}$, and construct maximum spanning subsets. Note that these maximum spanning subsets are defined in the same manner as before. Going forward, we will let $q = \prod_{i = 1}^u q_i$.

\begin{algorithm}[H]
    \caption{BuildMaxSpanningSubsetAbelian$(H)$}
        Initialize $T = \emptyset$. \\
        Let $k = 0$. \\
        \For{$i \in [m]$}{
        If the number of distinct codewords in the span of $H|_{T \cup \{i \}}$ is $\geq k$, then set $T = T \cup \{i \}$, and $k \leftarrow $ the number of distinct codewords in the span of $H|_{T \cup \{i \}}$.
}
\Return{$T$}
    \end{algorithm}

    \begin{algorithm}[H]
    \caption{ConstructSpanningSubsetsAbelian($H, t$)} \label{alg:constructSpanningSubsetsAbelian}
    \For{$i \in [t]$}{
    Let $T_i = \mathrm{BuildMaxSpanningSubsetAbelian}(H|_{\bar{T_1} \cap \dots \bar{T_{i-1}}})$.
    }
    \Return{$T_i: i \in [t]$}.
\end{algorithm}

Note that, as before, the correctness and efficient implementation of these algorithms follows from exactly the same reasoning as with codes over $\Z_q$, as these proofs rely on the contraction algorithm which has a direct analog. In particular, we are able to conclude the following:

\begin{claim}\label{clm:containsBadSetAppendix}
    For a generating matrix $G \in (\Z_{q_1} \times \dots \times \Z_{q_u})^{m \times n}$, any choice of $d$ and the set $S$ of bad rows (i.e. those guaranteed by \cref{cor:moduleKargerAbelian}) for $G$ and parameter $d$, for any disjoint maximum spanning subsets $T_1, \dots T_{2 d (\log(n) + \log(q))}$, we have that $S \subseteq T_1 \cup T_2 \dots \cup T_{2 d (\log(n) + \log(q))}$.
\end{claim}

\subsection{Weighted Decomposition}

Here, we introduce an analog of the weight decomposition step used for codes over $\Z_q$ which will instead be defined for codes over $(\Z_{q_1} \times \Z_{q_2} \dots \Z_{q_u})$. Consider the following procedure which operates on a code of length $m$ and at most $q^n$ distinct codewords in Algorithm \ref{alg:WeightClassDecompositionAppendix}. 

\begin{algorithm}
	\caption{WeightClassDecomposition$(\calC, \eps, \alpha)$}\label{alg:WeightClassDecompositionAppendix}
	Let $E_i$ be all coordinates of $\calC'$ that have weight between $[\alpha^{i-1}, \alpha^i]$. \\
	Let $\Dodd = E_1 \cup E_3 \cup E_5 \cup \dots$, and let $\Deven = E_2 \cup E_4 \cup E_6 \cup \dots$. \\
	\Return{$\Dodd, \Deven$}.
\end{algorithm}

Next, we prove some facts about this algorithm.

\begin{lemma}\label{clm:PreserveApproxWeightDecompAppendix}
	Consider a code $\calC$ with at most $q^n$ distinct codewords and length $n$. Let 
	\[
	\Dodd, \Deven = \mathrm{WeightClassDecomposition}(\calC, \eps, n).
	\] To get a $(1 \pm \eps)$-sparsifier for $\calC$, it suffices to get a $(1 \pm \eps)$ sparsifier to each of $\Dodd, \Deven$.
\end{lemma}

\begin{proof}
The creation of $\Dodd, \Deven$ forms a \emph{vertical} decomposition of the code $\calC'$. Thus, by \cref{clm:verticalDecomp}, if we have a $(1\pm \eps)$ sparsifier for each of $\Dodd, \Deven$, we have a $(1\pm \eps)$ sparsifier to $\calC$.
\end{proof}

Because of the previous claim, it is now our goal to create sparsifiers for $\Dodd, \Deven$. Without loss of generality, we will focus our attention only on $\Deven$, as the procedure for $\Dodd$ is exactly the same (and the proofs will be the same as well). At a high level, we will take advantage of the fact that 
\[
\Deven = E_2 \cup E_4 \cup \dots ,
\]
where each $E_i$ contains edges of weights $[\alpha^{i-1}, \alpha^i]$, for $\alpha = \frac{m^3}{\eps^3}$, where $m$ is the length of the code. Because of this, whenever a codeword $c \in \calC'$ is non-zero in a coordinate in $E_i$, we can effectively ignore all coordinates of lighter weights $E_{i-2}, E_{i-4}, \dots$. This is because any coordinate in $E_{\leq i-2}$ has weight at most a $\frac{\eps^3}{m^3}$ fraction of any single coordinate in $E_{i}$. Because there are at most $O(m)$ coordinates in $\calC'$, it follows that the total possible weight of all rows in $E_{\leq i-2}$ is still at most a $O(\eps / m)$ fraction of the weight of a single row in $E_{i}$. Thus, we will argue that when we are creating a sparsifier for codewords that are non-zero in a row in $E_i$, we will be able to effectively ignore all rows corresponding to $E_{\leq i-2}$. Thus, our decomposition is quite simple: we first restrict our attention to $E_i$ and create a $(1 \pm \eps)$ sparsifier for these rows. Then, we transform the remaining code such that only codewords which are all zeros on $E_i$ remain. We present this transformation below:

\begin{algorithm}
	\caption{SingleSpanDecomposition$(\Deven, \alpha, i)$}\label{alg:SingleSpanDecompositionAppendix}
	Let $E_i$ be all rows of $\Deven$ with weights between $\alpha^{i-1}$ and $\alpha^i$. \\
	Let $G$ be a generating matrix for $\Deven$. \\
	Store $G|_{E_i}$. \\
	Let $G' = G$. \\
	\While{$G'|_{E_i}$ is not all zero}
	{
		Find the first non-zero coordinate of $G'|_{E_i}$, call this $j$. \\
		Set $G' = \ContractAbelian(G', j)$. \\
}
	
	\Return{$G|_{E_i}$, $G'|_{\bar{E_i}}$}
\end{algorithm}

\begin{claim}\label{clm:conserveCodewordsAppendix}
	If the span of $G$ originally had $2^{n'}$ distinct codewords, and the span of $G|_{E_i}$ has $2^{n''}$ distinct codewords, then after \cref{alg:SingleSpanDecompositionAppendix}, the span of  $G'|_{\bar{E_i}}$ has $2^{n' - n''}$ distinct codewords.
\end{claim}

\begin{proof}
	After running the above algorithm, $G'$ is entirely $0$ on the rows corresponding to $E_i$, hence it follows that after running the algorithm, $G'$ and $G'|_{\bar{E_i}}$ have the same number of distinct codewords. Now, we will argue that the span of $G'$ has at most $2^{n' - n''}$ distinct codewords. Indeed, for any codeword $c$ in the span of $G'$, $c$ is also in the span of the original $G$. However, for this same $c$, in the original $G$ we could add any of the $2^{n'}$ distinct vectors which are non-zero on the rows corresponding to $E_i$. Thus, the span of $G$ must have at least $2^{n'}$ times as many distinct codewords as $G'$. This concludes the claim. 
\end{proof}

\begin{claim}
	For any codeword $c$ in the span of $G$, if $c$ is zero in the coordinates corresponding to $E_i$, then $c$ is still in the span of $G'$ after the contractions of \cref{alg:SingleSpanDecompositionAppendix}.
\end{claim}

\begin{proof}
	This follows from \cref{clm:stillInSpanAbelian}. If a codeword is $0$ in a coordinate which we contract on, then it remains in the span. Hence, if we denote by $c$ a codeword which is zero in all of the coordinates of $E_i$, then $c$ is still in the span after contracting on the coordinates in $E_i$.
\end{proof}

\begin{claim}\label{clm:recursivebreakdownAppendix}
	In order to get a $(1 \pm \eps)$ approximation to $\Deven$, it suffices to combine a $(1 \pm \eps / 2)$ approximation to $G|_{E_i}$ and a $(1 \pm \eps)$ approximation to $G'|_{\bar{E_i}}$. 
\end{claim}

\begin{proof}
	For any codeword $c \in \Deven$ which is non-zero on rows $E_i$, it suffices to get a $(1 \pm \eps/2)$ approximation to their weight on $G|_{E_i}$, as this makes up at least a $(1 \pm \eps / n)$ fraction of the overall weight of the codeword.
	
	For any codeword $c \in \Deven$ which is zero on rows $E_i$, then $c$ is still in the span of $G'$, and in particular, its weight when generated by $G$ is exactly the same as its weight in $G'_{\bar{E_i}}$ (as it is zero in the coordinates corresponding to $E_i$, we can ignore these coordinates). Hence, it suffices to get a $(1 \pm \eps)$ approximation to the weight of $c$ on $G'_{\bar{E_i}}$. Taking the union of these two sparsifiers will then yield a sparsifier for every codeword in the span of $\Deven$.
\end{proof}

\begin{algorithm}
    \caption{SpanDecomposition$(\Deven, \alpha)$}\label{alg:SpanDecompositionAppendix}
    Let $\Deven' = \Deven = E_2 \cup E_4 \cup E_6 \dots$. \\
    Let $S = \{ \}$.\\
    \While{$\Deven'$ is not empty}{
    Let $i$ be the largest integer such that $E_i$ is non-empty in $\Deven'$. \\
    Let $G|_{E_i}, G'|_{\bar{E_i}} =$ SingleSpanDecomposition$(\Deven', \alpha, i)$. \\
    Let $\Deven'$ be the span of $G'|_{\bar{E_i}}$, and let $H_i = G|_{E_i}$. \\
    Add $i$ to $S$.
    }
    \Return{$S, H_i$ for every $i \in S$}
\end{algorithm}

\begin{claim}\label{clm:conserveRankAppendix}
    Let $S, H_i$ be as returned by \cref{alg:SpanDecompositionAppendix}. Then, $\sum_{i \in S} \log(|\text{Span}(H_i) |) =  \log(|\text{Span}(\Deven)|)$.
\end{claim}

\begin{proof}
    This follows because from line 5 of \cref{alg:SpanDecompositionAppendix}. In each iteration, we store $G|_{E_i}$, and iterate on $G'|_{E_i}$. From \cref{clm:conserveCodewordsAppendix}, we know that
    \[
    (\text{number of distinct codewords in } G|_{E_i}) \cdot (\text{number of distinct codewords in } G'|_{\bar{E_i}})\]\[ = (\text{number of distinct codewords in } G),\]
    thus taking the log of both sides, we can see that the sum of the logs of the number of distinct codewords is preserved.
\end{proof}

\begin{lemma}\label{clm:OnlyApproxHiAppendix}
    Suppose we have a code of the form $\Deven$ created by \cref{alg:WeightClassDecompositionAppendix}. Then, if we run \cref{alg:SpanDecompositionAppendix} on $\Deven$, to get $S, (H_i)_{i \in S}$, it suffices to get a $(1 \pm \eps/2)$ sparsifier for each of the $H_i$ in order to get a $(1 \pm \eps)$ sparsifier for $\Deven$.
\end{lemma}

\begin{proof}
This follows by inductively applying \cref{clm:recursivebreakdownAppendix}. Let our inductive hypothesis be that getting a $(1 \pm \eps/2)$ sparsifier to each of codes returned of \cref{alg:SpanDecompositionAppendix} suffices to get a $(1 \pm \eps)$ sparsifier to the code overall. We will induct on the number of recursive levels that \cref{alg:SpanDecompositionAppendix} undergoes (i.e., the number of distinct codes returned by the algorithm). In the base case, we assume that there is only one level of recursion, and that \cref{alg:SpanDecompositionAppendix} simply returns a single code. Clearly then, getting a $(1 \pm \eps/2)$ sparsifier to this code suffices to sparsify the code overall. 

Now, we prove the claim inductively. Assume the algorithm returns $\ell$ codes. After the first iteration, we decompose $\Deven$ into $H_i = G|_{E_i}$ and $G'|_{\bar{E_i}}$. By \cref{clm:recursivebreakdownAppendix}, it suffices to get a $(1 \pm \eps/2)$ sparsifier to $H_i$, while maintaining a $(1 \pm \eps)$ sparsifier to $G'|_{\bar{E_i}}$. By invoking our inductive claim, it then suffices to get a $(1 \pm \eps/2)$ sparsifier for the $\ell -1$ codes returned by the algorithm on $G'|_{\bar{E_i}}$. Thus, we have proved our claim.
\end{proof}

\subsection{Dealing with Bounded Weights}

Let us consider any $H_i$ that is returned by \cref{alg:SpanDecompositionAppendix}, when called with $\alpha = m^3 / \eps^3$. By construction, $H_i$ will contain weights only in the range $[\alpha^{i-1}, \alpha^{i}]$ and will have at most $O(m)$ coordinates. In this subsection, we will show how we can turn $H_i$ into an unweighted code  with at most $\text{poly}(m / \eps)$ coordinates. First, note however, that we can simply pull out a factor of $\alpha^{i-1}$, and treat the remaining graph as having weights in the range of $[1, \alpha]$. Because multiplicative approximation does not change under multiplication by a constant, this is valid. Formally, consider the following algorithm:

\begin{algorithm}
    \caption{MakeUnweighted$(\calC, \alpha, i, \eps)$}\label{alg:MakeUnweightedAppendix}
    Divide all edge weights in $\calC$ by $\alpha^{i-1}$. \\
    Make a new unweighted code $\calC'$ by duplicating every coordinate $j$ of $\calC$ $\lfloor 10 w(j) / \eps \rfloor$ times. \\
    \Return{$\calC', \alpha^{i-1} \cdot \frac{\eps}{10}$}
\end{algorithm}

\begin{lemma}\label{clm:PreserveApproxUnweightedAppendix}
Consider a code $\calC$ with weights bounded in the range $[\alpha^{i-1}, \alpha^i]$. To get a $(1 \pm \eps)$ sparsifier for $\calC$ it suffices to return a $(1 \pm \eps / 10)$ sparsifier for $\calC' =$ MakeUnweighted$(\calC, \alpha, i, \eps)$ weighted by $\alpha^{i-1} \cdot \frac{\eps}{10}$.
\end{lemma}

\begin{proof}
    It suffices to show that $\calC'$ is $(1 \pm \eps/10)$ sparsifier for $\calC$, as our current claim will then follow by Claim \ref{clm:composingApproximations} (composing approximations). Now, to show that $\calC'$ is $(1 \pm \eps/10)$ sparsifier for $\calC$, we will use Claim \ref{clm:verticalDecomp} (vertical decomposition of a code), and show that in fact the weight contributed by every coordinate in $\calC$ is approximately preserved by the copies of the coordinate introduced in $\calC'$.
    
    Without loss of generality, let us assume that $i = 1$, as otherwise pulling out the factor of $\alpha^{i-1}$ in the weights clearly preserves the weights of the codewords. Indeed, for every coordinate $j$ in $\calC$, let $w(j)$ be the corresponding weight on this coordinate, and consider the corresponding $\lfloor 10 w(r) / \eps \rfloor$ coordinates in $\calC'$. We will show that the contribution from these coordinates in $\calC'$, when weighted by $\eps/10$, is a $(1 \pm \eps/10)$ approximation to the contribution from $j$. 

    So, consider an arbitrary coordinate $j$, and let its weight be $w(j)$. Then,
    \[
    \frac{10w}{\eps} - 1 \leq \lfloor 10 w(j) / \eps \rfloor \leq \frac{10w}{\eps}.
    \]

    When we normalize by $\frac{\eps}{10}$, we get that the combined weight of the new coordinates $w'$ satisfies 
    \[
    w - \eps/10 \leq w' \leq w.
    \]

    Because $w \geq 1$, it follows that this yields a $(1 \pm \eps/10)$ sparsifier, and we can conclude our statement.
\end{proof}

\begin{claim}\label{clm:boundUnweightedSupportAppendix}
    Suppose a code $\calC$ of length $m$ has weight ratio bounded by $\alpha$, and minimum weight $\alpha^{i-1}$. Then, MakeUnweighted$(\calC, \alpha, i, \eps)$ yields a new unweighted code of length $O(m \alpha / \eps)$.
\end{claim}

\begin{proof}
    Each coordinate is repeated at most $O(\alpha / \eps)$ times.
\end{proof}

\subsection{Sparsifiers for Codes of Polynomial Length}

In this section, we introduce an efficient algorithm for sparsifying codes. We will take advantage of the decomposition proved in \cref{cor:moduleKarger} in conjunction with the following claim:

\begin{claim}\label{clm:preserveWeightSubsampleAppendix}
    Suppose $\calC$ is a code with at most $q^{n}$ distinct codewords over $(\Z_{q_1} \times \Z_{q_2} \dots \Z_{q_u})$ (where $q = q_1 \cdot \dots q_u$), and let $b \geq 1$ be an integer such that for any integer $\alpha \geq 1$, the number of codewords of weight $\leq \alpha b$ is at most $\binom{n\log(q)}{\alpha } \cdot q^{\alpha + 1} \leq (qn)^{2\alpha}$. Suppose further that the minimum distance of the code $\calC$ is $b$. Then, sampling the $i$th coordinate of $\calC$ at rate $p_i \geq \frac{\log(n) \log(q) \eta}{b \eps^2}$ with weights $1/p_i$ yields a $(1 \pm \eps)$ sparsifier with probability $1 - 2^{-(0.19\eta - 110) \log n} \cdot n^{-101}$.
\end{claim}

\begin{proof}

    Consider any codeword $c$ of weight $[\alpha b / 2, \alpha b]$ in $\calC$. We know that there are at most $(qn)^{\alpha}$ codewords that have weight in this range. The probability that our sampling procedure fails to preserve the weight of $c$ up to a $(1 \pm \eps)$ fraction can be bounded by Claim \ref{clm:concentrationBound}. Indeed,
    \[
    \Pr[\text{fail to preserve weight of } c] \leq 2e^{-0.38 \cdot \eps^2 \cdot \frac{\alpha b}{2} \cdot \frac{\eta \log (n) \log(q)}{\eps^2 b}} = 2e^{-0.19 \alpha \eta \log (n)\log(q)}.
    \]
    Now, let us take a union bound over the at most $(qn)^{2\alpha}$ codewords of weight between $[\alpha b / 2, \alpha]$. Indeed,
    \begin{align*}
    \Pr[\text{fail to preserve any } c \text{ of weight } [\alpha b / 2, \alpha b]] &\leq 2^{2\alpha \log (qn)} \cdot 2e^{-0.19 \alpha \eta \log (n)\log(q)} \\
     & \leq 2^{\alpha \cdot (-0.19\eta + 2) \log (n)\log(q)} \\
     & \leq 2^{\alpha \cdot (-0.19\eta + 2) \log (n)} \\
     & \leq 2^{-(0.19\eta - 110) \alpha \log n} \cdot 2^{-108 \alpha \log n} \\
     & \leq 2^{-(0.19\eta - 110) \log n} \cdot n^{-108 \alpha},
    \end{align*}
    where we have chosen $\eta$ to be sufficiently large. Now, by integrating over $\alpha \geq 1$, we can bound the failure probability for any integer choice of $\alpha$ by $2^{-(0.19\eta - 110) \log n} \cdot n^{-101}$.
\end{proof}

Next, we consider \cref{alg:CodeDecompositionAppendix}:

\begin{algorithm}[H]
\caption{CodeDecomposition$(\calC, d)$}\label{alg:CodeDecompositionAppendix}
Let $T$ be $\cup_i T_i$ for $T_i$ the sets of coordinates returned by ConstructSpanningSubsetsAbelian$(\calC, 2d(\log(n) + \log(q)))$. \\
Let $\calC'$ be the code $\calC$ after removing the set of coordinates $T$. \\
\Return{$T, \calC'$}
\end{algorithm}

Intuitively, the set $T$ returned by \cref{alg:CodeDecompositionAppendix} contains all of the \say{bad} rows which were causing the violation of the codeword counting bound. We know that if we removed \emph{only} the true set of bad rows, denoted by $S$, then we could afford to simply sample the rest of the code at rate roughly $1/d$ while preserving the weights of all codewords. Thus, it remains to show that when we remove $T$ (a superset of $S$) that this property still holds. More specifically, we will consider the following algorithm:

\begin{algorithm}
    \caption{CodeSparsify$(\calC, n, \eps, \eta)$}\label{alg:CodeSparsifyAppendix}
    Let $m$ be the length of $\calC$. \\
    \If{$m \leq 100 \cdot n \cdot \eta \log^2(n)\log^2(q) / \eps^2$}{\Return{$\calC$}}
    Let $d = \frac{m \eps^2}{2\eta \cdot n \log^2 (n) \log^2(q)}$. \\
    Let $T, \calC' = \text{CodeDecomposition}(\calC, \sqrt{d} \cdot \eta \cdot \log(n) \log(q) / \eps^2)$.
    Let $\calC_1 = \calC|_T$.
    Let $\calC_2$ be the result of sampling every coordinate of $\calC'$ at rate $1 / \sqrt{d}$. \\
    \Return{$\mathrm{CodeSparsify}(\calC_1, n, \eps, \eta) \cup \sqrt{d} \cdot \mathrm{CodeSparsify}(\calC_2, n, \eps, \eta)$ }
\end{algorithm}

\begin{lemma}\label{clm:overallSpaceAppendix}
    In Algorithm \ref{alg:CodeSparsifyAppendix}, starting with a code $\calC$ of size $2d n \log^2 (n) \log^2(q) / \eps^2$, after $i$ levels of recursion, with probability $1 - 2^{i} \cdot 2^{-\eta n}$, the code being sparsified at level $i$, $\calC^{(i)}$ has at most 
    \[
    2(1 + 1 / 2 \log \log (n))^{i} \cdot d^{1/2^i} \cdot \eta \cdot n \log^2(n)\log^2(q) / \eps^2
    \]
    surviving coordinates.
\end{lemma}

\begin{proof}
    Let us prove the claim inductively. For the base case, note that in the $0$th level of recursion the number of surviving coordinates in $\calC^{(0)} = \calC$ is $d \cdot 2n \log^2 (n) \log^2(q) / \eps^2$, so the claim is satisfied trivially.

    Now, suppose the claim holds inductively. Let $\calC^{(i)}$ denote a code that we encounter in the $i$th level of recursion, and suppose that it has at most \[
    2(1 + 1 / 2 \log \log (n))^{i} \cdot d^{1/2^i} \cdot \eta \cdot n \log^2(n) \log^2(q) / \eps^2
    \]
    coordinates. Denote this number of coordinates by $\ell$. Now, if this number is smaller than $100 n \eta \log^2(n) \log^2(q) / \eps^2$, we will simply return this code, and there will be no more levels of recursion, so our claim holds vacuously. Instead, suppose that this number is larger than $100 n \eta \log^2 (n) \log^2(q) / \eps^2$, and let $d' = \frac{\ell \eps^2}{2\eta n \log^2(n) \log^2(q)} \leq (1 + 1 / 2 \log \log (n))^{i} \cdot d^{1/2^i}$.

    Then, we decompose $\calC^{(i)}$ into two codes, $\calC_1$ and $\calC_2$. $\calC_1$ is the restriction of $\calC$ to the set of disjoint maximum spanning subsets. By construction, we know that $T$ is constructed by calling ConstructSpanningSubsetsAbelian with parameter $\sqrt{d'}\eta \log(n) \log(q) / \eps^2$, and therefore
    \[
    |T| \leq 2 \sqrt{d'}n \eta \log^2(n) \log^2(q) / \eps^2 \leq 2 (1 + 1 / 2 \log \log (n))^{i} n \eta d^{1/2^{i+1}}\log^2(n) \log^2(q) / \eps^2 ,
    \]
    satisfying the inductive claim. 
   
    For $\calC_2$, we define random variables $X_1 \dots X_{\ell}$ for each coordinate in the support of $\calC_2$. $X_i$ will take value $1$ if we sample coordinate $i$, and it will take $0$ otherwise. Let $X= \sum_{i = 1}^{\ell} X_i$, and let $\mu = \E[X]$. Note that 
    \[
    \frac{\mu^2}{\ell} = \left ( \frac{\ell}{\sqrt{d'}} \right )^2 / {\ell} = \frac{\ell}{d'} \geq \eta \cdot n \cdot \log^2(n) \log^2(q) / \eps^2.
    \]

    Now, using Chernoff, \[
    \Pr[X \geq (1 + 1 / 2 \log \log (n)) \mu] \leq e^{\frac{-2}{4 \log^2 \log(n)} \cdot \eta \cdot n \cdot \log(n) \log(q) / \eps^2} \leq 2^{-\eta n}, 
    \]
    as we desire. Since $\mu = \ell / \sqrt{d'} \leq (1 + 1 / 2 \log \log (n))^{i} \cdot d^{1/2^{i+1}} \cdot \eta \cdot n\log^2(n) \log^2(q) / \eps^2$, we conclude our result. 
    
    Now, to get our probability bound, we also operate inductively. Suppose that up to recursive level $i-1$, all sub-codes have been successfully sparsified to their desired size. At the $i$th level of recursion, there are at most $2^{i-1}$ codes which are being probabilistically sparsified. Each of these does not exceed its expected size by more than the prescribed amount with probability at most $2^{-\eta n}$. Hence, the probability all codes will be successfully sparsified up to and including the $i$th level of recursion is at least $1 - 2^{i-1} 2^{-\eta n} - 2^{i-1}2^{-\eta n} = 1 - 2^i 2^{-\eta n}$. 
\end{proof}

\begin{lemma}\label{clm:goodApprox1IterAppendix}
For any iteration of Algorithm \ref{alg:CodeSparsifyAppendix} called on a code $\calC$, $\calC_1 \cup \sqrt{d} \cdot \calC_2$ is a $(1 \pm \eps)$ sparsifier to $\calC$ with probability at least $1 - 2^{-(0.19\eta - 110) \log n} \cdot n^{-101}$.
\end{lemma}

\begin{proof}
First, let us note that the set $T$ returned from \cref{alg:CodeDecompositionAppendix} is a superset of the bad set $S$ of rows guaranteed by \cref{cor:moduleKargerAbelian} (this follows from \cref{clm:containsBadSetAppendix}). Thus, we can equivalently view the procedure as producing three codes: $\calC|_S, \calC|_{T / S}$ and $\calC_{\bar{T}} = \calC'$. For our analysis, we will view this procedure in a slightly different light: we will imagine that first the algorithm removes exactly the bad set of rows $S$, yielding $\calC|_S$ and $\calC_{\bar{S}}$. Now, for this second code, $\calC_{\bar{S}}$, we know the code-word counting bound will hold, and in particular, random sampling procedure will preserve codeword weights with high probability. However, our procedure is \emph{not} uniformly sampling the coordinates in $\calC_{\bar{S}}$, because some of these coordinates are in $T / S$, and thus are preserved exactly (i.e. with probability $1$). For this, we will take advantage of the fact that preserving coordinates with probability $1$ is \emph{strictly better} than sampling at any rate $<1$. Thus, we will still be able to argue that the ultimate result $\calC_1 \cup \sqrt{d} \cdot \calC_2$ is a $(1 \pm \eps)$ sparsifier to $\calC$ with high probability. 

    As mentioned above, we start by noting that $\calC', \calC|_S, \calC_{T / S}$ form a \emph{vertical} decomposition of $\calC$. $\calC|_S$ is preserved exactly, so we do not need to argue concentration of the codewords on these coordinates. Hence, it suffices to show that $\sqrt{d} \cdot \calC_1 \cup \calC_{T / S}$ is a $(1 \pm \eps)$-sparsifier to $\calC' \cup \calC_{T / S}$. 

    To see that $\sqrt{d} \cdot \calC_1 \cup \calC_{T / S}$ is a $(1 \pm \eps)$-sparsifier to $\calC' \cup \calC_{T / S}$, first note that every codeword in $\calC' \cup \calC_{T / S}$ is of weight at least $\sqrt{d} \cdot \eta \cdot \log(n) \log(q) / \eps^2$. This is because if there were a codeword of weight smaller than this, there would exist a subcode of $\calC' \cup \calC_{T / S}$ with $2$ distinct codewords, and support bounded by $\sqrt{d} \cdot \eta \cdot \log(n) \log(q) / \eps^2$. But, because we have removed the set $S$ of bad rows, we know that there can be no such sub-code remaining in $\calC' \cup \calC_{T / S}$. Thus, every codeword in $\calC' \cup \calC_{T / S}$ is of weight at least $\sqrt{d} \cdot \eta \cdot \log(n) \log(q) / \eps^2$. 

    Now, we can invoke Claim \ref{clm:preserveWeightSubsampleAppendix} with $b = \sqrt{d} \eta \log(n) \log(q) / \eps^2$. Note that the hypothesis of Claim \ref{clm:preserveWeightSubsampleAppendix} is satisfied by virtue of our code decomposition. Indeed, we removed coordinates of the code such that in the resulting $\calC' \cup \calC_{T / S}$, for any $\alpha \geq 1$, there are at most $(qn)^{2\alpha}$ codewords of weight $\leq \alpha \sqrt{d} \eta \log(n) \log(q) / \eps^2$. Using the concentration bound of Claim \ref{clm:preserveWeightSubsampleAppendix} yields that with probability at least $1 - 2^{-(0.19\eta - 110) \log n} \cdot n^{-101}$, when samplin every coordinate at rate $\geq 1/ \sqrt{d}$ the resulting sparsifier for $\calC' \cup \calC_{T / S}$ is a $(1 \pm \eps)$ sparsifier, as we desire. Note that we are using the fact that every coordinate is sampled with probability $\geq 1/ \sqrt{d}$ (in particular, those in $T -S$ are sampled with probability $1$).
    
\end{proof}

\begin{corollary}\label{clm:overallCodeAccuracyAppendix}
    If Algorithm \ref{alg:CodeSparsifyAppendix} achieves maximum recursion depth $\ell$ when called on a code $\calC$, and $\eta > 600$, then the result of the algorithm is a $(1 \pm \eps)^{\ell}$ sparsifier to $\calC$ with probability $\geq 1 - (2^{\ell}-1) \cdot2^{-(0.19\eta - 110) \log n} \cdot n^{-101}$
\end{corollary}

\begin{proof}
    We prove the claim inductively. Clearly, if the maximum recursion depth reached by the algorithm is $0$, then we have simply returned the code itself. This is by definition a $(1 \pm \eps)^{0}$ sparsifier to itself.

    Now, suppose the claim holds for maximum recursion depth $i - 1$. We will show it holds for maximum recursion depth $i$. Let the code we are sparsifying be $\calC$. We break this into $\calC_1$, $\calC'$, and sparsify these. By our inductive claim, with probability $1 - (2^{i-1}-1) \cdot 2^{-(0.19\eta - 110) \log n} \cdot n^{-101}$ each of the sparsifiers for $\calC_1, \calC'$ are $(1 \pm \eps)^{i-1}$ sparsifiers. Now, by Lemma \ref{clm:goodApprox1IterAppendix} and our value of $\eta$, $\calC_1, \calC'$ themselves together form a $(1 \pm \eps)$ sparsifier for $\calC$ with probability $1 - 2^{-(0.19\eta - 110) \log n}\cdot n^{-101}$. So, by using Claim \ref{clm:composingApproximations}, we can conclude that with probability $1 - (2^i -1) \cdot 2^{-(0.19\eta - 110) \log n} \cdot n^{-101}$, the result of sparsifying $\calC_1, \calC'$ forms a $(1 \pm \eps)^i$ approximation to $\calC$, as we desire.
\end{proof}

We can then state the main theorem from this section:

\begin{theorem}\label{thm:codeSparsifyGeneralLengthAppendix}
    For a code $\calC$ over $(\Z_{q_1} \times \Z_{q_2} \dots \Z_{q_u})$ with at most $q^n$ distinct codewords, and length $m$, Algorithm \ref{alg:CodeSparsifyAppendix} creates a $(1 \pm \eps)$ sparsifier for $\calC$ with probability $1 - \log (m) \cdot 2^{-(0.19\eta - 110) \log n} \cdot n^{-100}$ with at most \[
    O(n \eta \log(n) \log^2(q) \log^2(m) (\log \log (m))^2 / \eps^2)
    \]
    coordinates.
\end{theorem}

\begin{proof}
    For a code of with $q^n$ distinct codewords, and length $m$, this means that our value of $d$ as specified in the first call to Algorithm \ref{alg:CodeSparsifyAppendix} is at most $m$ as well. As a result, after only $\log \log m$ iterations, $d = m^{1 / 2^{\log \log m}} = m^{1 / \log m} = O(1)$. So, by Corollary \ref{clm:overallCodeAccuracyAppendix}, because the maximum recursion depth is only $\log \log m$, it follows that with probability at least $1 - (2^{\log \log m} -1) \cdot 2^{-(0.19\eta - 110) \log n} \cdot n^{-101}$, the returned result from Algorithm \ref{alg:CodeSparsifyAppendix} is a $(1 \pm \eps)^{\log \log m}$ sparsifier for $\calC$.

    Now, by Lemma $\ref{clm:overallSpaceAppendix}$, with probability $\geq 1 - 2^{\log \log m} \cdot 2^{-\eta n} \geq 1 - \log (m) \cdot 2^{-(0.19\eta - 110) \log n} \cdot  2^{-n}$, every code at recursive depth $\log \log m$ has at most 
    \[
    (1 + 1 / 2 \log \log (n))^{\log\log m} \cdot m^{1/\log m} \cdot \eta \cdot n \log(n)\log(q) / \eps^2 = O(n \eta \log (n) \log^2(q) \cdot e^{\frac{\log \log m}{\log \log n}} / \eps^2)
    \]
    coordinates. Because the ultimate result from calling our sparsification procedure is the \emph{union} of all of the leaves of the recursive tree, the returned result has size at most 
    \[
    \log (m)  \cdot e^{\frac{\log \log m}{\log \log n}} \cdot O(n \eta \log (n)\log^2(q) / \eps^2) = O(n \eta \log(n) \log^2(q) \log^2(m) / \eps^2),
    \]
    with probability at least $1 - \log (m) \cdot 2^{-(0.19\eta - 110) \log n} \cdot n^{-101}$. 

    Finally, note that we can replace $\eps$ with a value $\eps' = \eps / 2 \log \log m$. Thus, the resulting sparsifier will be a $(1 \pm \eps')^{\log \log m} \leq (1 \pm \eps)$ sparsifier, with the same high probability.

    Taking the union bound of our errors, we can conclude that with probability $1 - \log (m) \cdot 2^{-(0.19\eta - 110) \log n} \cdot n^{-100}$, Algorithm \ref{alg:CodeSparsifyAppendix} returns a $(1 \pm \eps)$ sparsifier for $\calC$ that has at most 
    \[
    O(n \eta \log(n)\log^2(q) \log^2(m) (\log \log (m))^2 / \eps^2)
    \]
    coordinates.
\end{proof}

However, as we will address in the next subsection, this result is not perfect: 
\begin{enumerate}
    \item For large enough $m$, there is no guarantee that this probability is $\geq 0$ unless $\eta$ depends on $m$.
    \item For large enough $m$, $\log^2(m)$ may even be larger than $n$.
\end{enumerate}

\subsection{Final Algorithm}

Finally, we state our final algorithm in Algorithm \ref{alg:FinalCodeSparsifyAppendix}, which will create a $(1 \pm \eps)$ sparsifier for any code $\calC \subseteq (\Z_{q_1} \times \Z_{q_2} \dots \Z_{q_u})^m$ with $\leq q^n$ distinct codewords preserving only $\widetilde{O}(n \log^2(q) / \eps^2)$ coordinates. Roughly speaking, we start with a weighted code of arbitrary length, use the weight class decomposition technique, sparsify the decomposed pieces of the code, and then repeat this procedure now that the code will have a polynomial length. Ultimately, this will lead to the near-linear size complexity that we desire. We write a single iteration of this procedure below:

\begin{algorithm}
    \caption{FinalCodeSparsify$(\calC, \eps)$}\label{alg:FinalCodeSparsifyAppendix}
    Let $n$ be $\log_q(|\calC|)$. \\
    Let $m$ be the length of the code. \\
    Let $\alpha = (m/\eps)^3$, and $\Dodd, \Deven = $WeightClassDecomposition$(\calC, \eps, \alpha)$. \\
    Let $S_{\text{even}}, \{ H_{\text{even}, i} \} = $SpanDecomposition$(\Deven, \alpha)$. \\
    Let $S_{\text{odd}}, \{ H_{\text{odd}, i} \} = $SpanDecomposition$(\Dodd, \alpha)$. \\
    \For{$i \in S_{\text{even}}$}{
    Let $\widehat{H}_{\text{even}, i}, w_{\text{even}, i}=$ MakeUnweighted$({H}_{\text{even}, i}, \alpha, i, \eps/8)$. \\
    Let $\widetilde{H}_{\text{even}, i} = $CodeSparsify$(\widehat{H}_{\text{even}, i}, \log_q(|\mathrm{Span}(\widehat{H}_{\text{even}, i})|), \eps/80, 100 (\log(m / \eps) \log \log(q))^2)$.
    }
    \For{$i \in S_{\text{odd}}$}{
    Let $\widehat{H}_{\text{odd}, i}, w_{\text{odd}, i}=$ MakeUnweighted$({H}_{\text{odd}, i}, \alpha, i, \eps/8)$. \\
    Let $\widetilde{H}_{\text{odd}, i} = $CodeSparsify$(\widehat{H}_{\text{odd}, i}, \log_q(|\mathrm{Span}(\widehat{H}_{\text{odd}, i})|), \eps/80, 100 (\log(m/\eps) \log \log(q))^2)$.
    }
    \Return{$\bigcup_{i \in S_{\text{even}}} \left ( w_{\text{even}, i} \cdot \widetilde{H}_{\text{even}, i} \right ) \cup \bigcup_{i \in S_{\text{odd}}} \left ( w_{\text{odd}, i} \cdot \widetilde{H}_{\text{odd}, i} \right )$}
\end{algorithm}

First, we analyze the space complexity. WLOG we will prove statements only with respect to $\Deven$, as the proofs will be identical for $\Dodd$.

\begin{claim}\label{clm:singleCallSparseAppendix}
    Suppose we are calling Algorithm \ref{alg:FinalCodeSparsifyAppendix} on a code $\calC$ with $q^n$ distinct codewords. Let $\teveni = \log_q(\text{Span}((\widehat{H}_{\text{even}, i}))|)$ from each call to the for loop in line 5. 
    
    For each call $\widetilde{H}_{\text{even}, i} = $CodeSparsify$(\widehat{H}_{\text{even}, i}, \log_q(|\mathrm{Span}(\widehat{H}_{\text{even}, i})|), \eps/10, 100 (\log(n/\eps) \log \log(q))^2)$ in Algorithm \ref{alg:FinalCodeSparsifyAppendix}, the resulting sparsifier has
    \[
    O\left ( \teveni \log(\teveni) \log^4(m/\eps) \log^2(q) (\log \log (m/\eps) \log \log(q))^2 / \eps^2 \right )
    \]
    coordinates with probability at least $1 - \log(m/ \eps) \cdot 2^{-\Omega(\log^2(m / \eps) (\log \log (q))^2)}$.
\end{claim}

\begin{proof}
    We use several facts. First, we use Theorem \ref{thm:codeSparsifyGeneralLengthAppendix}. Note that the $m$ in the statement of Theorem \ref{thm:codeSparsifyGeneralLengthAppendix} is actually a $\text{poly}(m / \eps)$ because $\alpha = m^3 / \eps^3$, and we started with a weighted code of length $O(m)$. So, it follows that after using Algorithm \ref{alg:MakeUnweightedAppendix}, the support size is bounded by $O(m^4 / \eps^3)$. We've also added the fact that $\eta$ is no longer a constant, and instead carries $O((\log(m/\eps) \log \log(q))^2)$, and carried this through to the probability bound. 
\end{proof}

\begin{lemma}\label{clm:overallDevenSizeAppendix}
    In total, the combined number of coordinates over $i \in S_{\text{even}}$ of all of the $\widetilde{H}_{\text{even}, i}$ is at most $\widetilde{O}(n \log^4(m)\log^2(q) / \eps^2)$ with probability at least $1 - \log(m\log(q) / \eps) \cdot 2^{-\Omega(\log^2(m/ \eps) (\log \log (q))^2)}$.
\end{lemma}

\begin{proof}
    First, we use \cref{clm:conserveRankAppendix} to see that 
    \[
    \sum_{i \in S_{\text{even}}} \log_q( | \text{Span}(\widehat{H}_{\text{even}, i}) | ) \leq n.
    \]
    Thus, in total, the combined length (total number of coordinates preserved) of all the $\widetilde{H}_{\text{even}, i}$ is 
    \begin{align*}
        & \sum_{i \in S_{\text{even}}} \text{number of coordinates in }\widehat{H}_{\text{even}, i} \\
        &\leq \sum_{i \in S_{\text{even}}} O\left ( \teveni \log(\teveni) \log^4(m/\eps) \log^2(q) (\log \log (m/\eps) \log \log(q))^2 / \eps^2 \right ) \\
        & \leq \sum_{i \in S_{\text{even}}} (\teveni) \cdot \widetilde{O} \left (\log^4(m)  \log^2(q) / \eps^2 \right ) \\
        & = n \cdot \widetilde{O} \left (\log^4(m)  \log^2(q) / \eps^2 \right ) \\
        & = \widetilde{O}(n \log^4(m)\log^2(q) / \eps^2).
    \end{align*}

    To see the probability bound, we simply take the union bound over all at most $n$ distinct $\widetilde{H}_{\text{even}, i}$, and invoke Claim \ref{clm:singleCallSparseAppendix}.
\end{proof}

Now, we will prove that we also get a $(1 \pm \eps)$ sparsifier for $\Deven$ when we run Algorithm \ref{alg:FinalCodeSparsifyAppendix}.

\begin{lemma}\label{clm:DevenCorrectnessAppendix}
    After combining the $\widehat{H}_{\text{even}, i}$ from Lines 5-8 in Algorithm \ref{alg:FinalCodeSparsifyAppendix}, the result is a $(1 \pm \eps/4)$-sparsifier for $\Deven$ with probability at least $1 - \log(m\log(q) / \eps) \cdot 2^{-\Omega(\log^2(m/ \eps) (\log \log (q))^2)}$.
\end{lemma}

\begin{proof}
    
    We use Lemma \ref{clm:OnlyApproxHiAppendix}, which states that to sparsify $\Deven$ to a factor $(1 \pm \eps/4)$, it suffices to sparsify each of the $H_{\text{even}, i}$ to a factor $(1 \pm \eps / 8)$, and then combine the results.

    Then, we use Lemma \ref{clm:PreserveApproxUnweightedAppendix}, which states that to sparsify any $H_{\text{even}, i}$ to a factor $(1 \pm \eps / 8)$, it suffices to sparsify $\widehat{H}_{\text{even}, i}$ to a factor $(1 \pm \eps / 80)$, where again, $\widehat{H}_{\text{even}, i}$ is the result of calling Algorithm \ref{alg:MakeUnweightedAppendix}. Then, we must multiply $\widehat{H}_{\text{even}, i}$ by a factor $\alpha^{i-1} \cdot \eps / 10$.    
    
    Finally, the resulting code $\widehat{H}_{\text{even}, i}$ is now an unweighted code, whose length is bounded by $O(m^4 /\eps^3)$, with at most $q^\teveni$ distinct codewords. The accuracy of the sparsifier then follows from Theorem \ref{thm:codeSparsifyGeneralLengthAppendix} called with parameter $\eps / 80$.

    The failure probability follows from noting that we take the union bound over at most $n\log(q)$ $H_{\text{even}, i}$. By Theorem \ref{thm:codeSparsifyGeneralLengthAppendix}, our choice of $\eta$, and the bound on the length of the support being $O(m^4 / \eps^3)$, the probability bound follows.
\end{proof}

For Theorem \ref{thm:codeSparsifyGeneralLengthAppendix}, the failure probability is characterized in terms of the number of distinct codewords of the code that is being sparsified. However, when we call Algorithm \ref{alg:CodeSparsifyAppendix} as a sub-routine in Algorithm \ref{alg:FinalCodeSparsifyAppendix}, we have no guarantee that the number of distinct codewords is $\omega(q)$. Indeed, it is certainly possible that the decomposition in $H_i$ creates $n$ different codes, each with $q$ distinct codewords in their span. Then, choosing $\eta$ to only be a constant, as stated in Theorem \ref{thm:codeSparsifyGeneralLengthAppendix}, the failure probability could be constant, and taking the union bound over $n$ choices, we might not get anything meaningful. To amend this, instead of treating $\eta$ as a constant in Algorithm \ref{alg:CodeSparsifyAppendix}, we set $\eta = 100 (\log(m/\eps) \log \log(q))^2$, where now $m$ is the length of the original code $\calC$, \emph{not} in the current code that is being sparsified $H_i$. With this modification, we can then attain our desired probability bounds.

\begin{theorem}\label{thm:mainRestatedAppendix}
For any code $\calC$ with $q^n$ distinct codewords and length $m$ over $(\Z_{q_1} \times \Z_{q_2} \dots \Z_{q_u})$, Algorithm \ref{alg:FinalCodeSparsifyAppendix} returns a $(1 \pm \eps)$ sparsifier to $\calC$ with $\widetilde{O}(n \log^4(m)\log^2(q) / \eps^2)$ coordinates with probability $\geq 1 - 2^{-\Omega((\log(m/\eps) \log \log(q))^2)}$.
\end{theorem}

\begin{proof}
First, we use Lemma \ref{clm:PreserveApproxWeightDecompAppendix}. This Lemma states that in order to get a $(1 \pm \eps)$ sparsifier to a code $\calC$, it suffices to get a $(1 \pm \eps/4)$ sparsifier to each of $\Deven, \Dodd$, and then combine the results.

    Then, we invoke Lemma \ref{clm:DevenCorrectnessAppendix} to conclude that with probability $\geq 1 - 2^{-\Omega((\log(m/\eps) \log \log(q))^2)}$, Algorithm \ref{alg:FinalCodeSparsify} will produce $(1 \pm \eps/4)$ sparsifiers for $\Deven, \Dodd$.

    Further, to argue the sparsity of the algorithm, we use Lemma \ref{clm:overallDevenSizeAppendix}. This states that with probability $\geq 1 - 2^{-\Omega( (\log(m/\eps) \log \log(q))^2)}$, Algorithm \ref{alg:FinalCodeSparsifyAppendix} will produce code sparsifiers of size $\widetilde{O}(n \log^4(m) \log^2(q) / \eps^2)$ for $\Deven, \Dodd$. 

    Thus, in total, the failure probability is at most $2^{-\Omega((\log(m/\eps) \log \log(q))^2)}$, the total size of the returned code sparsifier is at most $\widetilde{O}(n \log^4(m)\log^2(q) / \eps^2)$, and the returned code is indeed a $(1 \pm \eps)$ sparsifier for $\calC$, as we desire.

    Note that the returned sparsifier may have some duplicate coordinates because of Algorithm \ref{alg:MakeUnweightedAppendix}. Even when counting duplicates of the same coordinate separately, the size of the sparsifier will be at most $\widetilde{O}(n \log^4(m) \log^2(q) / \eps^2)$. We can remove duplicates of coordinates by adding their weights to a single copy of the coordinate.
\end{proof}

\begin{claim}\label{clm:finalSparsifyRuntimeAppendix}
    Running \cref{alg:FinalCodeSparsifyAppendix} on a code of length $m$ with parameter $\eps$ takes time $\text{poly}(mn\log(q) / \eps)$.
\end{claim}

\begin{proof}
    Let us consider the constituent algorithms that are invoked during the execution of \cref{alg:FinalCodeSparsifyAppendix}. First, we consider weight class decomposition. This groups rows together by weight (which takes time $\widetilde{O}(\log(m))$). Next, we invoke SpanDecomposition, which then contracts on the rows in the largest weight class. Note that in the worst case, we perform $O(n \log(q))$ contractions, as each contraction reduces the number of codewords by a factor of $\geq 2$. Further, each contraction takes time $O(mn\log(q))$ as the total number of rows is bounded by $m$, and there are at most $n\log(q)$ columns in the generating matrix. Thus, the total runtime of this step is $\widetilde{O}(m n^2 \log^2(q))$.

    The next step is to invoke the algorithm MakeUnweighted. Because the value of $\alpha$ is $m^3 / \eps^3$, this takes time at most $O(m^4 / \eps^3)$ to create the new code with this many rows. 
    
    Finally, we invoke CodeSparsify on a code of length $\leq m^4 / \eps^3$ and with at most $q^{\teveni}$ distinct codewords. Note that there are $\polylog(m)$ nodes in the recursive tree that is built by CodeSparsify. Each such node requires removing the set $T$ which is a set of $\leq \widetilde{O}(\sqrt{m^4 / \eps^3}) = \widetilde{O}(m^2 / \eps^{1.5})$ maximum spanning subsets. Constructing each such subset (by \cref{clm:runtimeContraction}) takes time at most $O((m^4 \eps^3)^2 n \log(q))$. After this step, the subsequent random sampling is efficiently doable. Thus, the total runtime is bounded by $\text{poly}(mn\log(q) / \eps)$, as we desire.

\end{proof}

Note that creating codes of linear-size now simply requires invoking \cref{alg:FinalCodeSparsifyAppendix} two times (each with parameter $\eps/2$). Indeed, because the length of the code to begin with is $\leq q^n$, this means that after the first invocation, the resulting $(1 \pm \eps/2)$ sparsifier $\calC'$ that we get maintains $\leq \widetilde{O}(n^5 \log^6(q) / \eps^2)$ coordinates. In the second invocation, we get a $(1 \pm \eps/2)$-sparsifier $\calC''$ for $\calC'$, whose length is bounded by $\widetilde{O}(n \log^4(n^5 \log^6(q) / \eps^2) \log^2(q) / \eps^2) = \widetilde{O}(n \log^2(q) / \eps^2)$, as we desire.

Formally, this algorithm can be written as:
\begin{algorithm}
    \caption{Sparsify($\calC, \eps$)}\label{alg:SparsifyAppendix}
    $\calC' = $FinalCodeSparsify$(\calC, \eps/2)$. \\
    \Return{FinalCodeSparsify($\calC', \eps/2$)}
\end{algorithm}

\begin{theorem}
    \cref{alg:CodeSparsifyAppendix} returns a $(1 \pm \eps)$-sparsifier to $\calC$ of size $\widetilde{O}(n \log^2(q) / \eps^2)$ with probability $1 - 2^{-\Omega((\log(n/\eps) \log \log(q))^2)}$ in time $\text{poly}(mn\log(q) / \eps)$.
\end{theorem}

\begin{proof}
    Indeed, because the length of the code to begin with is $\leq q^n$, this means that after the first invocation, the resulting $(1 \pm \eps/2)$ sparsifier $\calC'$ that we get maintains $\leq \widetilde{O}(n^5 \log^6(q) / \eps^2)$ coordinates. In the second invocation, we get a $(1 \pm \eps/2)$-sparsifier $\calC''$ for $\calC'$, whose length is bounded by $\widetilde{O}(n \log^4(n^5 \log^6(q) / \eps^2) \log^2(q) / \eps^2) = \widetilde{O}(n \log^2(q) / \eps^2)$, as we desire. To see the probability bounds, note that $m \geq n$, and thus both processes invocations of FinalCodeSparsify succeed with probability $1 - 2^{-\Omega((\log(n/\eps) \log \log(q))^2)}$.

    Finally, to see that the algorithm is efficient, we simply invoke \cref{clm:finalSparsifyRuntimeAppendix} for each time we fun the algorithm. Thus, we get our desired bound. 
\end{proof}

\section{Sparsifying Binary Predicates over General Alphabets}\label{sec:GeneralAlphabet}

In this section, we show how to rederive the result of \cite{BZ20} using our framework. To do this, we first introduce a more general definition of an affine predicate.

\begin{definition}
    For a predicate $P: \Sigma^r \ra \zo$, we say that $P$ is an affine predicate if there exists a group $A$, $a_i, b \in A$, and $T_1, \dots T_r: \Sigma \ra \Z$ such that $\forall x \in \Sigma^r$
    \[
    P(x_1, \dots x_r) = 1 \iff \sum_{i = 1}^r T_i(x_i) a_i \neq b.
    \]
If the group $A$ is Abelian, then we say that $P$ is an {\em affine Abelian} predicate.
\end{definition}

Note that the purpose of the functions $T_i$ is to embed the alphabet into the integers. These functions \emph{do not} have to be linear in order to be sparsifiable. As an immediate consequence of \cref{thm:affine-abelian}, we can sparsify general alphabet, affine predicate $P$ over Abelian groups. 

\begin{theorem}\label{thm:affine-abelian-general}
    For any alphabet $\Sigma$, an affine Abelian predicate $P$ over Abelian group $A$, any instance $\Phi \in \CSP(P)$ on $n$ variables can be $(\eps, \widetilde{O}(n \log^2(|\Sigma|) / \eps^2))$-sparsified.
\end{theorem}

\begin{proof}[Proof sketch]
    By definition, there exists $a_i, b \in A$, $T_1, \dots T_r: \Sigma \ra \Z$ such that $\forall x \in \Sigma^r$
    \[
    P(x_1, \dots x_r) = 1 \iff \sum_{i = 1}^r T_i(x_i) a_i \neq b.
    \]

Now, let us create a generating matrix $G$ in $A^{m \times (n+1)}$ in the canonical way. Indeed, for each of the $[m]$ constraints, let the variables that the predicate is operating on be $x^{(1)}_j, \dots x^{(r)}_j$. In the $j$th row of the generating matrix, for the column corresponding to $x^{(i)}_j$, place the coefficient $a_i$. Finally, in the final column (the $n+1$st column), place the value $b$. It follows that for any linear combination of the columns $x \in \Z^n \circ 1$, the resulting codeword $Gx$ will have weight equal to the weight of the satisfied constraints on assignment $x$. Then, by sparsifying the code defined by $G$, this yields a sparsifier for our CSP instance $\Phi$. We conclude then by invoking \cref{thm:affine-abelian}. 

Note that even though the functions $T_i$ may not be linear, they still correspond to linear combinations of the columns of this generating matrix. Our sparsifier works for \emph{any} linear combination of the columns of the generating matrix, even if the coefficients $T_i(x_i)$ are calculated in a non-linear way.
\end{proof}

With this, we are able to explain our proof strategy. Indeed, given a general binary predicate $P: \Sigma^2 \ra \zo$, we view the predicate as a matrix in $\zo^{|\Sigma| \times |\Sigma|}$. For this matrix, the work of \cite{BZ20} showed that the predicate $P$ is sparsifiable if and only if there is no rectangle with corners forming an AND: i.e., a rectangle with corners having $3$ $0$'s and a single $1$. When this is the case, \cite{BZ20} showed that this yields an AND of arity $2$, and hence requires sparsifiers of quadratic size. On the other hand, when none of these rectangles are of this form, we can indeed sparsify the predicate.

\begin{lemma}
    Suppose a predicate $P: \Sigma^2 \ra \zo$ does not have a projection to AND. Then, any instance $\Phi \in \CSP(P)$ on $n$ variables admits an $(\eps, \widetilde{O}_{|\Sigma|}(n / \eps^2))$ sparsifier.
\end{lemma}

\begin{proof}
    As pointed out by Butti and \zivny~\cite{BZ20}, we note that if $P$ does not have a projection to AND then there must exist $t$ disjoint sets $S_1,\ldots,S_t \subseteq \Sigma$ and another family of $t$ disjoint sets $U_1,\ldots,U_t \subseteq \Sigma$ such that $P(a,b) = 1$ if and only if $(a,b) \not\in \cup_{i=1}^t S_i \times U_i$. We claim that this implies $P$ is an affine predicate. 

    We work over the group $A = \Z_{t+2}$. Next, we define the functions $T_1, T_2$ in our affine, Abelian predicate. We let $T_1,T_2: \Sigma \ra \Z$ be defined as $T_1(\sigma) = i$ if $\sigma \in S_i$, and $0$ otherwise. Likewise, we define $T_2(\sigma) = i$ if $\sigma \in U_i$, and $t+1$ otherwise. Further we set $a_1 = 1$, $a_2 = -1$ and $b=0$. This leads to the affine Abelian predicate $P'$ given by: 
    \[
    P'(x_1, x_2) = \mathbf{1}[T_1(x_1) - T_2(x_2) \neq 0 \mod t+2].
    \]

    One can verify that $\forall (x_1, x_2) \in \Sigma^2, P'(x_1, x_2) = P(x_1, x_2)$. Thus, by \cref{thm:affine-abelian-general}, we conclude the existence of an $(\eps, \widetilde{O}(n \log^2(|\Sigma|) / \eps^2))$ sparsifier.

\end{proof}

\section{Sparsifying Affine Predicates over Larger Alphabets}\label{sec:largerAlphabetAffine}

\subsection{More General Notions}

In a more general way, we can say:
\begin{definition}
    For a predicate $P: \Sigma^r \ra \zo$, we say that $P$ is a general affine predicate if there exists a constant $c$, a group $A$, $a_i, b \in A^c$, and $T_1, \dots T_r: \Sigma \ra \Z^c$ such that $\forall x \in \Sigma^r$
    \[
    P(x_1, \dots x_r) = 1 \iff \sum_{i = 1}^r \langle T_i(x_i), a_i \rangle \neq b.
    \]
If the group $A$ is Abelian, then we say that $P$ is a general {\em affine Abelian} predicate.
\end{definition}

Note that by the equivalence with the generating matrix, this will still be sparsifiable, as we can simply add $c$ columns for each variable. 

\subsection{Infinite Abelian Groups}

Note that we can extend our proof technique to sparsify affine abelian predicates that are defined over larger alphabets too.

Suppose that a predicate is affine only over an infinite Abelian group. I.e. $P(x) = \mathbf{1}[\sum_{i = 1}^r a_i x_i \neq b]$, for $a_i, b \in A$, and $A$ being an infinite group. First, we note that the relevant elements of $A$ will be finitely generated. I.e., the elements we care about $a_i, b \in H$, where $H$ is a subgroup of $A$ generated by $a_i,b$. This is because we only care about the subgroup generated by $a_1, \dots a_r, b$, and can effectively ignore everything else. We rewrite the predicate $P(x) = \mathbf{1}[\sum_{i = 1}^r a_i x_i -b \neq 0]$. Note that WLOG we can consider the case when $b = 0$, as otherwise we can simply make the predicate of arity $r+1$, and treat $b$ as the coefficient to another variable. 

Now, because the subgroup $H$ we care about is finitely generated, we can invoke the fundamental theorem of Abelian groups. This tells us that $H$ is isomorphic to a group of the form 
\[
\Z^{k_0} \times \Z_{p_1}^{k_1} \times \dots \times \Z_{p_{\ell}}^{k_{\ell}}.
\]

Let the new size $w = \sum_{j = 0}^{\ell} k_0$. This means that we can write $P$ as a predicate over a tuple of length $w$, where the $i$th entry in the tuple will be a constraint over one of the cyclic groups in the product. I.e., we can write 
\[
P(x) = \mathbf{1}[\sum_{i = 1}^r (a^{(i)}_1, \dots a^{(i)}_w)x_i].
\]

Now, note that this predicate is essentially taking the OR over each entry in the tuple. That is,
\[
P(x) = \mathbf{1}[\sum_{i = 1}^r a^{(i)}_1 x_i \neq 0] \vee \dots \vee \mathbf{1}[\sum_{i = 1}^r a^{(w)}_1 x_i \neq 0].
\]
Without loss of generality, we can then assume that $w \leq 2^{2^r}$. This is because the number of possible predicates on $r$ variables is at most $2^{2^r}$, so it follows that if we are taking an OR, there can be at most $2^{2^r}$ distinct predicates before we are forced to have repeating predicates. If any predicates are repeating, they are not contributing to the OR.

Now, we know that $P(x)$ can be written as the OR of at most $2^{2^r}$ predicates, where each predicate $P^{(i)}$ is of the form 
\[
P^{(i)} = \mathbf{1}[ \sum_{i = 1}^r a^{(i)}_1 x_i \neq 0],
\]
and algebra is done over \emph{some} cyclic group. 

It suffices for us to argue that we can represent each of these predicates $P^{(i)}$ over a smaller cyclic group $\Z_{p^*}$, where $p^*$ depends only on $r$. Indeed, to see this, consider any $P^{(i)}$, where the algebra is done over a finite cyclic group. We know that there is some set of assignments $S \subseteq \zo^r$ such that $P^{(i)}$ is $1$ for the assignments in $S$, and $0$ for the assignments outside $S$. We also know that there exists a constant $p$ such that $P^{(i)}$ is expressible as a linear equation of $\Z_p$. Now, let $p'$ be the smallest value of $p$ such that $P^{(i)}$ is expressible as a linear equation over $\Z_{p'}$. Then $p'$ must be bounded by a function of $r$. Next, consider the case when the algebra for $P^{(i)}$ is done over $\Z$. This means 
\[
P^{(i)} = \mathbf{1}[\sum_{i = 1}^r a_i x_i \neq 0].
\]
In particular, because we know $P^{(i)}$ is expressible over $\Z$, there must exist some choice of $a_1, \dots a_r$ such that $\max(|a_1|, \dots |a_r|) = a$ is as small as possible. Again, for this choice, $a$ must be bounded by some function of $r$ (denote this $f(r)$). Now, it follows that we can simply choose $p'$ to be the smallest prime larger than $2 r \cdot f(r)$. For this choice of $p'$, doing $\zo$ weighted arithmetic with coefficients $a_1, \dots a_r$ will be the same as doing arithmetic of $\Z_{p'}$. Further $p'$ will be bounded by a function of $r$. Thus, we have that $P$ is expressible as the OR of at most $2^{2^r}$ affine functions, each of which is doing arithmetic over a cyclic group bounded in size as a function of $r$. In particular, for $r$ constant, this leads to no overhead in the size of our sparsifiers.

\subsection{Lattice Perspective}

\begin{lemma}
    Suppose $B$ is a $d \times \ell$ matrix with entries in $\Z$, such that the magnitude of the largest sub-determinant is bounded by $M$, and $\rank(B) = k$. Then, every element of the lattice generated by the columns of $B$ is given exactly by the solutions to $d-k$ linear equations and $k$ modular equations. All coefficients of the linear equations are bounded in magnitude by $M$, and all modular equations are written modulo a single $M' \leq M$.
\end{lemma}

\begin{proof}
    First, if $\rank(B) = k < d$, this means that there exist $k$ linearly independent rows such that the remaining $d-k$ rows are linear combinations of these rows. Let us remove these rows for now, and focus on $B' \in \Z^{k \times \ell}$ where now the matrix has full row-rank. 

    It follows that for this matrix $B'$ we can create a new matrix $\hat{B}$ such that the lattice generated by $B'$ (denoted by $\calL(B') = \calL(\hat{B})$) and $B'$ is in Hermite Normal Form (HNF). In this form, $\hat{B}$ is lower triangular with the diagonal entries of $\hat{B}$ satisfying 
    \[
    \det(\hat{B}) = \prod_{i = 1}^k \hat{B}_{i,i} \leq \max_{k \times k \text{ subrectangle A}}\det(B'_A).
    \]
Further, all columns beyond the $k$th column will be all zeros, so we can remove these from the matrix.
    
    We can now define the \emph{dual} lattice to $\calL(B') = \calL(\hat{B}))$. For a lattice $\Lambda \subseteq \Z^k$, we say that 
    \[
    \dual(\Lambda) = \{x \in  \Q^k: \forall y \in \Lambda, \langle x, y \rangle \in Z\}.
    \]
    
    Here it is known that the dual is an exact characterization of the lattice $\Lambda$. I.e., any vector in $\Lambda$ will have integer-valued inner product with \emph{any} vector in the dual, while for any vector not in $\Lambda$, there exists a vector in the dual such that the inner-product is not integer valued. 

    Now, for our matrix $\hat{B}$, it is known that one can express the dual lattice to $\hat{B}$ as $\hat{D} = \hat{B} (\hat{B}^T \hat{B})^{-1}$. As a result, it must be the case that $\hat{D} \subseteq \Z^k / \det(\hat{B})$. If a vector $x$ of length $k$ is not in $\calL(\hat{B})$, it must be the case that there exists a column $y$ of $\hat{D}$ such that $\langle x, y \rangle \notin \Z$. Otherwise, if the inner-product with every column is in $\Z$, it follows that for any vector in the dual, the inner product would also be in $\Z$, as we can express any vector in the dual as an integer linear combination of columns in $\hat{D}$. Thus, it follows that membership of a vector $x$ in $\calL(\hat{B})$ can be tested exactly by the $k$ equations $\forall i \in [k]: \langle x, \hat{D}_i \rangle \in \Z$. Now, because every entry of $\hat{D}$ has denominator dividing $\det(\hat{B})$, it follows that we can scale up the entire equation by $\det(\hat{B})$. Thus, an equivalent way to test if $x \in \calL(\hat{B})$ is by checking if $\forall i \in [k]: \langle x, \det(\hat{B}) \cdot \hat{D}_i \rangle = 0 \mod \det(\hat{B})$. Now, all the coefficients of these equation are integers, and we are testing whether the sum is $0$ modulo an integer. Thus, we can test membership of any $k$-dimensional integer vector in $\calL(\hat{B})$ with $k$ modular equations over $\det{\hat{B}} \leq M$.

    The above argument gives a precise way to characterize when the restriction of a $d$ dimensional vector to a set of coordinates corresponding with linearly independent rows in $B$ is contained in the lattice generated by these same rows of $B$. It remains to show that we can also characterize when the dependent coordinates (i.e. coordinates corresponding to the rows that are linearly dependent on these rows) are contained in the lattice. Roughly speaking, the difficulty here arises from the fact that we are operating with a non-full dimensional lattice. I.e., there exist directions that one can continue to travel in $\Z^n$ without ever seeing another lattice point. In this case, we do not expect to be able to represent membership in the lattice with a modular linear equation, as these modular linear equations rely on periodicity of the lattice. 
    
    Instead, here we rely on the fact that for any of the rows of $B$ that are linearly dependent, we know that there is a way to express it as a linear combination of the set of linearly independent rows. WLOG, we will assume the first $k$ rows $r_1, \dots r_k$ are linearly independent, and we are interested in finding $c_i$ such that $\sum_{i = 1}^k c_i r_i = r_{k+1}$. Now, consider any subset $A$ of $k$ linearly independent columns amongst these $k$ rows. We denote the corresponding restriction of the rows to these columns by $r_i^{(A)} \in \zo^k$. It follows that if we want a linear combination of these rows such that $\sum_{i = 1}^k c_i r_i^{(A)} = r_{k+1}^{(A)}$, we can express this a constraint of the form $M^{(A)} c = (r_{k+1}^{(A)})^T$, where we view the $i$th column of $M$ as being the (transpose of) $r_i^{(A)}$ and $c$ as being the vector of values $c_1, \dots c_k$. Using Cramer's rule, we can calculate that $c_i = \det(M^{(A)}_i) / \det(M^{(A)})$, where $M^{(A)}_i$ is defined to be the matrix $M^{(A)}$ with the $i$th column replaced by $(r_{k+1}^{(A)})^T$. In particular, this means that we can express $r_{k+1}^{(A)} = \sum_{i = 1}^k \det(M^{(A)}_i) / \det(M^{(A)}) \cdot r_i^{(A)}$, and because $A$ corresponds to a set of linearly independent columns, it must also be the case that $r_{k+1} = \sum_{i = 1}^k \det(M^{(A)}_i) / \det(M^{(A)}) \cdot r_i$. We can re-write this as an integer linear equation by expressing $r_{k+1} \cdot \det(M^{(A)}) = \sum_{i = 1}^k \det(M^{(A)}_i) \cdot r_i$. This means that for any valid vector $x \in \Z^d$ expressable as a linear combination of the columns of $B$, it must be the case that $x_{k+1} \cdot \det(M^{(A)}) = \sum_{i = 1}^k \det(M^{(A)}_i) \cdot x_i$/

    We can repeat the above argument for each of the $d-k$ linearly independent rows. Let $j$ denote the index of a row linearly dependent on the first $k$ rows. It follows that $x_j \cdot \det(M^{(A)}) = \sum_{i = 1}^k \det(M^{(A),j}_i) \cdot x_i$, where now $M^{(A),j}_i$ is the $d\times d$ matrix $M^{(A)}$ where the $i$th column has been replaced with $(r_j^{(A)})^T$. Note that every coefficient that appears in these equations above is of the form $\det(C)$ where $C$ is a $d \times d$ submatrix of $B$. It follows that each of these coefficients is bounded in magnitude by $M$, where $M$ is again defined to be the maximum magnitude of the determinant of any square sub-matrix. 

    To conclude, we argue that for any vector $x \in \calL(B)$, $x$ satisfies all of the above linear equations and modular equations. The first part of the proof showed that amongst the set of linearly independent rows $S$, the dual of the lattice exactly captures when $x_S$ is in the lattice generated by $B_S$. That is, if $x_S$ is generated by $B_S$, then $x_S$ satisfies the above modular linear equations, while if $x_S$ is not in the span of $B_S$, then $x_S$ does not satisfy the modular linear equations. Now, if $x_S$ does not satisfy the modular linear equations, this is already a witness to the fact that $x$ is not in the $\calL(B)$. But, if $x_S$ is in the span of $B_S$, then if $x$ is in the span of $B$, it must also be the case that the coordinates of $x_{\bar{S}}$ satisfy the exact same linear dependence on the $x_S$ that $B_{\bar{S}}$ has on $B_S$. This is captured by our second set of linear equations.
\end{proof}

\begin{theorem}
     Let $P: \Sigma^r \ra \zo$ be a predicate over an arbitrary alphabet of arity $r$. Let $S = P^{-1}(0) \subseteq \Sigma^r$ denote the unsatisfying assignments of $P$, and let $\hat{S} \subseteq \zo^{|\Sigma|r}$ denote the lifted version of $S$ where we map $\sigma \in \Sigma$ to a vector $v \in \zo^{|\Sigma|}$ such that $v_{\sigma} = 1$, and is $0$ otherwise. If $\hat{S}$ is closed under integer valued linear combinations, then CSPs with predicate $P$ on $n$ variables are sparsifiable to size $\widetilde{O}(n\cdot |\Sigma|^4\cdot r^4 / \eps^2)$.
\end{theorem}

\begin{proof}
Let us create a matrix $B \in \zo^{|\Sigma|r \times |\hat{S}|}$ where the $i$th column of $B$ is the $i$th element of $\hat{S}$. Let $k$ be the rank of $B$. It follows that for any assignment $x \in \zo^{|\Sigma|r}$, we can exactly express the membership of $x$ in $\calL(B)$ with $d$ modular linear equations, and $|\Sigma|r - d$ linear equations. I.e., $x$ is in $\calL(B)$ if and only if all of these equations are satisfied. Note that the $d$ modular linear equations are all over modulus $M \leq \max_{k \in [|\Sigma|r], k \times k \text{ subrectangle } A} \det(B'_A) \leq (|\Sigma|r)^{|\Sigma|r}$. Likewise, the integer linear equations also all have coefficients $\leq (|\Sigma|r)^{|\Sigma|r}$. It follows that because $x \in \zo^{|\Sigma|r}$, we can choose a prime $p$ such that $p \geq 2 \cdot |\Sigma|r \cdot (|\Sigma|r)^{|\Sigma|r}$. Now, for any of the integer linear equations of the form $c_1 x_1 + \dots c_k x_k - c_{k+1} x_{k+1}$, it will be the case that for $x \in \in \zo^{|\Sigma|r}$,
\[
c_1 x_1 + \dots c_k x_k - c_{k+1} x_{k+1} = 0 \iff c_1 x_1 + \dots c_k x_k - c_{k+1} x_{k+1} = 0 \mod p.
\]

This is because the expression on the left can never be as large as $p$ or $-p$ since we chose $p$ to be sufficiently large.

Thus, we can express $x \in \zo^{|\Sigma|r}$ as being in the lattice $\calL(B)$ if and only if all $|\Sigma|r$ modular equations are $0$. This is then the OR of $|\Sigma|r$ modular equations, which can be expressed over the Abelian group $A = \Z_M \times \Z_m \dots \times Z_m \times Z_p \dots \times \Z_p$, where there are $k$ copies of $Z_m$, and $|\Sigma|r$ copies of $\Z_p$. It follows that $|A| \leq \left ( 2 \cdot |\Sigma|r \cdot (|\Sigma|r)^{|\Sigma|r} \right )^{|\Sigma|r} = (2|\Sigma|r)^{|\Sigma|r} \cdot (|\Sigma|r)^{|\Sigma|^2 r^2}$. Thus, for any CSP on predicate $P$ of the above form on $n$ variables, we can $1 \pm \eps$ sparsify $P$ to size $\widetilde{O}(n\cdot |\Sigma|^4\cdot r^4 / \eps^2)$.
\end{proof}

\section{Non-Affine Predicates with no Projections to $\AND$}\label{sec:separatingAffineAND}

In this section, we present a predicate which provably is not affine, and yet still does not have any projection to an $\AND$ of arity $2$. This shows that there is necessarily a separation between the techniques we have for creating sparsifiers versus showing lower-bounds. We will do this by specifically constructing a predicate which is not affine (in particular, there will exist a linear combination of unsatisfying assignments which yields a \emph{satisfying} assignment).

Indeed, consider the following predicate $P: \zo^9 \ra \zo$:
\begin{enumerate}
    \item $P(000000000) = 0$. 
    \item $P(111111000) = 0$.
    \item $P(111000111) = 0$.
    \item $P(110001001) = 0$.
    \item $P(101010010) = 0$.
    \item For all other assignments $x$, $P(x) = 1$.
\end{enumerate}

In particular, note that the unsatisfying assignments of $P$ are \emph{not} closed under integer linear combinations. We can consider 
\[
1 \cdot (000000000) + 1 \cdot (111111000) + 1 \cdot (111000111) -1 \cdot (110001001) -1 \cdot (101010010) = (011100100).
\]

Thus, we immediately get the following claim:

\begin{claim}
    $P$ is not expressible as an affine predicate over \emph{any} Abelian group.
\end{claim}

\begin{proof}
    Any predicate $P$ which is expressible as an affine predicate over an Abelian group must have its unsatisfying assignments closed under integer linear combinations. Indeed, let $y_1, \dots y_{\ell} \in P^{-1}(0)$, and $\alpha_1, \dots \alpha_{\ell}$ be such that $\sum_{j = 1}^{\ell} \alpha_j y_j \in \zo^r$ (when addition is done over $\Z$). We claim then that $P(\sum_{j = 1}^{\ell} \alpha_j y_j) = 0$ also. 

    This follows simply from $P$ being an affine Abelian predicate. It must be the case that $P(b_1, \dots b_r) = \mathbf{1}[\sum_i a_i b_i \neq 0]$, for some $a_i \in A$, where $A$ is an Abelian group. Then, 
    \[
    P(\sum_{j = 1}^{\ell} \alpha_j y_j) = \mathbf{1}[\sum_i a_i (\sum_{j = 1}^{\ell} \alpha_j y_j)_i \neq 0] = \mathbf{1}[\sum_{j = 1}^{\ell} \alpha_j \sum_i a_i (y_j)_i \neq 0].
    \]
    But, because each $y_j \in P^{-1}(0)$, it must be the case that $\sum_i a_i (y_j)_i = 0$. Thus, the entire sum must be $0$, so we can conclude that $P(\sum_{j = 1}^{\ell} \alpha_j y_j) = 0$.
\end{proof}

It remains to prove that the above predicate has no projection to $\AND_2$.

\begin{claim}
    $P$ has no projection to $\AND_2$.
\end{claim}

\begin{proof}[Proof sketch]
    Note that if there is a projection to $\AND$ this implies there is some restriction of the variables $y_1, \dots y_9$ which yields $3$ unsatisfying assignments and $1$ satisfying assignment. One can verify that for any set of $3$ unsatisfying assignments $y_1, y_2, y_3$ for $P$, they can be captured exactly with an affine predicate (i.e., one can construct an affine predicate $\hat{P}$ which is unsatisfied if and only if the inputs are $z_1$, $z_2$, or $z_3$). But, any predicate which is affine does not have a projection to $\AND_2$. Thus, there is no projection of $\hat{P}$ which yields an $\AND_2$, and consequently no restriction of the variables $y_1, \dots y_9$ for which there are $\geq 3$ surviving unsatisfying assignments.
\end{proof}

\section{Symmetric Predicates with No $\AND_3$ and No Degree $2$ Polynomial}\label{sec:symmetricSeparationPolyAND}

Consider the predicate $P: \zo^r \ra \zo$ (for instance, with $r = 20$) such that $P(x) = 0$ if $|x| = 0 \mod 6$ or $|x| = 1 \mod 6$, and otherwise, $P(x) = 1$. Clearly, $P$ is a symmetric predicate, as $P$ depends only on the number of $1$'s in $x$. Next, we will show that $P$ does not have a projection to $\AND_3$.

\begin{claim}
    $P$ does not have a projection to $\AND_3$.
\end{claim}

\begin{proof}
    Recall that we define a projection as a fixing $\pi$ of the variables $x_1, \dots x_r$ such that each $x_i$ maps to $0, 1, Y_1, Y_2, Y_3$ (or the negation). In particular, in order to get an $\AND$ of arity $3$, it must be the case that $P(\pi(x_1), \dots \pi(x_r)) = \AND(Y_1, Y_2, Y_3)$. Let us consider any such restriction $\pi$, and suppose that it sends $\alpha_0$ of the $x_i$'s to $0$, $\alpha_1$ of the $x_i$'s to $1$, $\alpha_{Y_i}$ to $Y_i$, and $\alpha_{1- Y_i}$ to $1 - Y_i$.
    
    Also, recall that because $P$ is symmetric, we can equivalently create the predicate $P_0: [r] \ra \zo$, such that $P(x) = P_0(|x|)$. In terms of the $Y_i$'s then, we have that 
    \[
    P(\pi(x_1), \dots \pi(x_r)) = P_0(\alpha_0 \cdot 0 + \alpha_1 \cdot 1 + \alpha_{Y_1} \cdot Y_1 + \alpha_{1 - Y_1} (1 - Y_1) + \alpha_{Y_2} \cdot Y_2 + \alpha_{1 - Y_2} (1 - Y_2)+ \alpha_{Y_3} \cdot Y_3 + \alpha_{1 - Y_3} (1 - Y_3))
    \]
    \[
    = P_0(\alpha_1 + \alpha_{1 - Y_1} + \alpha_{1 - Y_2} + \alpha_{1 - Y_3} + (\alpha_{Y_1} - \alpha_{1 - Y_1}) \cdot Y_1 + (\alpha_{Y_2} - \alpha_{1 - Y_2}) \cdot Y_2 + (\alpha_{Y_3} - \alpha_{1 - Y_3}) \cdot Y_3)
    \]
    \[
    = P_0(a + bY_1 + cY_2 + dY_3),
    \]
    for some choice of constants $a,b,c,d$. WLOG let us assume that $a = 0$ or $1$, since the predicate is periodic. 
    
    Now, for $Y_1 = Y_2 = Y_3 = 0$, we know that the expression must evaluate to $0$, as $\AND(0,0,0) = 0$. Thus, $P_0(a) = 0$. Further, when exactly $1$ or exactly $2$ of $Y_1, Y_2, Y_3$ are $1$, the expression must also be $0$. Hence, $P_0(a + b) = P_0(a + c) = P_0(a + d) = P_0(a + b + c) = P_0(a + c + d) = P_0(a + b + d) = 0$, yet $P_0(a + b + c + d) = 1$.

    We consider cases, based on whether $a = 0 $ or $1$:
    \begin{enumerate}
        \item $a = 0$. Then $P_0(b) = P_0(c) = P_0(d) = P_0(b+c) = P_0(c+d) = P_0(b+d) = 0$. In particular, all of $b, c, d$ must be $0,1 \mod 6$. In fact, at most one of them can be $1 \mod 6$, as if $2$ are $1 \mod 6$, then their sum would be $2 \mod 6$, and $P_0$ would evaluate to $1$. But, if at most one of them is $1 \mod 6$, then their sum is also either $0,1 \mod 6$, and hence $P_0(b+c+d) = 0$, so it doesn't simulate $\AND_3$.
        \item $a = 1$. Then $P_0(1+b) = P_0(1+c) = P_0(1+d) = P_0(1+b+c) = P_0(1+c+d) = P_0(1+b+d) = 0$. In particular, all of $b, c, d$ must be $-1,0 \mod 6$. In fact, at most one of them can be $-1 \mod 6$, as if $2$ are $-1 \mod 6$ (say $b,c$), then $1+b+c = -1 \mod 6$, and $P_0$ would evaluate to $1$. But, if at most one of them is $-1 \mod 6$, then their sum is also either $0,1 \mod 6$, and hence $P_0(1+b+c+d) = 0$, so it doesn't simulate $\AND_3$.
    \end{enumerate}
    This concludes the proof. 
\end{proof}

Simultaneously, there is no canonical way to express $P$ as a polynomial. Indeed, what we would like to do is write $P$ as the product of two linear functions, one which is $0$ if and only if $x = 0 \mod 6$, and the other which is $0$ if and only if $x = 1 \mod 6$. That is, we would like to simply sparsify the predicate 
\[
P'(x) = \mathbf{1}[(|x|) \cdot (|x|-1) \neq 0 \mod 6].
\]

However, this predicate unfortunately \emph{does not} capture the behavior of $P$. For instance, when $|x| = 4$, we get that $P'(x) = 0$, whereas $P(x) = 1$.

We present this more formally below, due to \ifanon Anonymous\else  Swastik Kopparty\fi:

\begin{claim}\label{clm:symm-mod-6}
    There is no symmetric degree $2$ polynomial $f$ such that for $|x| = 0, 1 \mod 6$ $f(x) = 0$, and otherwise $f(x) = 1$.
\end{claim}

\begin{proof}
    Indeed, because $f$ is symmetric, $f$ depends only on the hamming weight of $x$. Because $f$ is degree $2$, $f$ takes the form $a \cdot \binom{|x|}{2} + b \cdot |x| + c$ (over some group). When $|x| = 0$, we know $f$ must be $0$, which means that $c = 0$. Likewise, when $|x| = 1$, $f$ must also be $0$, which means that $b = 0$. So, $f$ can only be a polynomial of the form $f(x) = a \cdot \binom{|x|}{2}$. We also have that for any $x$, then for any $y: |y| = |x| + 6$, $f(y) = f(x)$. Plugging this in yields that $f(y) = a \cdot \binom{|x|+6}{2} = a \cdot (\binom{|x|}{2} + 6|x| + 15) = a \cdot \binom{|x|}{2} = f(x)$. Thus, it must be the case that over any group we consider, $a \cdot (6|x| + 15) = 0$ (for all $|x|$), which immediately implies that $3a = 0$. By cases then, if we assume $2a = 0$ it must be that $a = 0$, and the polynomial is identically $0$, which does not work. Otherwise, the output of $f$ only has a period of $3$, as it depends only on whether $\binom{|x|}{2} \mod 3 = 0, 1, 2$, and thus fails to capture our predicate.
\end{proof}

To conclude, the predicate has a period of $6$ and therefore seemingly requires a period of $6$ (i.e., over $\Z_6$) in whichever polynomial we use to represent it. At the same time, because the period is composite, the polynomial over $\Z_6$ has extra zeros, meaning we do not accurately capture the behavior of the predicate $P$.
\end{document}